\documentclass[12pt]{iopart}

\usepackage[utf8]{inputenc}
\usepackage{amssymb,graphicx,hyperref}
\usepackage[varg]{txfonts}

\renewcommand{\d}{\mathrm{d}}
\newcommand{\ii}{\mathrm{i}}
\newcommand{\tensor}[3]{#1^{#2}_{\phantom{#2}#3}}
\newcommand{\cvector}[1]{\left(\begin{array}{c}#1\end{array}\right)}

\graphicspath{./newFigs/}

\begin{document}

\title{Gravitational Lensing}
\author{Matthias Bartelmann}
\address{Zentrum für Astronomie der Universität Heidelberg, Institut für Theoretische Astrophysik, Albert-Ueberle-Str.~2, 69120 Heidelberg, Germany}

\begin{abstract}

\noindent Gravitational lensing has developed into one of the most powerful tools for the analysis of the dark universe. This review summarises the theory of gravitational lensing, its main current applications and representative results achieved so far. It has two parts. In the first, starting from the equation of geodesic deviation, the equations of thin and extended gravitational lensing are derived. In the second, gravitational lensing by stars and planets, galaxies, galaxy clusters and large-scale structures is discussed and summarised.

\end{abstract}

\section{From general relativity to gravitational lensing}

\subsection{Preliminaries}

We assume throughout that a valid model for the geometry and the evolution of the universe is given by a Friedmann-Lemaître-Robertson-Walker model with cosmological parameters narrowly constrained by numerous cosmological observations. The observational evidence for this cosmological standard model has recently been reviewed elsewhere \cite{BA10.1}. Here, we just summarise its main parameters in Tab.~\ref{tab:01}, adapted from \cite{KO10.1}.

\begin{table}[ht]
\caption{Main cosmological parameters, adapted from \cite{KO10.1}. The Hubble constant $h$ is dimension-less and defined by $H_0=100\,h\,\mathrm{km\,s^{-1}\,Mpc^{-1}}$. ``CMB only'' means parameters obtained from the analysis of the 7-year WMAP cosmic microwave background data, ``CMB, BAO and $H_0$'' takes additional constraints from baryonic acoustic oscillations and external measurements of the Hubble constant into account. See \cite{KO10.1} for detail.}
\label{tab:01}
\hfill\begin{tabular}{|l|l|rcl|rcl|}
\hline
\multicolumn{2}{|l}{Parameter} & \multicolumn{3}{|c}{CMB only} & \multicolumn{3}{|c|}{CMB, BAO and $H_0$} \\
\hline
Hubble constant & $h$ &
$0.710$ & $\pm$ & $0.025$ & $0.704$ & $^+_-$ & $^{0.013}_{0.014}$ \\
baryon density parameter & $\Omega_\mathrm{b, 0}$ &
$0.0449$ & $\pm$ & $0.0028$ & $0.0456$ & $\pm$ & $0.0016$ \\
cold dark matter density & $\Omega_\mathrm{cdm, 0}$ &
$0.222$ & $\pm$ & $0.026$ & $0.227$ & $\pm$ & $0.014$ \\
cosmological constant & $\Omega_{\Lambda, 0}$ &
$0.734$ & $\pm$ & $0.029$ & $0.728$ & $^+_-$ & $^{0.015}_{0.016}$ \\
power-spectrum normalisation & $\sigma_8$ &
$0.801$ & $\pm$ & $0.030$ & $0.809$ & $\pm$ & $0.024$ \\
\hline
\end{tabular}
\end{table}

\subsection{Equation of geodesic deviation}

Gravitational lensing studies the effects of light deflection on the appearance of cosmic objects. Many approaches to gravitational lensing exist, some of them heuristic, others rather formal mathematically; see \cite{SC92.1, WA98.1, NA99.1, ST04.1, SC06.1} for examples.

We discuss gravitational lensing here under three main assumptions which are underlying the entire treatment. First, we shall remain within the framework of general relativity. Second, we shall assume that gravitationally lensing matter inhomogeneities have weak gravitational fields in the sense that their Newtonian gravitational potential is small, $\Phi\ll c^2$. Third, the sources of the potential are assumed to move slowly with respect to the mean cosmic flow, such that peculiar velocities are small compared to the speed of light. The assumption of weak, slowly moving gravitational lenses is well valid in all astrophysical applications except for light propagation near compact objects, which is not covered by this review. Lensing by moving and rotating astrophysical bodies has been discussed in the literature and generally been found to be negligibly small \cite{BI83.1, IB83.1, SE03.1, HE05.3, SC06.2}. Within the frame of these assumptions, gravitational lensing can be considered as a complete theory whose theoretical aspects are fully developed, including the mathematics of singularities in the lens mapping (cf.~the textbook by \cite{PE01.1}).

\begin{figure}[ht]
  \hfill\includegraphics[width=0.5\hsize]{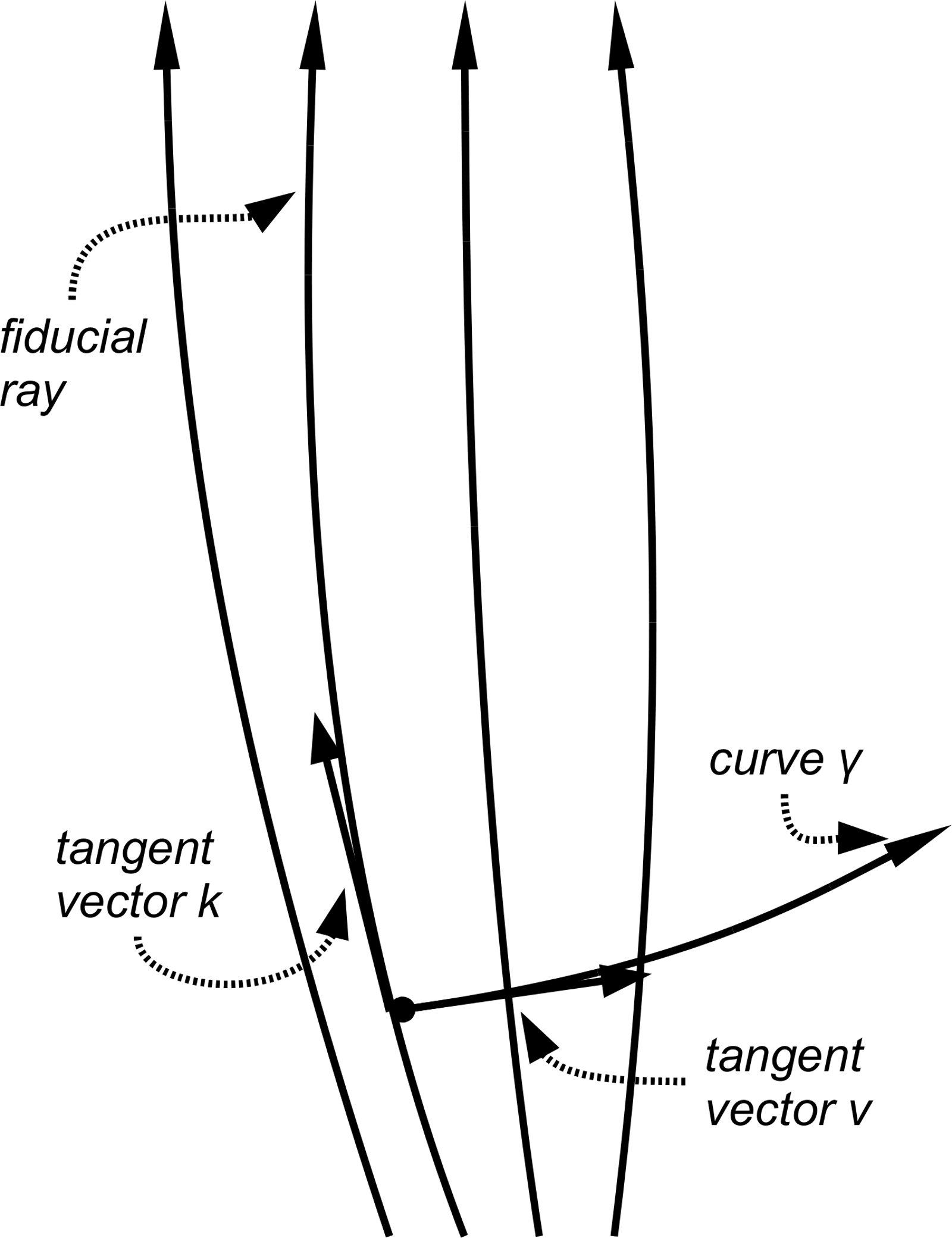}
\caption{A congruence of null geodesics with the fiducial light ray, a curve $\gamma$ connecting rays in the bundle, and the tangent vectors to $\gamma$ and the fiducial ray.}
\label{fig:01}
\end{figure}

Our approach here begins with a congruence of null geodesics (see Fig.~\ref{fig:01}), modelling a bundle of light rays in the approximation of geometrical optics within general relativity. We select one of the light rays as a fiducial ray and parameterise it by an affine parameter $\lambda$ which we shall later specify (cf.~Fig.~\ref{fig:01}). The tangent vector to the fiducial ray is
\begin{equation}
  \tilde k^\mu=\frac{\d x^\mu}{\d\lambda}\;,
\label{eq:01-1}
\end{equation}
and we choose to normalise $\tilde k$ such that its projection on the four-velocity $u_\mathrm{obs}$ of a freely falling observer is unity,
\begin{equation}
  \langle\tilde k, u_\mathrm{obs}\rangle=1\;.
\label{eq:01-2}
\end{equation}
Since the wave vector $k$ of a light ray, when projected on $u_\mathrm{obs}$, gives the (negative\footnote{We adopt the signature $-, +, +, +$ for the metric.}) frequency $\omega_\mathrm{obs}$ measured by the observer, we must choose $\lambda$ such that
\begin{equation}
  \tilde k=\frac{k}{\omega_\mathrm{obs}}=\frac{k}{|\langle k, u_\mathrm{obs}\rangle|}\;.
\label{eq:01-3}
\end{equation}
When the fiducial ray is connected with a neighbouring ray by a curve $\gamma(\sigma)$ with tangent vector $\partial_\sigma\gamma=v$, the equation of geodesic deviation or Jacobi equation
\begin{equation}
  \nabla^2_{\tilde k}v=R(\tilde k, v)\tilde k
\label{eq:01-4}
\end{equation}
quantifies how the vector $v$ changes along the fiducial ray in response to the curvature.

To evaluate this equation, we introduce a two-dimensional screen perpendicular to the four-velocity $u_\mathrm{obs}$ of the freely-falling observer and to the normalised tangent vector $\tilde k$ of the fiducial light ray, i.e.~a screen in the 3-space of the observer perpendicular to the light ray. This screen is spanned by the vectors $E_{1,2}$, which are parallel-transported along the fiducial ray,
\begin{equation}
  \nabla_{\tilde k}E_i=0\;.
\label{eq:01-5}
\end{equation}
Denoting with $v^{1,2}$ the components of $v$ on this screen, the equation of geodesic deviation turns into
\begin{equation}
  \nabla_{\tilde k}^2\cvector{v^1\\ v^2}=\mathcal{T}\cvector{v^1\\ v^2}
\label{eq:01-6}
\end{equation}
\cite{SE94.1}, where the optical tidal matrix
\begin{equation}
  \mathcal{T}=\left(\begin{array}{cc}
    \mathcal{R}+\Re(\mathcal{F})&\Im(\mathcal{F})\\\Im(\mathcal{F})&\mathcal{R}-\Re(\mathcal{F})
  \end{array}\right)
\label{eq:01-7}
\end{equation}
with the components
\begin{eqnarray}
  \mathcal{R}&=&-\frac{1}{2}R_{\alpha\beta}\tilde k^\alpha\tilde k^\beta+
  \frac{1}{2}C_{\alpha\beta\gamma\delta}\epsilon^\alpha\tilde k^\beta\tilde k^\gamma\epsilon^{*\delta}\;,
  \nonumber\\
  \mathcal{F}&=&
  \frac{1}{2}C_{\alpha\beta\gamma\delta}\epsilon^\alpha\tilde k^\beta\tilde k^\gamma\epsilon^\delta
\label{eq:01-8}
\end{eqnarray}
appears. It contains the Weyl curvature specified in Eq.~(\ref{eq:01-35}) below, and the complex vector $\epsilon=E_1+\ii E_2$.

We continue with the fundamental assumption that we can split the optical tidal matrix into two contributions,
\begin{equation}
  \mathcal{T}=\mathcal{T}_\mathrm{bg}+\mathcal{T}_\mathrm{cl}\;,
\label{eq:01-9}
\end{equation}
where the first is due to the background, homogeneous and isotropic space-time of a Friedmann-Lemaître-Robertson-Walker model, and the second is due to local inhomogeneities described by their Newtonian gravitational potential. This assumption is justified by our fundamental assumptions laid out in the beginning, i.e.~that the peculiar velocities of any inhomogeneities with respect to the comoving flow of the background matter are small compared to the light speed, and that their gravitational fields are weak in the sense that their Newtonian potential $\Phi$ is small, $\Phi/c^2=:\phi\ll1$.

\subsection{Background contribution}

We first work out the background contribution $\mathcal{T}_\mathrm{bg}$ to the tidal matrix. According to our assumption, the background has the Friedmann-Lemaître-Robertson-Walker metric with spatial curvature $K$,
\begin{equation}
  \d s^2=a^2(\eta)\left[-\d\eta^2+\d w^2+f_K^2(w)\d\Omega^2\right]\;,
\label{eq:01-10}
\end{equation}
where $\eta$ is the conformal time related to the cosmic time $t$ by $a\d\eta=c\d t$, and $f_K(w)$ is the comoving angular-diameter distance as a function of the comoving radial distance $w$,
\begin{equation}
  f_K(w)=\left\{\begin{array}{ll}
    \frac{1}{\sqrt{K}}\sin\left(\sqrt{K}w\right) & K>0 \\
    w & K=0 \\
    \frac{1}{\sqrt{-K}}\sinh\left(\sqrt{-K}w\right) & K<0 \\
  \end{array}\right.\;.
\label{eq:01-11}
\end{equation}
In the homogeneous contribution to the optical tidal matrix, the Weyl curvature must vanish because of the symmetry of the background model. Since $\tilde k$ is a null vector, we can further write
\begin{equation}
  R_{\alpha\beta}\tilde k^\alpha\tilde k^\beta=G_{\alpha\beta}\tilde k^\alpha\tilde k^\beta=
  \frac{8\pi G}{c^4}T_{\alpha\beta}\tilde k^\alpha\tilde k^\beta\;.
\label{eq:01-12}
\end{equation}
Here, Einstein's field equations enter for the first time. In alternative, metric theories of gravity, the relation between the global properties of light propagation and the matter-energy content of the universe will change, but the description of light deflection by local inhomogeneities in weak gravitational fields remains valid to some degree, as specified in Subsect.~\ref{subsec-1.5} below.

Inserting the energy-momentum tensor of an ideal fluid and assuming negligible pressure, $p=0$, we have
\begin{equation}
  T_{\alpha\beta}\tilde k^\alpha\tilde k^\beta=\rho c^2\langle u, \tilde k\rangle^2\;.
\label{eq:01-13}
\end{equation}
Now, the projection of $k$ on $u$ is the frequency of the light as measured by an observer co-moving with the fluid. Since we normalise $k$ according to Eq.~(\ref{eq:01-3}) by the frequency $\omega_\mathrm{obs}$ measured by us as observers, we must have
\begin{equation}
  |\langle\tilde k, u\rangle|=\frac{|\langle k, u\rangle|}{\omega_\mathrm{obs}}=
  \frac{\omega}{\omega_\mathrm{obs}}=1+z
\label{eq:01-14}
\end{equation}
according to Eq.~(\ref{eq:01-3}), where $z$ is the redshift of the fluid with respect to the observer. For a pressure-less fluid, $\rho=\rho_0(1+z)^3$ with the density $\rho_0$ measured by the observer, and thus
\begin{equation}
  \mathcal{R}=-\frac{4\pi G}{c^2}\rho_0(1+z)^5\;.
\label{eq:01-15}
\end{equation}
The optical tidal matrix of the homogeneous and isotropic background is thus
\begin{equation}
  \mathcal{T}_\mathrm{bg}=-\mathcal{R}\mathcal{I}_2\;,
\label{eq:01-16}
\end{equation}
where $\mathcal{I}_2$ is the unit matrix in two dimensions.

To satisfy Eqs.~(\ref{eq:01-1}) and (\ref{eq:01-14}), we now have to choose $\lambda$ such that $|\langle\tilde k, u\rangle|=1+z$. Assuming that peculiar velocities can be neglected and that we can thus project $\tilde k$ on the mean flow velocity $u^\mu=\delta^\mu_0$, we must have
\begin{equation}
  \langle\tilde k, u\rangle=\left\langle\frac{\d x}{\d\lambda}, u\right\rangle=\frac{\d x^0}{\d\lambda}=
  \frac{c\d t}{\d\lambda}=1+z=a^{-1}\;,
\label{eq:01-17}
\end{equation} 
and thus $\d\lambda=ac\d t=a^2\d\eta$. The equation of geodesic deviation for the smooth background then reads
\begin{equation}
  \nabla_{\tilde k}^2v^i=\frac{\d^2v^i}{\d\lambda^2}=
  \mathcal{T}^{\phantom{\mathrm{bg}}\,i}_{\mathrm{bg}\,j}v^j=\mathcal{R}v^i\;,
\label{eq:01-18}
\end{equation}
where we have used in the first equality that the basis vectors $E_i$ spanning the screen are parallel-transported along the fiducial ray, see Eq.~(\ref{eq:01-5}).

We now replace $\d\lambda$ by $\d w$, the comoving radial distance element, using that
\begin{equation}
  \d w=\d\eta=a^{-2}\d\lambda\quad\Rightarrow\quad\d\lambda=a^2\d w
\label{eq:01-19}
\end{equation}
for light, since $\d s=0$ in Eq.~(\ref{eq:01-10}). Introducing further the \textit{co-moving} bundle dimensions $v^i/a$, their propagation with $w$ is given by
\begin{equation}
  \frac{\d^2}{\d w^2}\left(\frac{v^i}{a}\right)=a^2\frac{\d}{\d\lambda}\left(v^{i\prime}a-v^ia'\right)=
  a^2\left(v^{i\prime\prime}a-v^ia''\right)\;,
\label{eq:01-20}
\end{equation}
where the primes denote derivatives with respect to $\lambda$. Moreover, since $\d\lambda=ac\d t$, we have
\begin{equation}
  a'=\frac{\d a}{\d\lambda}=\frac{1}{ca}\frac{\d a}{\d t}=\frac{\dot a}{ca}
\label{eq:01-21}
\end{equation}
and
\begin{equation}
  a''=\frac{\d a'}{\d\lambda}=\frac{1}{ac}\frac{\d a}{\d t}\frac{\d a'}{\d a}=
  \frac{1}{c^2}\frac{\dot a}{a}\frac{\d}{\d a}\frac{\dot a}{a}=\frac{1}{2c^2}\frac{\d}{\d a}\left(\frac{\dot a}{a}\right)^2\;.
\label{eq:01-22}
\end{equation}
We can now insert Friedmann's equation
\begin{equation}
  \left(\frac{\dot a}{a}\right)^2=\frac{8\pi G}{3}\frac{\rho_0}{a^3}+\frac{\Lambda}{3c^2}-\frac{Kc^2}{a^2}
\label{eq:01-23}
\end{equation}
for $(\dot a/a)^2$ to find
\begin{equation}
  a''=-\frac{4\pi G}{c^2}\frac{\rho_0}{a^4}+\frac{K}{a^3}\;.
\label{eq:01-24}
\end{equation}
Note that the cosmological-constant term drops out. Loosely related in this context, arguments that the cosmological constant might measurably affect gravitational lensing by isolated objects \cite{RI07.1} were refuted by \cite{SI10.1}. The propagation equation (\ref{eq:01-20}) for the comoving bundle dimensions in the homogeneous and isotropic universe thus becomes
\begin{equation}
  \frac{\d^2}{\d w^2}\left(\frac{v^i}{a}\right)=a^3v^{i\prime\prime}+
  \frac{4\pi G}{c^2}\frac{\rho_0}{a^2}v^i-K\frac{v^i}{a}=-K\frac{v^i}{a}
\label{eq:01-25}
\end{equation}
where Eqs.~(\ref{eq:01-15}) and (\ref{eq:01-18}) were inserted in the last step, or
\begin{equation}
  \left(\frac{\d^2}{\d w^2}+K\right)\frac{v^i}{a}=0\;.
\label{eq:01-26}
\end{equation}

\subsection{Clump contribution}

In presence of local perturbations characterised by the Newtonian gravitational potential $\Phi=\phi c^2\ll c^2$, we can write the perturbed Friedmann-Lemaître-Robertson-Walker line element as
\begin{equation}
  \d s^2=a^2(\eta)\left[-(1+2\phi)\d\eta^2+(1-2\phi)\left(\d w^2+f_K^2(w)\d\Omega^2\right)\right]
\label{eq:01-27}
\end{equation}
if the perturbations move slowly with respect to the mean cosmic flow (cf.~the more general remarks on light propagation in weakly perturbed metrics in the following Subsect.~\ref{subsec-1.5}). Our next assumption is that the perturbations are well localised, i.e.~that their spatial extent is much smaller than the curvature scale of the background universe. Then, we can consider the perturbation as being locally embedded into flat space and approximate $f_K(w)\approx w$. Consequently, we shall first study light propagation in the comoving Newtonian metric
\begin{equation}
  \d\tilde s^2=-(1+2\phi)\d\eta^2+(1-2\phi)\d\vec w^2
\label{eq:01-28}
\end{equation}
and later combine it with light propagation in the background universe.

The line element Eq.~(\ref{eq:01-28}) suggests introducing the dual basis
\begin{equation}
  \theta^0=(1+\phi)\d\eta\;,\quad\theta^i=(1-\phi)\d w^i\;,
\label{eq:01-29}
\end{equation} 
in which the metric becomes Minkowskian. The following calculation is linear in the sense that terms of higher than first order in $\phi$ are neglected, as are time derivatives of the potential. By Cartan's first structure equation, the connection forms are
\begin{equation}
  \tensor{\omega}{0}{i}=\phi_i\theta^0\;,\quad\tensor{\omega}{i}{j}=-\phi_j\theta^i+\phi_i\theta^j\;,
\label{eq:01-30}
\end{equation}
where $\phi_j\equiv\partial_j\phi$ and partial derivatives are taken with respect to comoving Cartesian coordinates. Cartan's second structure equation gives the curvature forms
\begin{equation}
  \tensor{\Omega}{0}{i}=\phi_{ik}\theta^k\wedge\theta^0\;,\quad
  \tensor{\Omega}{i}{j}=-\phi_{jk}\theta^k\wedge\theta^i+\phi_{ik}\theta^k\wedge\theta^j\;,
\label{eq:01-31}
\end{equation}
which yield the only non-vanishing elements
\begin{eqnarray}
  &&R_{0i0j}=\phi_{ij}\;,\quad R_{ijij}=\phi_{ii}+\phi_{jj}\;,\nonumber\\
  &&R_{1213}=\phi_{23}\;,\quad R_{1223}=-\phi_{13}\;,\quad R_{1323}=\phi_{12}
\label{eq:01-32}
\end{eqnarray}
of the Riemann tensor. The Ricci tensor and the Ricci scalar are
\begin{equation}
  R_{\alpha\beta}=\vec\nabla^2\phi\,\mathcal{I}_4\;,\quad R=2\vec\nabla^2\phi\;,
\label{eq:01-33}
\end{equation}
respectively, where $\mathcal{I}_4$ is the unit matrix in four dimensions. The Einstein tensor is
\begin{equation}
  G_{\alpha\beta}=\vec\nabla^2\phi\,\delta^0_\alpha\delta^0_\beta\;,
\label{eq:01-34}
\end{equation}
with $\vec\nabla^2=\partial^i\partial_i$. The Weyl curvature, which we need for the clump contribution to the optical tidal matrix, is
\begin{equation}
  C_{\alpha\beta\gamma\delta}=R_{\alpha\beta\gamma\delta}-g_{\alpha[\gamma}R_{\delta]\beta}+
  g_{\beta[\gamma}R_{\delta]\alpha}+\frac{R}{3}g_{\alpha[\gamma}g_{\delta]\beta}\;,
\label{eq:01-35}
\end{equation}
and has the only non-vanishing components
\begin{eqnarray}
  &&C_{0i0j}=\phi_{ij}-\frac{1}{3}\vec\nabla^2\phi\eta_{ij}\;,\quad
  C_{ijij}=\phi_{ii}+\phi_{jj}-\frac{2}{3}\vec\nabla^2\phi\;,\nonumber\\
  &&C_{1213}=\phi_{23}\;,\quad C_{1223}=-\phi_{13}\;,\quad C_{1323}=\phi_{12}
\label{eq:01-36}
\end{eqnarray}
here. We can now evaluate the clump contribution first in the comoving metric given by Eq.~(\ref{eq:01-28}). With the components Eq.~(\ref{eq:01-32}) of the  Ricci and Eq.~(\ref{eq:01-36}) of the Weyl tensors, we find
\begin{equation}
  \mathcal{R}_\mathrm{cl}=-\vec\nabla^2\phi\;,\quad
  \mathcal{F}_\mathrm{cl}=-\left(\phi_{11}-\phi_{22}\right)-2\ii\phi_{12}\;,
\label{eq:01-37}
\end{equation}
hence the clump contribution to the optical tidal matrix becomes
\begin{equation}
  \mathcal{T}_\mathrm{cl}=
  -2\left(\begin{array}{cc}\phi_{11}&\phi_{12}\\\phi_{12}&\phi_{22}\end{array}\right)\;,\quad
  \left(\mathcal{T}_\mathrm{cl}\right)_{ij}=-2\partial_i\partial_j\phi\;.
\label{eq:01-38}
\end{equation}

Let now $\phi^{(0)}$ be the gravitational potential passed by the fiducial ray. Then, the potential gradient can be expanded as
\begin{equation}
  \partial_i\left(\phi-\phi^{(0)}\right)=\partial_i\delta\phi=\partial_j\partial_i\phi\,\Big|_0x^j=
  -\frac{1}{2}\left(\mathcal{T}_\mathrm{cl}\right)_{ij}x^j\;.
\label{eq:01-39}
\end{equation} 
We can thus bring the equation of geodesic deviation for the clump contribution into the form
\begin{equation}
  \frac{\d^2x^i}{\d\lambda^2}=\frac{\d^2x^i}{\d w^2}=-2\partial^i\delta\phi\;,
\label{eq:01-40}
\end{equation}
where the first equality follows from $\d\lambda=\d w$ in the local approximation Eq.~(\ref{eq:01-28}) to the metric. Since the absolute value of the potential is of no importance, we can rename the potential difference $\delta\phi$ to the potential $\phi$. Finally, we must be aware that the bundle dimensions $x^i$ evaluated in the local frame are comoving bundle dimensions $v^i/a$ in the cosmological frame. 

If we now combine the global, homogeneous and the local, clump contributions expressed by Eqs.~(\ref{eq:01-40}) and (\ref{eq:01-26}), we find the inhomogeneous propagation equation for the comoving bundle dimensions $x^i$
\begin{equation}
  \left(\frac{\d^2}{\d w^2}+K\right)x^i=-2\partial^i\phi\;.
\label{eq:01-41}
\end{equation}

\subsection{Remarks on light propagation in other metric theories of gravity}
\label{subsec-1.5}

At this point, it is worth noticing that part of the results derived so far remain valid beyond general relativity. In a metric theory of gravity, the most general, lowest-order deviation from the Minkowski space-time is
\begin{equation}
  \d\tilde s^2=-(1+2\phi)\d\eta^2+(1+2\psi)\d\vec w^2
\label{eq:01-41a}
\end{equation}
with the two otherwise unspecified Bardeen potentials $\phi$ and $\psi$ with $\phi, \psi\ll1$. Since null geodesics are conformally invariant, light propagation can be described by the null geodesics of the conformally transformed metric
\begin{equation}
  \d\tilde s^2=(1+2\phi)\left[-\d\eta^2+(1+2\tilde\psi)\d\vec w^2\right]\;,
\label{eq:01-41b}
\end{equation}
with $\tilde\psi=\psi-\phi$. The geodesic equation of the conformally transformed metric (\ref{eq:01-41b}) is identical to (\ref{eq:01-40}) with $-2\phi$ replaced by $\tilde\psi$. Locally, light deflection is thus governed by the difference between the Bardeen potentials which, in general relativity in absence of anisotropic stresses, is $\tilde\psi=-2\phi$. The gravitational field equations only enter through the relation between the Bardeen potentials and the matter-energy content of the space-time, which also governs the global, cosmological properties of light propagation. Thus, in any metric theory of gravity which locally contains special relativity, light deflection by localised, weak gravitational fields is determined by $\tilde\psi$. Alternative field equations will change global light propagation and the way how matter and energy source $\tilde\psi$.

\subsection{The lens mapping and the lensing potential}

The Greens function $G(w, w')$ of the operator $\d^2/\d w^2+K$ is
\begin{eqnarray}
  G(w, w')&=&\frac{1}{\sqrt{K}}\sin\left[\sqrt{K}(w-w')\right]\Theta(w-w')\nonumber\\
  &=&f_K(w-w')\Theta(w-w')\;,
\label{eq:01-42}
\end{eqnarray}
where we have introduced the comoving angular diameter distance $f_K(w)$ of the Friedmann metric, Eq.~(\ref{eq:01-10}). The appropriate boundary conditions for the solution of Eq.~(\ref{eq:01-41}) are
\begin{equation}
  x^i\,\Big|_{w=0}=0\;,\quad\left.\frac{\d x^i}{\d w}\,\right|_{w=0}=\theta^i\;,
\label{eq:01-43}
\end{equation}
because both rays start at the observer and enclose an angle $(\theta^1, \theta^2)$ there. Thus, the solution is
\begin{equation}
  x^i(w)=f_K(w)\theta^i-2\int_0^w\d w'f_K(w-w')\partial^i\phi\left(x^j(w'), w'\right)\;.
\label{eq:01-44}
\end{equation}

In line with the Newtonian approximation, $\phi\ll c^2$, we assume small deflection angles and use Born's approximation to integrate along the unperturbed light path $x^i(w')\approx f_K(w')\theta^i$ in Eq.~(\ref{eq:01-44}). Since numerical simulations show that Born's approximation is typically very well satisfied \cite{DO05.1, SH06.1, HI09.1}, we shall adopt it throughout. See \cite{BE10.1} for a formidable calculation to second order. Integrating from the observer to a source at the comoving radial distance $w_\mathrm{s}$ and defining $x^i(w_\mathrm{s})=f_K(w_\mathrm{s})\beta^i$, we find
\begin{equation}
  \beta^i=\theta^i-2\int_0^{w_\mathrm{s}}\d w'\frac{f_K(w_\mathrm{s}-w')}{f_K(w_\mathrm{s})}
  \partial^i\phi\left(f_K(w')\theta^j, w'\right)\;.
\label{eq:01-45}
\end{equation}
We define the reduced deflection angle by
\begin{equation}
  \alpha^i(\theta^j):=2\int_0^{w_\mathrm{s}}\d w'\frac{f_K(w_\mathrm{s}-w')}{f_K(w_\mathrm{s})}
  \partial^i\phi\left(f_K(w')\theta^j, w'\right)
\label{eq:01-46}
\end{equation}
and obtain the \textit{lens equation}
\begin{equation}
  \beta^i=\theta^i-\alpha^i(\theta^j)\;.
\label{eq:01-47}
\end{equation} 
Introducing derivatives with respect to angular coordinates $\theta^j$ on the sky by
\begin{equation}
  \partial_x=\frac{\partial_\theta}{f_K(w)}\;,
\label{eq:01-48}
\end{equation}
we can write the reduced deflection angle as the angular gradient of an effective lensing potential
\begin{equation}
  \psi(\theta^j)=2\int_0^{w_\mathrm{s}}\d w'\frac{f_K(w_\mathrm{s}-w')}{f_K(w')f_K(w_\mathrm{s})}
  \phi\left(f_K(w')\theta^j, w'\right)\;.
\label{eq:01-49}
\end{equation}
The derivatives with respect to $\theta$ have to be interpreted as covariant derivatives on the sphere if the curvature of the sphere becomes important. Then, different suitable and convenient bases are used. We shall later return to this issue. For the purposes of gravitational lensing, the flat-sky approximation is often, but not always justified. Unless stated otherwise, partial derivatives are taken with respect to angular coordinates from now on.

\subsection{Local properties of the lens mapping}

Having arrived at this point, it may be useful for clarification to describe lensing without reference to specific coordinates on the sphere. The lensing potential $\psi$ assigns a number to each point on the observer's sky,
\begin{equation}
  \psi: \mathbb{S}^2\to\mathbb{R}\;,\quad p\mapsto\psi(p)\;.
\label{eq:01-50}
\end{equation}
The gradient of $\psi$ defines a vector field $\alpha$ on $\mathbb{S}^2$ by
\begin{equation}
  \alpha^\flat=\d\psi\;,
\label{eq:01-51}
\end{equation}
where $\alpha^\flat$ is the 1-form dual to $\alpha$. The codifferential of $\alpha^\flat$ is the Laplace-de-Rham operator of $\psi$ and defined to be twice the so-called convergence
\begin{equation}
  \delta\alpha^\flat=\left({\ast\d}{\ast\d}\right)\psi=2\kappa\;.
\label{eq:01-52}
\end{equation}
The lens mapping defines a map $\varphi$ from the observer's sky to a sphere on which the sources are located,
\begin{equation}
  \varphi: \mathbb{S}^2\to\mathbb{S}^2\;,\quad p\mapsto\varphi(p)\;.
\label{eq:01-53}
\end{equation}
Its differential $D\varphi$ describes how sources are locally deformed under the lens mapping. In absence of scattering, absorption or emission, the phase-space distribution function $f$ satisfies Liouville's theorem. This implies that $f\propto\omega^{-3}I(\omega)$ is constant along null geodesics, where $\omega$ and $I(\omega)$ are the frequency and the specific intensity of the light. If the frequency is unchanged by the lensing mass distribution, $I(\omega)$ is constant and the flux from the source is changed only because the lens mapping changes the solid angle under which the source appears. Thus, lensing causes the magnification
\begin{equation}
  \mu=\left|\det(D\varphi)\right|^{-1}\;.
\label{eq:01-54}
\end{equation}
Points $p_\mathrm{crit}$ where the lens mapping $\varphi$ is singular, $\det(D\varphi)(p_\mathrm{crit})=0$, are called critical points. They form closed curves, the so-called critical curves. Their images $\varphi(p_\mathrm{crit})$ under the lens mapping are called caustics.

\begin{figure}[ht]
  \hfill\includegraphics[width=0.7\hsize]{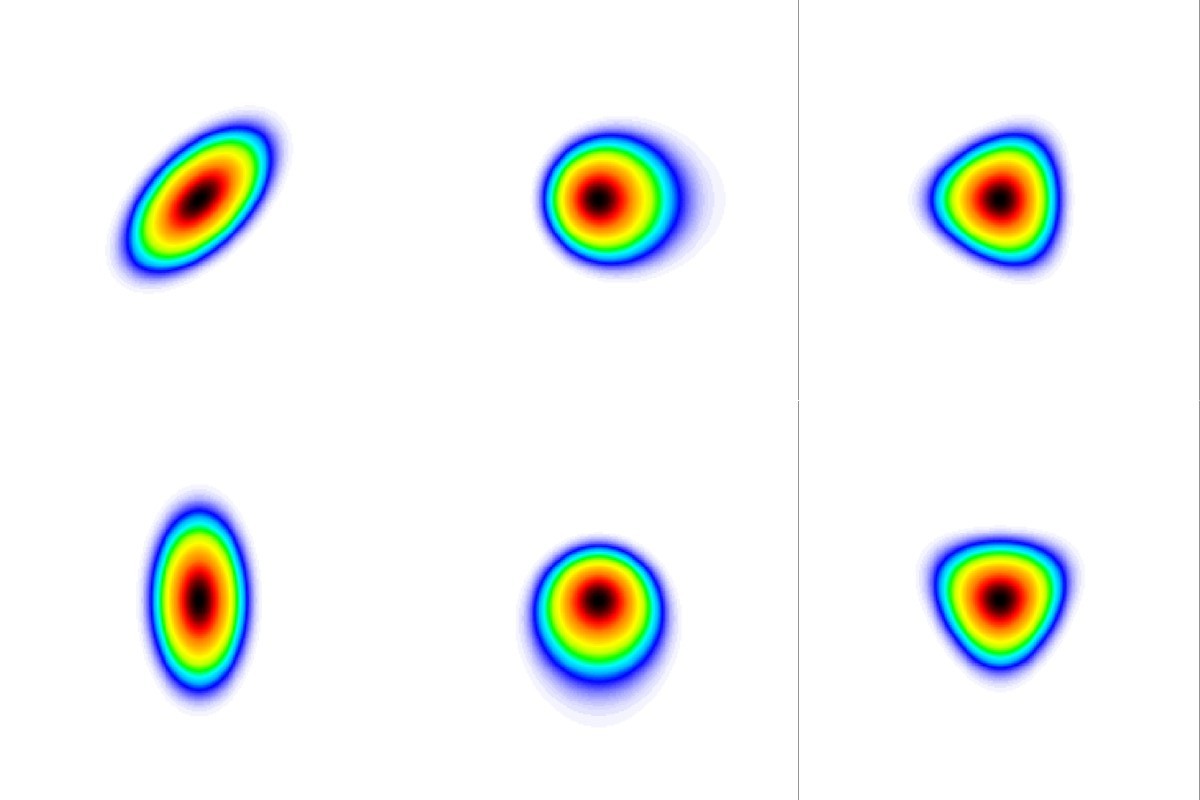}
\caption{The effect of shear $\gamma$ (left column), first flexion $F$ (middle column) and second flexion $G$ (right column) on an idealised circular image.}
\label{fig:02}
\end{figure}

If we now introduce charts $h_1: \mathbb{S}^2\supset U\to U'\subset\mathbb{R}^2$ with $p\in U$ and $h_2: \mathbb{S}^2\supset V\to V'\subset\mathbb{R}^2$ with $\varphi(p)\in V$, we can locally span $U'$ and $V'$ by Cartesian coordinates $(\theta^1, \theta^2)$ and $(\beta^1, \beta^2)$, respectively, and express the Jacobian matrix $D\varphi$ by
\begin{equation}
  (D\varphi)^i_j=\frac{\partial\beta^i}{\partial\theta^j}=\delta^i_j-\frac{\partial\alpha^i}{\partial\theta^j}=
  \delta^i_j-\frac{\partial^2\psi}{\partial\theta_i\partial\theta^j}=:\delta^i_j-\psi^i_j\;.
\label{eq:01-55}
\end{equation}
It is convenient and instructive to separate $D\varphi$ into its trace and a trace-free part,
\begin{equation}
  (D\varphi)^i_j=\left(\begin{array}{cc}1-\kappa&0\\0&1-\kappa\end{array}\right)-
  \left(\begin{array}{cc}\gamma_1&\gamma_2\\\gamma_2&-\gamma_1\end{array}\right)\;,
\label{eq:01-56}
\end{equation}
where
\begin{equation}
  \kappa=\frac{1}{2}\tensor{\psi}{i}{i}=\frac{1}{2}\partial^i\partial_i\psi
\label{eq:01-57}
\end{equation}
is the convergence introduced as the codifferential (divergence) of $\alpha^\flat$ in Eq.~(\ref{eq:01-52}), and
\begin{equation}
  \gamma_1=\frac{1}{2}\left(\tensor{\psi}{1}{1}-\tensor{\psi}{2}{2}\right)\;,\quad
  \gamma_2=\tensor{\psi}{1}{2}=\tensor{\psi}{2}{1}
\label{eq:01-58}
\end{equation}
are the components of the shear
\begin{equation}
  \gamma=\gamma_1+\ii\gamma_2\;.
\label{eq:01-59}
\end{equation} 
While the convergence is responsible for stretching a source isotropically under the lens mapping, the shear is responsible for its distortion (see Fig.~\ref{fig:02}). In fact, a circular source of unit radius is mapped into an elliptical image with semi-major and semi-minor axes
\begin{equation}
  A=\left(1-\kappa-|\gamma|\right)^{-1}\;,\quad B=\left(1-\kappa+|\gamma|\right)^{-1}\;.
\label{eq:01-60}
\end{equation}
This distortion allows the systematic detection of gravitational lensing. The relative axis ratio of elliptically distorted images, the ellipticity $\epsilon$, is the so-called reduced shear
\begin{equation}
  \epsilon=\frac{A-B}{A+B}=\frac{|\gamma|}{1-\kappa}=|g|\;.
\label{eq:01-61}
\end{equation}
This illustrates that not the shear $\gamma$ can be directly measured from image distortions, but only the reduced shear $g$. This is because the absolute size of a source is typically unknown and image ellipticities are invariant under transformations
\begin{equation}
  D\varphi\to\lambda(D\varphi)\quad\hbox{with}\quad\lambda\ne0\;,
\label{eq:01-62}
\end{equation} 
because they change only the size of the image but not its shape and leave the reduced shear unchanged. The convergence is transformed by Eq.~(\ref{eq:01-62}) as
\begin{equation}
  \kappa\to1-\lambda(1-\kappa)\;.
\label{eq:01-63}
\end{equation}
With $\lambda=1-\delta\kappa$ and $\delta\kappa\ll1$, this transformation is approximately
\begin{equation}
  \kappa\to\kappa+\delta\kappa
\label{eq:01-64}
\end{equation}
if $\kappa\ll1$. Then, it corresponds to adding a sheet of constant surface-mass density $\delta\kappa$ to the lens, whence it has been called the mass-sheet degeneracy \cite{GO85.1, GO88.1}.

In the coordinates introduced by the charts $h_{1,2}$ before Eq.~(\ref{eq:01-55}), the magnification is
\begin{equation}
  \mu=\left|\det(D\varphi)\right|^{-1}=\frac{1}{(1-\kappa)^2-\gamma_1^2-\gamma_2^2}\;.
\label{eq:01-65}
\end{equation}
In these coordinates, the critical curves are given by
\begin{equation}
  (1-\kappa)^2-\gamma_1^2-\gamma_2^2=0\;.
\label{eq:01-66}
\end{equation}

Third-order derivatives of the lensing potential, termed gravitational flexion \cite{GO05.1, BA06.1}, may become measurable in the near future. Since partial derivatives of $\psi$ commute, only four of the original eight components of $\partial_i\partial_j\partial_k\psi=\psi_{ijk}$ are independent, e.g.~$\psi_{111}$, $\psi_{112}$, $\psi_{122}$ and $\psi_{222}$. They are commonly combined into the first flexion $F$ with the two components
\begin{equation}
  F_1=\frac{1}{2}\left(\psi_{111}+\psi_{122}\right)\;,\quad F_2=\frac{1}{2}\left(\psi_{112}+\psi_{222}\right)
\label{eq:01-64a}
\end{equation}
and the second flexion $G$ with
\begin{equation}
  G_1=\frac{1}{2}\left(\psi_{111}-3\psi_{122}\right)\;,\quad G_2=\frac{1}{2}\left(3\psi_{112}-\psi_{222}\right)\;.
\label{eq:01-64b}
\end{equation}

Applying the local definition of the convergence in Eq.~(\ref{eq:01-57}) to the lensing potential Eq.~(\ref{eq:01-49}), using Eq.~(\ref{eq:01-48}), we find
\begin{equation}
  \kappa\left(x^j\right)=\int_0^{w_\mathrm{s}}\d w'\,\frac{f_K(w')f_K(w_\mathrm{s}-w')}{f_K(w_\mathrm{s})}\,
  \partial_{x^i}\partial^{x^i}\phi\left(x^j, w'\right)\;,
\label{eq:01-67}
\end{equation}
where the partial derivatives of $\phi$ under the integral are to be taken with respect to the co-moving coordinates $x^i$ perpendicular to the line-of-sight. We can, however, augment this two-dimensional Laplacian by its component along the line-of-sight,
\begin{equation}
  \partial_{x^i}\partial^{x^i}\phi=\nabla^2\phi=\frac{4\pi G}{c^2}\rho\;,
\label{eq:01-68}
\end{equation}
because the derivative with respect to the line-of-sight direction vanishes after the line-of-sight integration in Eq.~(\ref{eq:01-67}) if the potential is localised. Inserting Eq.~(\ref{eq:01-68}) into Eq.~(\ref{eq:01-67}) shows that the convergence is the suitably scaled and geometrically weighed surface-mass density of the lensing matter inhomogeneities,
\begin{equation}
  \kappa(\theta^i)=\frac{4\pi G}{c^2}\int_0^{w_\mathrm{s}}\d w'\,
  \frac{f_K(w')f_K(w_\mathrm{s}-w')}{f_K(w_\mathrm{s})}\,\rho\left(f_K(w')\theta^i, w'\right)\;.
\label{eq:01-69}
\end{equation} 

\section{Thin gravitational lenses}

\subsection{Introduction and axially-symmetric lenses}

In many astrophysically relevant applications of gravitational lensing, the lensing mass distribution is very thin compared to the cosmological distances between observer, lens, and source. We can then introduce the thin-lens approximation, replacing the Newtonian potential $\phi$ by its line-of-sight projection
\begin{equation}
  \phi\to\delta(w-w_\mathrm{d})\int\d w'\phi=:\delta(w-w_\mathrm{d})\phi^{(2)}\;,
\label{eq:03-1}
\end{equation}
located at the co-moving radial distance $w_\mathrm{d}$ from the observer. From now on, unless stated otherwise, we shall assume that the spatial curvature vanishes, $k=0$, allowing us to identify $f_k(w)=w$. Then, the lens mapping Eq.~(\ref{eq:01-45}) becomes
\begin{equation}
  \beta^i=\theta^i-2\frac{w_\mathrm{d}(w_\mathrm{s}-w_\mathrm{d})}{w_\mathrm{s}}\,
  \partial^i\phi^{(2)}\left(w_\mathrm{d}\theta^j\right)\;,
\label{eq:03-2}
\end{equation}
where the partial derivative is again to be taken with respect to the angular coordinates $\theta^i$. The lensing potential Eq.~(\ref{eq:01-49}) simplifies to
\begin{equation}
  \psi(\theta^i)=2\frac{w_\mathrm{s}-w_\mathrm{d}}{w_\mathrm{d}w_\mathrm{s}}
  \phi^{(2)}\left(w_\mathrm{d}\theta^i\right)\;,
\label{eq:03-3}
\end{equation}
and the deflection angle simplifies accordingly. The convergence Eq.~(\ref{eq:01-69}) becomes accordingly
\begin{equation}
  \kappa(\theta^i)=\frac{4\pi G}{c^2}
  \frac{(w_\mathrm{s}-w_\mathrm{d})w_\mathrm{d}}{w_\mathrm{s}}
  \int\d w'\rho\left(w'\theta^i, w'\right)\;.
\label{eq:03-4}
\end{equation} 
The line-of-sight projection of the mass density $\rho$ is the surface-mass density
\begin{equation}
  \Sigma(\theta^i)=\int\d w'\rho\left(w'\theta^i, w'\right)\;,
\label{eq:03-5}
\end{equation}
and the quantity $\Sigma_\mathrm{cr}$ defined by
\begin{equation}
  \Sigma_\mathrm{cr}^{-1}:=\frac{4\pi G}{c^2}
  \frac{(w_\mathrm{s}-w_\mathrm{d})w_\mathrm{d}}{w_\mathrm{s}}
\label{eq:03-6}
\end{equation}
is called the critical surface mass density. With these definitions, the convergence of a thin lens is
\begin{equation}
  \kappa(\theta^i)=\frac{\Sigma(\theta^i)}{\Sigma_\mathrm{cr}}\;.
\label{eq:03-7}
\end{equation}

\begin{figure}[ht]
  \hfill\includegraphics[width=0.5\hsize]{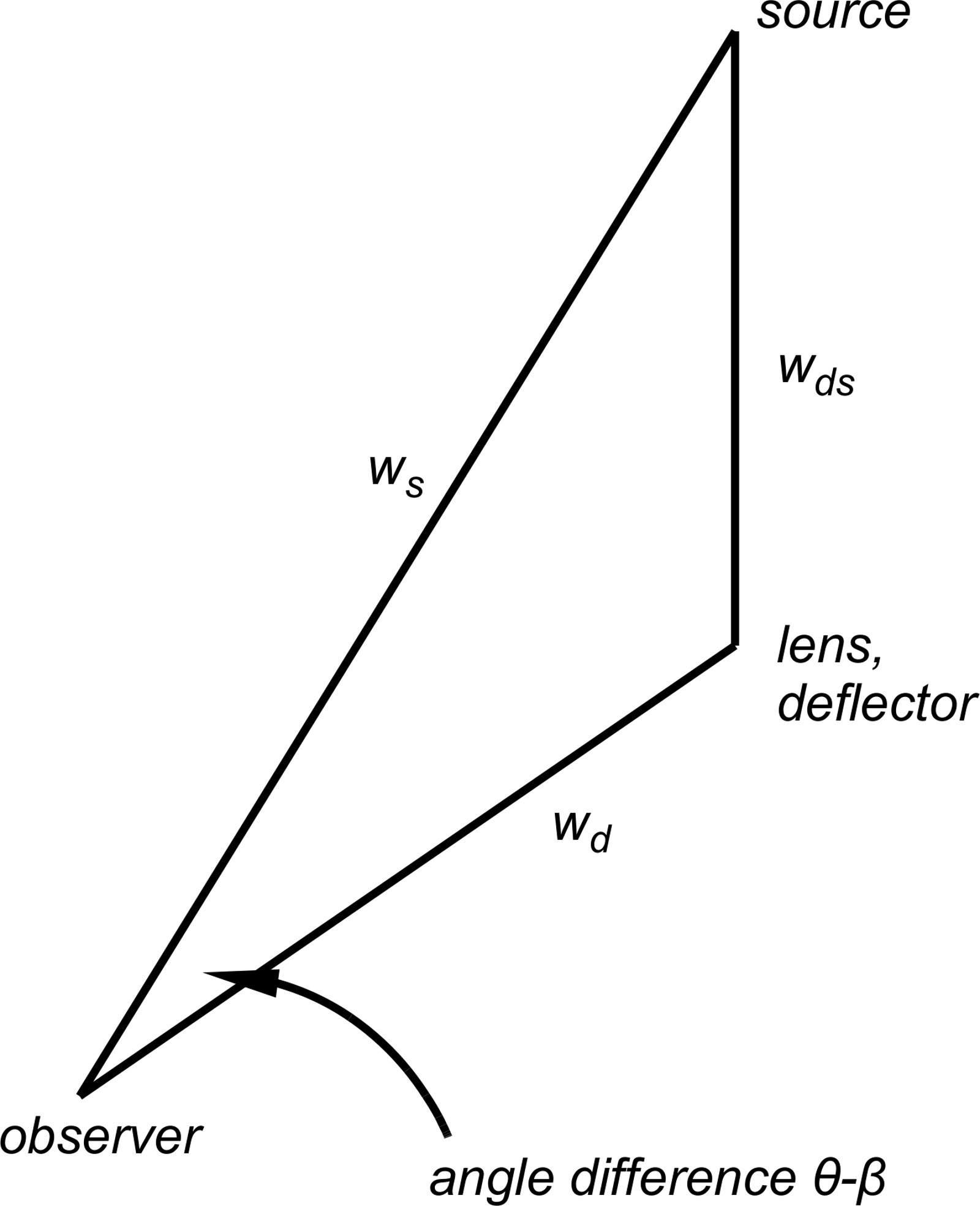}
\caption{Illustration of the co-moving undeflected and deflected light paths and their length difference (see Eq.~(\ref{eq:03-9}).}
\label{fig:03}
\end{figure}

The light-travel time along the deflected light path is longer than along the undeflected path for two reasons. First, the deflected path is geometrically longer, and second, there is a gravitational time delay due to the local gravitational potential causing the deflection. The geometrical time delay is most easily worked out in the conformally static, spatially flat metric
\begin{equation}
  \d\tilde s^2=-\d\eta^2+\d\vec w^2\;,
\label{eq:03-8}
\end{equation}
but the following derivation can be straightforwardly generalised to the spatially curved cases. We construct a triangle between the source, the deflection point and the observer and introduce its side lengths $w_\mathrm{d}$ from the observer to the deflection point, $w_\mathrm{ds}$ from the deflection point to the source, and $w_\mathrm{s}$ from the observer to the source (see Fig.~\ref{fig:03}). From the cosine law, we know that
\begin{equation}
  w_\mathrm{ds}^2=w_\mathrm{s}^2+w_\mathrm{d}^2-2w_\mathrm{s}w_\mathrm{d}\cos(\theta-\beta)\approx
  (w_\mathrm{s}-w_\mathrm{d})^2+w_\mathrm{s}w_\mathrm{d}(\theta-\beta)^2\;,
\label{eq:03-9}
\end{equation}
where the small-angle approximation for the cosine was used in the last step. In the same approximation, we can rewrite this equation as
\begin{equation}
  w_\mathrm{ds}\approx w_\mathrm{s}-w_\mathrm{d}+
  \frac{w_\mathrm{s}w_\mathrm{d}}{2(w_\mathrm{s}-w_\mathrm{d})}(\theta-\beta)^2
\label{eq:03-10}
\end{equation}
and invert it to approximate
\begin{equation}
  w_\mathrm{s}-w_\mathrm{d}\approx w_\mathrm{ds}-w_\mathrm{d}w_\mathrm{s}(\theta-\beta)^2
\label{eq:03-10a}
\end{equation}
at the same order in $(\theta-\beta)$. The co-moving geometrical difference between the deflected and the undeflected path lengths is thus
\begin{equation}
  \Delta w=w_\mathrm{d}+w_\mathrm{ds}-w_\mathrm{s}\approx
  \frac{w_\mathrm{s}w_\mathrm{d}}{2(w_\mathrm{s}-w_\mathrm{d})}(\theta-\beta)^2\approx
  \frac{w_\mathrm{s}w_\mathrm{d}}{w_\mathrm{ds}}\frac{(\theta-\beta)^2}{2}\;.
\label{eq:03-11}
\end{equation}
Since the conformal time delay $\Delta\eta=\Delta w$ is very small compared to the Hubble time, we can write the cosmological time delay $\Delta t_\mathrm{geo}=\Delta\eta/c$ or
\begin{equation}
  \Delta t_\mathrm{geo}(\theta)=\frac{\Delta w}{c}=
  \frac{w_\mathrm{s}w_\mathrm{d}}{c\,w_\mathrm{ds}}\frac{(\theta-\beta)^2}{2}\;.
\label{eq:03-12}
\end{equation}

At the same time, the light experiences the gravitational time delay \cite{SH66.1}
\begin{equation}
  \Delta t_\mathrm{grv}(\theta)=-2\int\d w'\,\frac{\phi(\theta w', w')}{c}
\label{eq:03-13}
\end{equation}
which, when expressed in terms of the lensing potential of a thin lens (\ref{eq:03-3}), can be written as
\begin{equation}
  \Delta t_\mathrm{grv}(\theta)=-\frac{1}{c}\frac{w_\mathrm{s}w_\mathrm{d}}{w_\mathrm{ds}}\psi(\theta)\;.
\label{eq:03-14}
\end{equation}
The combination of both the geometrical and the gravitational time delays gives the total time delay
\begin{equation}
  \Delta t(\theta)=\frac{1}{c}\frac{w_\mathrm{s}w_\mathrm{d}}{w_\mathrm{ds}}\left(
    \frac{(\theta-\beta)^2}{2}-\psi(\theta)
  \right)\;.
\label{eq:03-15}
\end{equation}
The stationary points of $\Delta t$, where $\partial_i\Delta t=0$, satisfy the lens equation. This illustrates Fermat's principle in gravitational lensing: Images are formed where the light-travel time is stationary. As an immediate corollary, this implies the odd-number theorem: non-singular thin lenses produce an odd number of images. In absence of the lens, $\Delta t$ is a paraboloid in $\theta$ for fixed $\beta$, which has a single minimum where one image is formed. Further stationary points are added in pairs as local maxima and saddle points. The total number of images must thus be odd.

A further important result is that the light-travel time scales like an effective distance $w_\mathrm{s}w_\mathrm{d}/w_\mathrm{ds}$, which is itself proportional to the Hubble distance $c/H_0$. The time delay between different gravitationally lensed images is thus proportional to the Hubble time $H_0^{-1}$. This allows the Hubble constant to be derived from measured time delays in multiple-image gravitational-lens systems, if $\psi$ and $\beta$ are known, e.g.~from lens modelling \cite{RE64.1}.

Models for thin gravitational lenses are typically constructed by fixing either the effective lensing potential $\psi$ or the convergence $\kappa$. Since the two are related by the Poisson equation $2\kappa=\vec\nabla^2\psi$ in two dimensions, the lensing potential can easily be obtained by means of the Greens function of the Laplacian in two dimensions,
\begin{equation}
  \psi(\vec\theta)=\frac{1}{\pi}\int\d^2\vec\theta'\,\kappa(\vec\theta')\ln\left|\vec\theta-\vec\theta'\right|\;.
\label{eq:03-16}
\end{equation}

The situation simplifies considerably for axially symmetric lens models. In this case, the lensing potential from Eq.~(\ref{eq:03-16}) is
\begin{equation}
  \psi(\theta)=2\left[
    \ln\theta\int_0^\theta\theta'\d\theta'\,\kappa(\theta')+\int_\theta^\infty\theta'\d\theta'\,\ln\theta'\kappa(\theta')
  \right]\;,
\label{eq:03-17}
\end{equation}
giving the deflection angle
\begin{equation}
  \alpha(\theta)=\psi'=\frac{m(\theta)}{\theta}\quad\hbox{with}\quad
  m(\theta)=2\int_0^\theta\theta'\d\theta'\,\kappa(\theta')\;.
\label{eq:03-18}
\end{equation}
Since $m$ is an area integral over the scaled surface mass density of the lens, it is proportional to the lens' mass. The primes denote derivatives with respect to the angular radius $\theta$. Using the definition of the convergence,
\begin{equation}
  \kappa=\frac{1}{2}\nabla^2\psi=\frac{1}{2}\left(\psi''+\frac{\psi'}{\theta}\right)=
  \frac{1}{2}\left(\psi''+\frac{\alpha}{\theta}\right)\;,
\label{eq:03-19}
\end{equation}
we can write the components of the shear
\begin{equation}
  \gamma_1=\frac{\cos2\phi}{2}\left(\psi''-\frac{\psi'}{\theta}\right)\;,\quad
  \gamma_2=\frac{\sin2\phi}{2}\left(\psi''-\frac{\psi'}{\theta}\right)
\label{eq:03-20}
\end{equation}
in the form
\begin{equation}
  \gamma_1=\cos2\phi\left(\kappa-\frac{\alpha}{\theta}\right)\;,\quad
  \gamma_2=\sin2\phi\left(\kappa-\frac{\alpha}{\theta}\right)\;.
\label{eq:03-21}
\end{equation}

The Jacobi determinant
\begin{equation}
  \det(D\varphi)=
  (1-\kappa)^2-\gamma_1^2-\gamma_2^2=(1-\kappa)^2-\left(\kappa-\frac{\alpha}{\theta}\right)^2=
  \left(1-2\kappa+\frac{\alpha}{\theta}\right)\left(1-\frac{\alpha}{\theta}\right)
\label{eq:03-22}
\end{equation} 
can be brought into the form
\begin{equation}
  \det(D\varphi)=\left(1-\frac{\d}{\d\theta}\frac{m(\theta)}{\theta}\right)\left(1-\frac{m(\theta)}{\theta^2}\right)
\label{eq:03-23}
\end{equation}
by means of Eq.~(\ref{eq:03-18}). Axially symmetric lenses thus have critical curves where either the slope of enclosed dimension-less mass $m(\theta)$ divided by $\theta$ equals unity,
\begin{equation}
   \frac{\d}{\d\theta}\frac{m(\theta)}{\theta}=1\;,
\label{eq:03-24}
\end{equation}
or where the enclosed mass equals $\theta^2$,
\begin{equation}
  m(\theta)=\theta^2\;.
\label{eq:03-25}
\end{equation}
Because of the preferred directions of distortion occuring there, critical curves described by Eqs.~(\ref{eq:03-24}) and (\ref{eq:03-25}) are called radial and tangential, respectively.

\subsection{Simple lens models}

\begin{figure}[ht]
  \hfill\includegraphics[width=0.7\hsize]{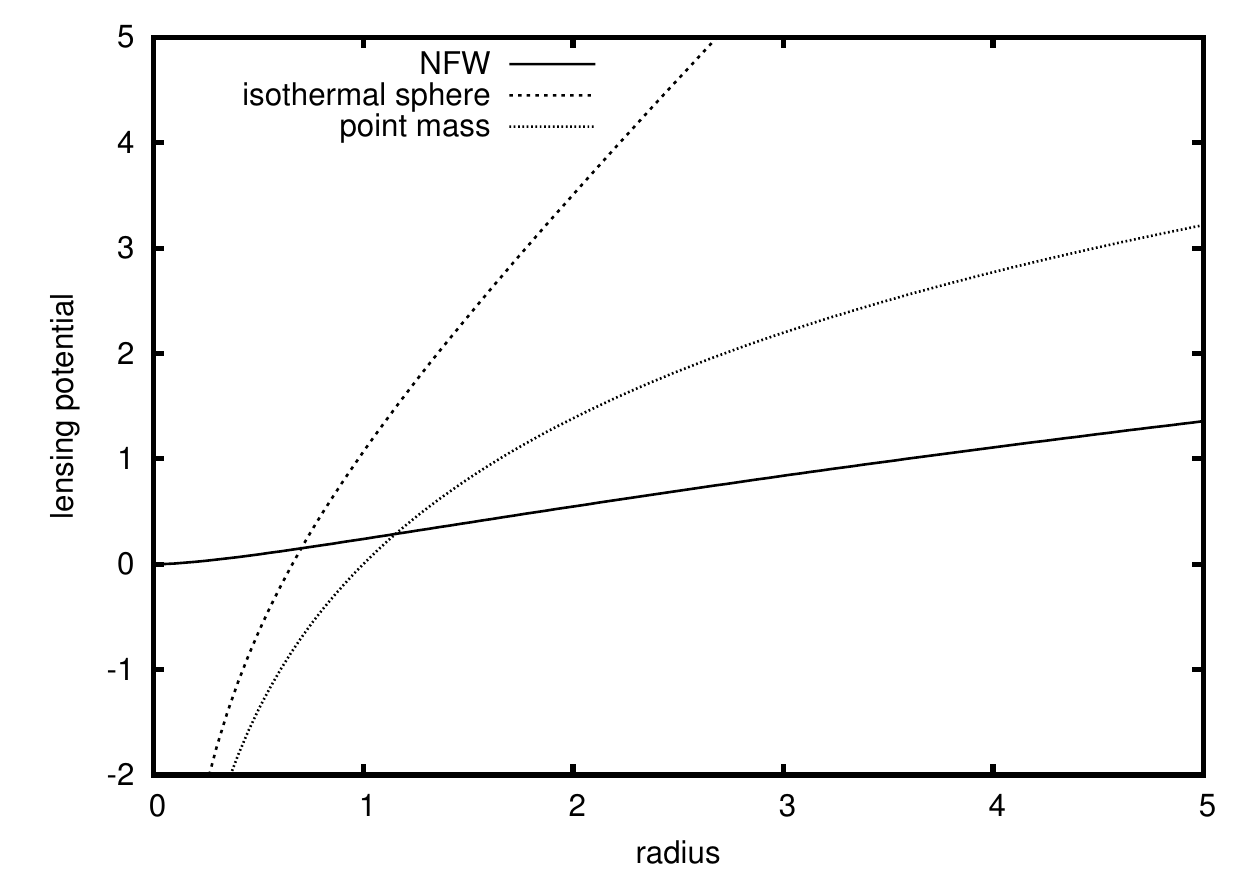}
\caption{Radial dependence of three axially-symmetric lensing potentials introduced here.}
\label{fig:04}
\end{figure}

We begin with the instructive case of a point mass $M$ at the coordinate origin, whose mass density is
\begin{equation}
  \rho(\vec x)=M\delta(\vec x)\;.
\label{eq:03-26}
\end{equation}
Thus, its convergence is
\begin{equation}
  \kappa(\vec\theta)=\frac{4\pi GM}{c^2}
  \frac{(w_\mathrm{s}-w_\mathrm{d})w_\mathrm{d}}{w_\mathrm{s}}
  \delta(w_\mathrm{d}\vec\theta)\;,
\label{eq:03-27}
\end{equation}
and the lensing potential turns out to be
\begin{equation}
  \psi(\vec\theta)=\frac{4GM}{c^2}
  \frac{w_\mathrm{s}-w_\mathrm{d}}{w_\mathrm{d}w_\mathrm{s}}\ln|\vec\theta|\;.
\label{eq:03-28}
\end{equation}
Together with other lensing potentials, this is shown in Fig.~\ref{fig:04}. With the deflection angle
\begin{equation}
  \alpha^i=\partial^i\psi=\frac{4GM}{c^2}
  \frac{w_\mathrm{s}-w_\mathrm{d}}{w_\mathrm{d}w_\mathrm{s}}
  \frac{\theta^i}{\theta^2}\;,
\label{eq:03-29}
\end{equation}
the lens equation turns into
\begin{equation}
  \beta=\theta-\frac{4GM}{c^2\theta}
  \frac{w_\mathrm{s}-w_\mathrm{d}}{w_\mathrm{d}w_\mathrm{s}}\;,
\label{eq:03-30}
\end{equation}
which is effectively one-dimensional because of the axisymmetry of the lens. Since this is a quadratic equation for $\theta$, it always has two solutions unless the source is placed exactly behind the lens, i.e.~at $\beta=0$. Such sources are mapped into rings with radius
\begin{equation}
  \theta_\mathrm{E}=\left(\frac{4GM}{c^2}
  \frac{w_\mathrm{s}-w_\mathrm{d}}{w_\mathrm{d}w_\mathrm{s}}\right)^{1/2}\;,
\label{eq:03-31}
\end{equation}
the so-called Einstein radius. With this definition, the lens equation (\ref{eq:03-30}) can be written
\begin{equation}
  y=x-\frac{1}{x}\;,
\label{eq:03-32}
\end{equation}
where $y=\beta/\theta_\mathrm{E}$ and $x=\theta/\theta_\mathrm{E}$. It has the two solutions
\begin{equation}
  x_\pm=\frac{1}{2}\left(y\pm\sqrt{y^2+4}\right)\;.
\label{eq:03-33}
\end{equation}
The determinant of the Jacobi matrix is
\begin{equation}
  \det(D\varphi)=1-\frac{1}{x^4}\;,
\label{eq:03-34}
\end{equation}
thus the critical curve is a circle with radius $x=1$, while the caustic degenerates to the point $y=0$. The two images at $x_\pm$ are magnified by
\begin{equation}
  \mu_\pm=\frac{y^2+2}{2y\sqrt{y^2+4}}\pm\frac{1}{2}\;,
\label{eq:03-35}
\end{equation}
hence their total magnification is given by
\begin{equation}
  \mu=|\mu_+|+|\mu_-|=\frac{y^2+2}{y\sqrt{y^2+4}}\;.
\label{eq:03-36}
\end{equation} 
This point-mass model and collections thereof are commonly used for studying the gravitational lensing effects of stars and planets.

A model frequently used for a simple description of galaxy-scale lenses starts from the density profile
\begin{equation}
  \rho(r)=\frac{\sigma_v^2}{2\pi G}\frac{1}{r^2+r_\mathrm{c}^2}\;,
\label{eq:03-37}
\end{equation}
which is characterised by a velocity dispersion $\sigma_v$ and a core radius $r_\mathrm{c}$. This density profile is called the non-singular isothermal sphere. It reproduces the approximately flat rotation curves of stars orbiting in galaxies and is thus considered appropriate for modelling gravitational lensing by galaxies. Its surface mass density is
\begin{equation}
  \Sigma(\vec\theta)=\frac{\sigma_v^2}{2Gw_\mathrm{d}}
  \frac{1}{\sqrt{\theta_\mathrm{c}^2+\theta^2}}\;,
\label{eq:03-38}
\end{equation}
where $\theta_\mathrm{c}=r_\mathrm{c}/w_\mathrm{d}$ is the angular core radius. The convergence can be written in the form
\begin{equation}
  \kappa(\theta)=\frac{\kappa_0}{\sqrt{\theta^2+\theta_\mathrm{c}^2}}\;,\quad
  \kappa_0=2\pi\frac{\sigma_v^2}{c^2}\frac{w_\mathrm{s}-w_\mathrm{d}}{w_\mathrm{s}}\;.
\label{eq:03-39}
\end{equation}
The effective lensing potential,
\begin{equation}
  \psi(\theta)=2\kappa_0\left(
    \sqrt{\theta^2+\theta_\mathrm{c}^2}-
    \theta_\mathrm{c}\ln\frac{\theta_\mathrm{c}+\sqrt{\theta^2+\theta_\mathrm{c}^2}}{\theta}
  \right)\;,
\label{eq:03-40}
\end{equation}
can be approximated by
\begin{equation}
  \psi(\theta)\approx2\kappa_0\sqrt{\theta^2+\theta_\mathrm{c}^2}
\label{eq:03-41}
\end{equation}
for $\theta\gg\theta_\mathrm{c}$. The deflection angle
\begin{equation}
  \alpha(\theta)=\frac{2\kappa_0\sqrt{\theta^2+\theta_\mathrm{c}^2}}{\theta}
\label{eq:03-42}
\end{equation}
becomes constant for $\theta_\mathrm{c}\to0$. In the same limit, the singular isothermal sphere has two images at
\begin{equation}
  \theta_\pm=\beta\pm2\kappa_0
\label{eq:03-43}
\end{equation}
if $\beta<2\kappa_0$ and a single image at $\theta_+$ otherwise.

Numerical simulations show that density profiles of dark-matter haloes are well approximated by
\begin{equation}
  \rho(r)=\frac{\rho_\mathrm{s}}{x(1+x)^2}\;,\quad x:=\frac{r}{r_\mathrm{s}}\;.
\label{eq:03-44}
\end{equation}
This density profile has the convergence \cite{BA96.1}
\begin{equation}
  \kappa(\theta)=\frac{2\kappa_\mathrm{s}}{x^2-1}
  \left[1-\frac{2}{\sqrt{1-x^2}}\mathrm{arctanh}\sqrt{\frac{1-x}{1+x}}\right]
\label{eq:03-45}
\end{equation} 
and the lensing potential
\begin{equation}
  \psi(\theta)=4\kappa_\mathrm{s}
  \left[\frac{1}{2}\ln^2\frac{x}{2}-2\mathrm{arctanh}^2\sqrt{\frac{1-x}{1+x}}\right]\;,\quad
  \kappa_\mathrm{s}:=\frac{\rho_\mathrm{s}r_\mathrm{s}}{\Sigma_\mathrm{cr}}\;.
\label{eq:03-46}
\end{equation}

Peculiar cases are the matter sheet of constant convergence $\kappa_0$, whose potential is
\begin{equation}
  \psi(\theta)=\frac{\kappa_0}{2}\theta^2\;,
\label{eq:03-47}
\end{equation}
and the potential of a constant shear $(\gamma_1, \gamma_2)$ with vanishing convergence $\kappa=0$,
\begin{equation}
  \psi(\theta)=\frac{\gamma_1}{2}(\theta_1^2-\theta_2^2)+\gamma_2\theta_1\theta_2\;.
\label{eq:03-48}
\end{equation}

Concluding this brief discussion of axially-symmetric, simple lens models, it should be emphasised that these should be seen as building blocks of more realistic and complete lens models. Often, real lens systems can be modelled only by embedding simple models into external shear, e.g.~by adding a potential of the form Eq.~(\ref{eq:03-48}), by elliptical distortions of formerly symmetric models, or by combining multiple simple components into one lens system. Note that, since we are working in linearised gravity, the superposition principle holds and the potentials of individual lenses may be added.

\subsection{Reconstruction of mass distributions}

Localised lenses can usually be well described in the flat-sky approximation, i.e.~on tangent planes to the sky. In such planes, we can expand the lensing potential into a Fourier series and write the relations Eq.~(\ref{eq:01-57}) and Eq.~(\ref{eq:01-58}) in the algebraic form
\begin{equation}
  \hat\kappa=-\frac{l^2}{2}\hat\psi\;,\quad
  \hat\gamma_1=-\frac{l_1^2-l_2^2}{2}\hat\psi\;,\quad
  \hat\gamma_2=-l_1l_2\hat\psi\;.
\label{eq:03-49}
\end{equation}
Hats above symbols denote Fourier transforms, and $\vec l$ is the two-dimensional wave vector. The shear components $\gamma_1$ and $\gamma_2$ can be measured from the distortions of lensed images, while the scaled surface mass density $\kappa$ is to be recovered. The second and third relations (\ref{eq:03-49}) yield the two independent estimators
\begin{equation}
  \hat\kappa_1=\frac{l^2}{l_1^2-l_2^2}\hat\gamma_1\;,\quad
  \hat\kappa_2=\frac{l^2}{2l_1l_2}\hat\gamma_2
\label{eq:03-50}
\end{equation}
for the convergence. In the linear combination
\begin{equation}
  \kappa_\mathrm{est}=\lambda\hat\kappa_1+(1-\lambda)\hat\kappa_2\;,
\label{eq:03-51}
\end{equation}
the weight $\lambda$ can now be chosen such that the variance
\begin{equation}
  \sigma_\kappa^2=\left\langle\left(\hat\kappa_\mathrm{est}-\langle\hat\kappa\rangle\right)^2\right\rangle
\label{eq:03-52}
\end{equation}
is minimised with respect to $\lambda$. Setting
\begin{equation}
  \frac{\partial\sigma^2_\kappa}{\partial\lambda}=0\;,
\label{eq:03-53}
\end{equation}
using that the shear $\hat\gamma$ should be isotropic on average, $\langle\hat\gamma_1^2\rangle=\langle\hat\gamma_2^2\rangle$, that its components should be independent, $\langle\hat\gamma_1\hat\gamma_2\rangle=0$, and that the average convergence should vanish, $\langle\hat\kappa\rangle=0$, gives the minimum-variance weights
\begin{equation}
  \lambda=\frac{(l_1^2-l_2^2)^2}{l^4}\;,\quad1-\lambda=\frac{2l_1^2l_2^2}{l^4}
\label{eq:03-54}
\end{equation}
and thus the minimum-variance convergence estimator
\begin{equation}
  \hat\kappa=\frac{l_1^2-l_2^2}{l^2}\,\hat\gamma_1+\frac{2l_1l_2}{l^2}\,\hat\gamma_2=
  \cos2\varphi\hat\gamma_1+\sin2\varphi\hat\gamma_2=\Re\left(\e^{-2\ii\varphi}\hat\gamma\right)\;,
\label{eq:03-55}
\end{equation}
where $\varphi$ is the polar angle of the wave vector $\vec l$, and we refer to Eq.~(\ref{eq:01-59}) for the definition of the complex shear. Therefore, in Fourier space, the convergence is obtained from the shear by multiplication with a phase factor. In real space, this corresponds to the convolution
\begin{equation}
  \kappa=\int\mathcal{D}(\vec\theta-\vec\theta')\gamma(\vec\theta')\,\d^2\theta'\;,\quad
  \mathcal{D}(\vec\theta)=\frac{\theta_1^2-\theta_2^2+2\ii\theta_1\theta_2}{\theta^2}\;.
\label{eq:03-56}
\end{equation}
This opens one important approach to mass reconstruction \cite{KA93.1}.

\section{Cosmological weak lensing}

\subsection{Weak-lensing power spectra}

The lensing potential given in Eq.~(\ref{eq:01-49}) is the fundamental quantity for studying gravitational lensing beyond the thin-lens approximaton. We first expand the Newtonian gravitational potential into three-dimensional Fourier modes with respect to co-moving spatial coordinates $\vec x$,
\begin{equation}
  \phi(\vec x)=\int\frac{\d^3q}{(2\pi)^3}\hat\phi(\vec q)\e^{\ii\vec q\cdot\vec x}\;,
\label{eq:02-1}
\end{equation}
introduce the power spectrum
\begin{equation}
  \left\langle\hat\phi(\vec q)\hat\phi^*(\vec q')\right\rangle=(2\pi)^3\delta(\vec q-\vec q')P_\phi(q)\;,
\label{eq:02-2}
\end{equation}
and find the angular correlation function
\begin{equation}
  \left\langle\psi(\vec\theta)\psi(\vec\theta')\right\rangle=
  \int_0^{w_\mathrm{s}}\d w\int_0^{w_\mathrm{s}}\d w'
  \frac{w_\mathrm{s}-w}{w_\mathrm{s}w}\frac{w_\mathrm{s}-w'}{w_\mathrm{s}w'}\,
  \int\frac{\d^3q}{(2\pi)^3}P_\phi(q)\e^{\ii\vec q\cdot(\vec x-\vec x^{\,\prime})}
\label{eq:02-3}
\end{equation}
for the effective lensing potential. The average in Eq.~(\ref{eq:02-2}) is taken over all Fourier modes with fixed wave number $q$ contained in the accessible volume. In Eq.~(\ref{eq:02-3}), $\vec x=(w\vec\theta, w)$ and $\vec x^{\,\prime}=(w'\vec\theta', w')$. We now expand the Fourier basis into spherical harmonics, using
\begin{equation}
  \e^{\ii\vec q\cdot\vec x}=4\pi\sum_{lm}\ii^lj_l(qw)Y_{lm}^*(\vec\theta_x)Y_{lm}(\vec\theta_q)\;,
\label{eq:02-4}
\end{equation}
where $\vec\theta_x$ and $\vec\theta_q$ are the direction angles of $\vec x$ and $\vec q$, respectively. The integration over the angle $\vec\theta_q$ can then be carried out by means of the orthogonality relation of the spherical harmonics, and (\ref{eq:02-3}) turns into
\begin{eqnarray}
  \left\langle\psi(\vec\theta)\psi(\vec\theta')\right\rangle&=&
  \frac{2}{\pi}\sum_{lm}\int_0^{w_\mathrm{s}}\d w\int_0^{w_\mathrm{s}}\d w'
  \frac{w_\mathrm{s}-w}{w_\mathrm{s}w}\frac{w_\mathrm{s}-w'}{w_\mathrm{s}w'}\,\nonumber\\&\times&
  \int_0^\infty q^2\d q\,P_\phi(q)\,j_l(qw)j_l(qw')\,Y_{lm}^*(\vec\theta)Y_{lm}(\vec\theta')\;.
\label{eq:02-5}
\end{eqnarray}
Expanding the effective lensing potential into spherical harmonics, we can write
\begin{equation}
  \psi(\vec\theta)=\sum_{lm}\psi_{lm}Y_{lm}\;,\quad
  \psi_{lm}=\int\d^2\theta\,\psi(\vec\theta)Y_{lm}^*(\vec\theta)
\label{eq:02-6}
\end{equation}
and define the angular power spectrum by
\begin{equation}
  \left\langle\psi_{lm}\psi_{l'm'}^*\right\rangle=\delta_{ll'}\delta_{mm'}C_l^\psi\;.
\label{eq:02-7}
\end{equation}
With (\ref{eq:02-5}), this gives
\begin{equation}
  C_l^\psi=\frac{2}{\pi}\int_0^{w_\mathrm{s}}\d w\int_0^{w_\mathrm{s}}\d w'
  \frac{w_\mathrm{s}-w}{w_\mathrm{s}w}\frac{w_\mathrm{s}-w'}{w_\mathrm{s}w'}\,
  \int_0^\infty q^2\d q\,P_\phi(q)\,j_l(qw)j_l(qw')\;.
\label{eq:02-8}
\end{equation}

This expression allows a further simplification if the typical scales in the large-scale structure, $2\pi/q$, are much smaller than the cosmological distances, $w$ and $w'$. Then, $qw\gg2\pi$, and the spherical Bessel functions $j_l(qw)$ and $j_l(qw')$ vary very quickly compared to $P_\phi(q)$. We can then use
\begin{equation}
  \int_0^\infty q^2\d q\,j_l(qw)j_l(qw')=\frac{\pi}{2w^2}\delta(w-w')\;,
\label{eq:02-9}
\end{equation}
evaluate $P_\phi(q)$ at $q=l/w$ where the spherical Bessel functions $j_l(qw)$ peak, and approximate
\begin{equation}
  C_l^\psi=\int_0^{w_\mathrm{s}}\d w\left(\frac{w_\mathrm{s}-w}{w_\mathrm{s}w^2}\right)^2\,
  P_\phi\left(\frac{l}{w}\right)\;.
\label{eq:02-10}
\end{equation}
This expresses Limber's \cite{LI53.1, LI54.1} approximation.

Since the Newtonian potential $\phi$ is given by the density contrast $\delta$ in terms of the Poisson equation
\begin{equation}
  \vec\nabla^2\phi=\frac{3}{2a}\frac{H_0^2}{c^2}\Omega_{\mathrm{m}0}\delta\;,
\label{eq:02-11}
\end{equation}
the power spectra of the potential and of the density contrast are related by
\begin{equation}
  P_\phi(k)=\frac{9}{4a^2}\left(\frac{H_0}{c}\right)^4\Omega_{\mathrm{m}0}^2\,\frac{P_\delta(k)}{k^4}\;,
\label{eq:02-12}
\end{equation}
and the power spectrum of the effective lensing potential becomes
\begin{equation}
  C_l^\psi=\frac{9}{4}\left(\frac{H_0}{c}\right)^4\Omega_{\mathrm{m}0}^2\frac{1}{l^4}\,
  \int_0^{w_\mathrm{s}}\d w\left(\frac{w_\mathrm{s}-w}{w_\mathrm{s}a(w)}\right)^2\,
  P_\delta\left(\frac{l}{w}\right)\;.
\label{eq:02-13}
\end{equation}
For the convergence $\kappa$, we use $\partial_i\partial^i\,Y_{lm}(\vec\theta)=l(l+1)Y_{lm}(\vec\theta)$ in Eq.~(\ref{eq:02-6}) and find, for $l\gg1$,
\begin{equation}
  C_l^\kappa=\frac{9}{4}\left(\frac{H_0}{c}\right)^4\Omega_{\mathrm{m}0}^2\,
  \int_0^{w_\mathrm{s}}\d w\left(\frac{w_\mathrm{s}-w}{w_\mathrm{s}a(w)}\right)^2\,
  P_\delta\left(\frac{l}{w}\right)\;.
\label{eq:02-14}
\end{equation}
As we shall see, this is identical to the power spectrum of the shear $\gamma$, $C_l^\gamma=C_l^\kappa$. Before we proceed, we need a digression to clarify the treatment of spin-weighed fields on the sphere.

\subsection{Basis vectors and spin-$s$ fields on the sphere}

So far, when we had to introduce coordinates, we used the flat-sky approximation and kept the treatment restricted to local domains in which ordinary, Cartesian coordinates could be used. This is often adequate, but it is instructive and sometimes necessary to extend the domain such that the curvature of the sky can no longer be neglected.

There are two bases commonly used on the sphere. One is the orthonormal coordinate basis
\begin{equation}
  e_1=\partial_\theta\;,\quad e_2=\frac{\partial_\phi}{\sin\theta}
\label{eq:02-15}
\end{equation}
with the dual basis
\begin{equation}
  \theta^1=\d\theta\;,\quad\theta^2=\sin\theta\d\phi\;.
\label{eq:02-16}
\end{equation}
In this basis, the metric has the components $g=\mathrm{diag}(1, 1)$. Often, the helicity basis
\begin{equation}
  e_\pm=\frac{1}{\sqrt{2}}\left(e_1\mp\ii e_2\right)
\label{eq:02-17}
\end{equation}
and its dual basis
\begin{equation}
  \theta^\pm=\frac{1}{\sqrt{2}}\left(\theta^1\pm\ii\theta^2\right)
\label{eq:02-18}
\end{equation}
are more convenient because the basis vectors are eigenvectors under rotations $\mathcal{R}(\varphi)$ on the sphere,
\begin{equation}
  \mathcal{R}(\varphi)e_\pm=\e^{\mp\ii\varphi}e_\pm\;.
\label{eq:02-19}
\end{equation}
Since the metric has the peculiar form
\begin{equation}
  g_{ij}=\left(\begin{array}{cc}0&1\\1&0\end{array}\right)=g^{ij}
\label{eq:02-20}
\end{equation}
in this basis, pulling the indices $\pm$ up or down changes their signs. The only non-vanishing connection forms are
\begin{equation}
  \omega^+_+=-\omega^-_-=-\frac{\cot\theta}{\sqrt{2}}\left(\theta^+-\theta^-\right)\;,
\label{eq:02-21}
\end{equation}
from which covariant derivatives are easily constructed.

Because its basis vectors are eigenvectors under rotations, the helicity basis is particularly useful for representations of spin-weighed quantities. A function ${_sf}$ is said to have spin $s$ if
\begin{equation}
  {_sf}\to\e^{-\ii s\varphi}{_sf}
\label{eq:02-22}
\end{equation}
under right-handed rotations of the coordinate frame by an angle $\varphi$. Let $\mathcal{T}$ be a tensor field of rank $s$ defined everywhere on $\mathbb{S}^2$. It defines a spin-$s$ field on $\mathbb{S}^2$ when applied to $e_-$, and a spin-($-s$) field when applied to $e_+$,
\begin{equation}
  {_st}=\mathcal{T}(e_-, \ldots, e_-)\;,\quad{_{-s}t}=\mathcal{T}(e_+, \ldots, e_+)\;.
\label{eq:02-23}
\end{equation}
The covariant derivative $\nabla\mathcal{T}$ of $\mathcal{T}$ is a rank-($s+1$) tensor field which defines a spin-($s+1$) field
\begin{equation}
  \eth{_st}=-\sqrt{2}(\nabla\mathcal{T})(e_-, \ldots, e_-, e_-)
\label{eq:02-24}
\end{equation}
and a spin-($-s-1$) field
\begin{equation}
  \eth^\dagger{_{-s}t}=-\sqrt{2}(\nabla\mathcal{T})(e_+, \ldots, e_+, e_+)\;.
\label{eq:02-25}
\end{equation}
The operator $\eth$ is called ``edth'' \cite{NE66.1}. It raises the spin by unity, while its adjoint $\eth^\dagger$ lowers the spin by unity.

Spin-$s$ fields on the sphere can be decomposed into the spin-weighed spherical harmonics ${_sY_{lm}}$ defined by
\begin{equation}
  \nabla_\pm({_sY_{lm}})=\sqrt{\frac{(l\pm s)(l\mp s+1)}{2}}{_{s\mp1}Y_{lm}}
\label{eq:02-26}
\end{equation}
such that
\begin{equation}
  \nabla_\pm^2Y_{lm}=\frac{1}{2}\sqrt{\frac{(l+2)!}{(l-2)!}}\,{_{\mp2}Y_{lm}}\;,\quad
  \nabla_+\nabla_-Y_{lm}=-\frac{l(l+1)}{2}\,Y_{lm}=\nabla_-\nabla_+Y_{lm}\;.
\label{eq:02-27}
\end{equation}
The covariant derivatives $\nabla_{\pm}$ are short-hand notations for $\nabla_{e_\pm}$.

\subsection{$E$ and $B$ modes}

Let now $\mathcal{P}$ be a symmetric, trace-free, rank-$2$ tensor field on $\mathbb{S}^2$. In the coordinate basis $\{e_1, e_2\}$, $\mathcal{P}$ may have the components
\begin{equation}
  \mathcal{P}_{ij}=\left(\begin{array}{cc}Q&U\\U&-Q\\\end{array}\right)\;.
\label{eq:02-28}
\end{equation}
This tensor field defines the spin-($\mp2$) fields
\begin{equation}
  \mathcal{P}(e_\pm, e_\pm)=\left(\mathcal{P}_{ij}\theta^i\otimes\theta^j\right)(e_\pm,e_\pm)=
  \frac{1}{2}\left(Q\mp\ii U\right)=:{_{\mp2}p}\;.
\label{eq:02-29}
\end{equation}
The shear tensor $\Gamma$ in the coordinate basis $\{e_1, e_2\}$ introduced in Eq.~(\ref{eq:01-56}) is an example for such a tensor field, with $\gamma_1=Q$ and $\gamma_2=U$. Another example is the linear polarisation tensor, where $Q$ and $U$ represent the usual Stokes parameters.

The fields $_{\pm2}p$ are thus conveniently decomposed into spin-($\pm2$) spherical harmonics,
\begin{equation}
  {_{\pm2}p}=\frac{1}{2}\left(Q\pm\ii U\right)=\sum_{lm}p_{\pm2, lm}\,{_{\pm2}Y_{lm}}\;,
\label{eq:02-30}
\end{equation}
with expansion coefficients $p_{\pm2, lm}$. Applying $\nabla_-^2$ to ${_{-2}p}$ will raise the spin from $-2$ to zero,
\begin{equation}
  \nabla_-^2{_{-2}p}=:q_-=\frac{1}{2}\sum_{lm}p_{-2, lm}\sqrt{\frac{(l+2)!}{(l-2)!}}\,Y_{lm}
\label{eq:02-31}
\end{equation}
according to Eq.~(\ref{eq:02-27}), while applying $\nabla_+^2$ to ${_2p}$ will lower the spin from $2$ to zero,
\begin{equation}
  \nabla_+^2{_2p}=:q_+=\frac{1}{2}\sum_{lm}p_{2, lm}\sqrt{\frac{(l+2)!}{(l-2)!}}\,Y_{lm}\;.
\label{eq:02-32}
\end{equation}
Since the two fields $q_\pm$ both have spin-$0$, they are independent of the orientation of the coordinate frame in which they are measured. Now we have, from Eqs.~(\ref{eq:02-31}) and (\ref{eq:02-32}),
\begin{equation}
  q_{\pm, lm}=\int\d\Omega\,q_\pm Y_{lm}^*=\frac{1}{2}p_{\pm2, lm}\,\sqrt{\frac{(l+2)!}{(l-2)!}}
\label{eq:02-33}
\end{equation}
and thus
\begin{equation}
  p_{\pm2, lm}=2\sqrt{\frac{(l-2)!}{(l+2)!}}\,q_{\pm, lm}\;.
\label{eq:02-34}
\end{equation}

Imagine now observing the celestial sphere once inside-out and once outside-in. North will remain North, but East and West will be interchanged by this change of viewpoint, i.e.~the coordinate basis $\{e_1, e_2\}$ will transform into $\{e_1, -e_2\}$. Since
\begin{equation}
  Q=\mathcal{P}(e_1, e_1)\;,\quad U=\mathcal{P}(e_1, e_2)\;,
\label{eq:02-35}
\end{equation}
with the tensor Eq.~(\ref{eq:02-28}),
\begin{equation}
  Q\to Q\;,\quad U\to-U
\label{eq:02-36}
\end{equation}
under this parity transformation.

Since the tensor components $Q$ and $U$ can be combined from the spin-$\pm2$ fields ${_{\pm2}p}$ as
\begin{equation}
  Q=\left({_2}p+{_{-2}p}\right)\;,\quad
  U=-\ii\left({_2p}-{_{-2}p}\right)\;,
\label{eq:02-37}
\end{equation}
the linear combinations of spherical-harmonic coefficients
\begin{equation}
  a_{E, lm}=-\left(p_{2, lm}+p_{-2, lm}\right)\;,\quad
  a_{B, lm}=-\ii\left(p_{2, lm}-p_{-2, lm}\right)
\label{eq:02-38}
\end{equation}
define a parity-conserving $E$-mode and a parity-changing $B$-mode. The terminology reminds of the electric and magnetic fields because the electric field, as a gradient field, is invariant under parity changes, while the magnetic field, as a curl field, is not. A simple example for a $B$-mode is a cyclone on Earth: a satellite looking at it from above will see it with opposite parity as a sailor looking at it from below.

With Eq.~(\ref{eq:02-34}), we obtain the spherical-harmonic coefficients
\begin{eqnarray}
  a_{E, lm}&=&-2\sqrt{\frac{(l-2)!}{(l+2)!}}\left(q_{+, lm}+q_{-, lm}\right)\;,\nonumber\\
  a_{B, lm}&=&-2\ii\sqrt{\frac{(l-2)!}{(l+2)!}}\left(q_{+, lm}-q_{-, lm}\right)\;.
\label{eq:02-39}
\end{eqnarray}
from Eq.~(\ref{eq:02-38}). They define the $E$ and $B$ modes
\begin{equation}
  E(\vec\theta)=\sum_{lm}a_{E, lm}Y_{lm}(\vec\theta)\;,\quad
  B(\vec\theta)=\sum_{lm}a_{B, lm}Y_{lm}(\vec\theta)\;.
\label{eq:02-40}
\end{equation}
While this decomposition is completely general for spin-($\pm2$) fields on $\mathbb{S}^2$, we know that
\begin{equation}
  {_{\pm2}p}=\nabla_\mp^2\psi
\label{eq:02-41}
\end{equation}
for gravitational lensing. Decomposing the lensing potential into spherical harmonics,
\begin{equation}
  \psi=\sum_{lm}\psi_{lm}Y_{lm}\;,
\label{eq:02-42}
\end{equation}
we find
\begin{equation}
  \nabla_{\pm}^2\psi=\frac{1}{2}\sum_{lm}\sqrt{\frac{(l+2)!}{(l-2)!}}\psi_{lm}\;{_{\mp2}Y_{lm}}\;,
\label{eq:02-43}
\end{equation}
and thus, by comparison with Eq.~(\ref{eq:02-30}),
\begin{equation}
  p_{\pm2, lm}=\frac{1}{2}\sqrt{\frac{(l+2)!}{(l-2)!}}\psi_{lm}\;.
\label{eq:02-44}
\end{equation}
Therefore, $a_{B, lm}=0$, showing that lensing alone cannot create a $B$-mode distortion pattern.

Simplifications are possible in the flat-sky approximation, when we can set $\sin\theta=1$ in Eq.~(\ref{eq:02-15}). Then, we introduce Cartesian coordinates $\theta_1=\theta$ and $\theta_2=\phi$ and have
\begin{equation}
  \nabla_{1,2}=\partial_{1,2}\;,\quad\nabla_{\pm}=\frac{1}{\sqrt{2}}\left(\partial_1\mp\ii\partial_2\right)\;,\quad
  \eth=\partial_1+\ii\partial_2\;,\quad\eth^\dagger=\partial-\ii\partial_2\;.
\label{eq:02-45}
\end{equation}
The decomposition into spherical harmonics can be replaced by Fourier transforms,
\begin{equation}
  f(\vec\theta)=\sum_{lm}f_{lm}Y_{lm}\to\int\frac{\d^2l}{(2\pi)^2}\hat f(\vec l)\,\e^{\ii\vec l\cdot\vec\theta}\;,
\label{eq:02-46}
\end{equation}
such that
\begin{equation}
  \nabla_\pm^2f(\vec\theta)=
  -\int\frac{\d^2l}{(2\pi)^2}\left(l_1^2-l_2^2\mp2\ii l_1l_2\right)\hat f(\vec l)\e^{\ii\vec l\cdot\vec\theta}
\label{eq:02-47}
\end{equation}
for any field $f(\vec\theta)$ on the sphere. Identifying now $f$ with $Q\pm\ii U$ in Eq.~(\ref{eq:02-47}) shows that
\begin{equation}
  q_{\pm, lm}\to-\frac{1}{2}\left(l_1^2-l_2^2\right)\hat Q+l_1l_2\hat U\mp
  \ii\left[l_1l_2\hat Q+\frac{1}{2}\left(l_1^2-l_2^2\right)\hat U\right]\;.
\label{eq:02-48}
\end{equation}
The flat-sky approximation is valid on small angular scales only, i.e.~for $l\gg1$. In this limit,
\begin{equation}
  \sqrt{\frac{(l-2)!}{(l+2)!}}\approx\frac{1}{l^2}\;,
\label{eq:02-49}
\end{equation}
and Eqs.~(\ref{eq:02-34}) and (\ref{eq:02-39}) give
\begin{equation}
  a_{E, lm}\to\frac{l_1^2-l_2^2}{l^2}\hat Q-2\frac{l_1l_2}{l^2}\hat U\;,\quad
  a_{B, lm}\to2\frac{l_1l_2}{l^2}\hat Q+\frac{l_1^2-l_2^2}{l^2}\hat U\;.
\label{eq:02-50}
\end{equation}
Introducing the angle $\varphi$ between $\vec l$ and $\vec\theta$, we thus find
\begin{equation}
  \hat E=\hat Q\cos2\varphi-\hat U\sin2\varphi\;,\quad
  \hat B=\hat Q\sin2\varphi+\hat U\cos2\varphi
\label{eq:02-51}
\end{equation}
for the Fourier-transformed $E$ and $B$ modes in the flat-sky approximation.

\subsection{Shear correlation functions}

After this detour through spin-$s$ fields on the sphere, we can now return to the treatment of shear correlation functions on the sky. Clearly, the shear components depend on the local coordinate frame and must be parallel-transported along well-defined curves when they are to be compared at two different locations on the sky. To avoid complications and ambiguities, it is much more appropriate to first combine them into spin-$0$ fields which are independent of the local coordinate orientations, and then compute their correlation properties. The preceding discussion tells us how to do this in a systematic way.

Locally, we can return to the flat-sky approximation and combine the shear components $\gamma_{1,2}$ into the spin-$\pm2$ quantities
\begin{equation}
  \gamma=\gamma_1+\ii\gamma_2\;,\quad\gamma^*=\gamma_1-\ii\gamma_2\;.
\label{eq:02-52}
\end{equation}
Expressed by the edth operator in the flat-sky approximation $\eth=\partial_1+\ii\partial_2$,
\begin{equation}
  \kappa=\frac{1}{2}\eth^\dagger\eth\psi\;,\quad\gamma=\frac{1}{2}\eth^2\psi\;,\quad
  F=\frac{1}{2}\eth\eth^\dagger\eth\psi\;,\quad G=\frac{1}{2}\eth^3\psi\;,
\label{eq:02-52a}
\end{equation}
which shows immediately that the $F$ and $G$ flexions as defined in Eqs.~(\ref{eq:01-64a}) and (\ref{eq:01-64b}) are spin-1 and spin-3 fields, respectively.

\begin{figure}[ht]
  \hfill\includegraphics[width=0.7\hsize]{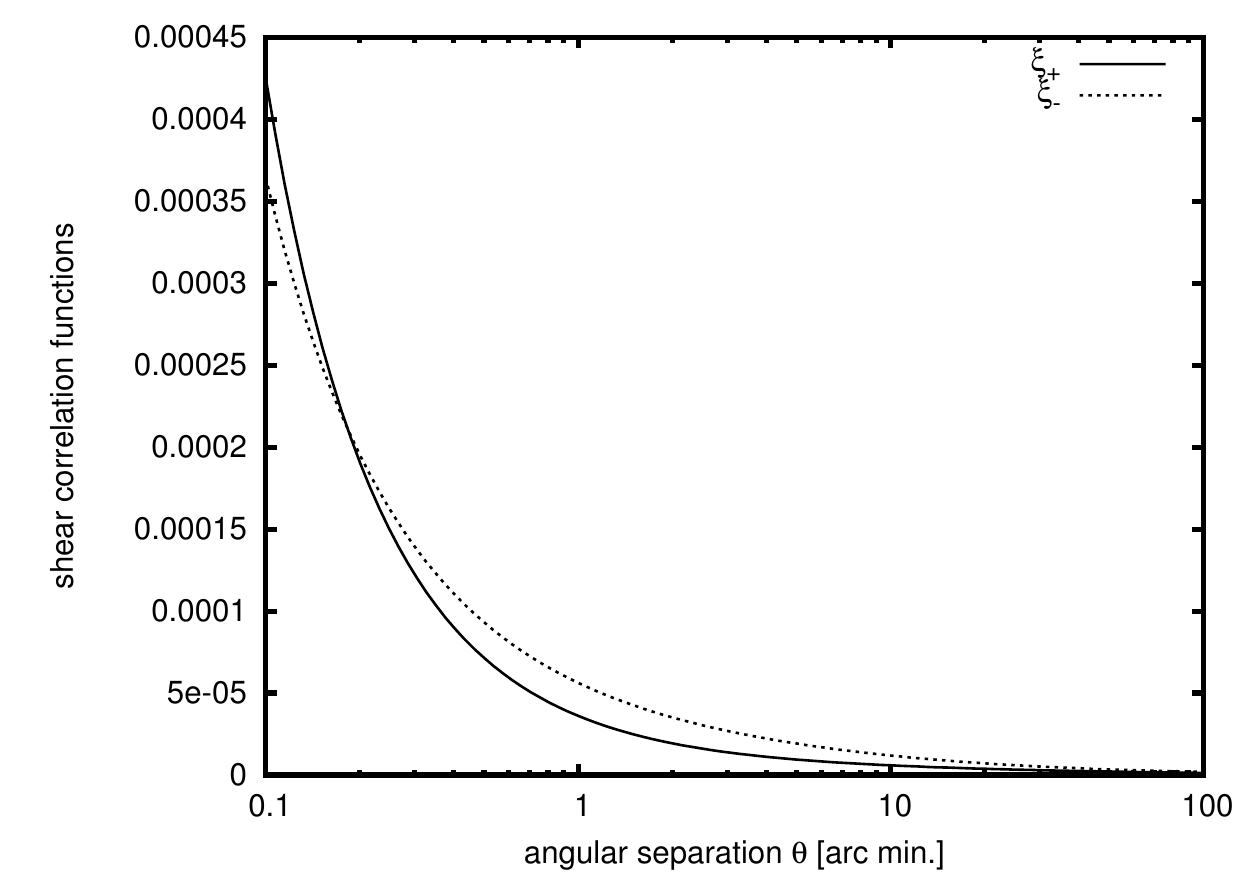}
\caption{Shear correlation functions $\xi_\pm(\theta)$ as functions of the angular separation $\theta$ in arc minutes. The standard cosmological parameters given in Tab.~\ref{tab:01} are assumed, and the source redshift is $z_\mathrm{s}=1$.}
\label{fig:05}
\end{figure}

The (negative) determinant of the shear tensor,
\begin{equation}
  -\det\Gamma=\gamma_1^2+\gamma_2^2=\gamma\gamma^*=|\gamma|^2\;,
\label{eq:02-53}
\end{equation}
has spin-$0$ and is thus independent of the orientation of the local coordinate frame. In the flat-sky approximation, we can Fourier transform the shear and find
\begin{equation}
  \hat\gamma\hat\gamma^*=\hat\gamma_1^2+\hat\gamma_2^2=\frac{l^4}{4}\hat\psi\hat\psi^*=
  \hat\kappa\hat\kappa^*\;,
\label{eq:02-54}
\end{equation}
which shows immediately that the shear power spectrum defined in this way is identical to the convergence power spectrum,
\begin{equation}
  C_l^\gamma=\left\langle\hat\gamma\hat\gamma^*\right\rangle=C_l^\kappa\;,
\label{eq:02-55}
\end{equation}
which was already given in Eq.~(\ref{eq:02-14}). Other shear power spectra can be defined by converting the shear into spin-$0$ fields in alternative ways. The $E$ and $B$ modes defined in Eq.~(\ref{eq:02-40}) and expressed in the flat-sky approximation in Eq.~(\ref{eq:02-51}) have spin-$0$ by construction. According to Eq.~(\ref{eq:02-51}), their products are given by
\begin{eqnarray}
  \hat E\hat E^*&=&\cos^22\varphi\hat\gamma_1\hat\gamma_1^*+
  \sin^22\varphi\hat\gamma_2\hat\gamma_2^*\quad\mbox{and}\nonumber\\
  \hat B\hat B^*&=&\sin^22\varphi\hat\gamma_1\hat\gamma_1^*+
  \cos^22\varphi\hat\gamma_2\hat\gamma_2^*\;,
\label{eq:02-56}
\end{eqnarray}
where
\begin{eqnarray}
  \hat\gamma_1\hat\gamma_1^*=\frac{(l_1^2-l_2^2)^2}{4}\hat\psi\hat\psi^*=
  \cos^22\varphi\hat\kappa\hat\kappa^*\quad\mbox{and}\nonumber\\
  \hat\gamma_2\hat\gamma_2^*=(l_1l_2)^2\hat\psi\hat\psi^*=
  \sin^22\varphi\hat\kappa\hat\kappa^*\;.
\label{eq:02-57}
\end{eqnarray}
Obviously,
\begin{equation}
  \hat E\hat E^*+\hat B\hat B^*=
  \hat\kappa\hat\kappa^*=\hat\gamma\hat\gamma^*\quad\mbox{and}\quad
  \hat E\hat E^*-\hat B\hat B^*=\cos4\varphi\hat\gamma\hat\gamma^*\;.
\label{eq:02-58}
\end{equation}
The two-point correlation functions of convergence and shear are given by the inverse Fourier transforms
\begin{equation}
  \xi_\kappa(\theta)=\int_0^\infty\frac{l\d l}{2\pi}C_l^\kappa\mbox{J}_0(l\theta)=\xi_\gamma(\theta)
\label{eq:02-59}
\end{equation}
or the redundant definitions
\begin{equation}
  \xi_\pm(\theta)=\left\{
  \begin{array}{ll}\displaystyle
    \int_0^\infty\frac{l\d l}{2\pi}C_l^\kappa\mbox{J}_0(l\theta) & (+) \\\\
    \displaystyle
    \int_0^\infty\frac{l\d l}{2\pi}C_l^\kappa\mbox{J}_4(l\theta) & (-) \\
  \end{array}\right.\;.
\label{eq:02-60}
\end{equation}
Examples for these shear correlation functions are shown in Fig.~\ref{fig:05}.

\subsection{Gravitational lensing of the CMB temperature}

Gravitational lensing of the CMB has the effect that the relative temperature fluctuation \textit{observed} in direction $\vec\theta$,
\begin{equation}
  \tau_\mathrm{obs}(\vec\theta)=\frac{T(\vec\theta)-\bar T}{\bar T}\;,
\label{eq:02-72}
\end{equation}
is the \textit{intrinsic} temperature fluctuation in the \textit{deflected} direction $\vec\theta-\vec\alpha$,
\begin{equation}
  \tau_\mathrm{obs}(\vec\theta)=\tau[\vec\theta-\vec\alpha(\vec\theta)]\;.
\label{eq:02-73}
\end{equation}

Since typical scales in $\tau$ are substantially larger than those in $\vec\alpha$, we can Taylor-expand Eq.~(\ref{eq:02-73}) and write
\begin{equation}
  \tau_\mathrm{obs}(\vec\theta)=\tau(\vec\theta)-\vec\alpha\cdot\vec\nabla\tau(\vec\theta)+
  \frac{1}{2}\frac{\partial^2\tau}{\partial\theta_i\partial\theta_j}\alpha_i\alpha_j\;.
\label{eq:02-74}
\end{equation}
These terms are more easily expressed in Fourier space. Since the deflection angle is the gradient of the lensing potential Eq.~(\ref{eq:01-49}),
\begin{eqnarray}
  \vec\alpha&=&\int\frac{\d^2l}{(2\pi)^2}(\ii\vec l\hat\psi)\e^{\ii\vec l\cdot\vec\theta}
  \quad\mbox{and}\nonumber\\
  \vec\nabla\tau(\vec\theta)&=&
  \int\frac{\d^2l}{(2\pi)^2}(\ii\vec l\hat\tau)\e^{\ii\vec l\cdot\vec\theta}
\label{eq:02-75}
\end{eqnarray}
in terms of the Fourier transforms $\hat\psi$ and $\hat\tau$. Consequently, the scalar product $\vec\alpha\cdot\vec\nabla\tau$ can be written as
\begin{equation}
  \vec\alpha\cdot\vec\nabla\tau=-\int\frac{\d^2l_1}{(2\pi)^2}\int\frac{\d^2l_2}{(2\pi)^2}
  \vec l_1\cdot\vec l_2\hat\psi(\vec l_1)\hat\tau(\vec l_2)
  \e^{\ii(\vec l_1+\vec l_2)\cdot\vec\theta}\;,
\label{eq:02-76}
\end{equation}
which has the Fourier transform
\begin{equation}
  -\int\frac{\d^2l_1}{(2\pi)^2}\,
  \vec l_1\cdot(\vec l-\vec l_1)\hat\psi(\vec l_1)\hat\tau(\vec l-\vec l_1)\;.
\label{eq:02-77}
\end{equation}
Likewise, we can write
\begin{eqnarray}
  \frac{1}{2}\frac{\partial^2\tau}{\partial\theta_i\partial\theta_j}\alpha_i\alpha_j&=&
  \frac{1}{2}\int\frac{\d^2l_1}{(2\pi)^2}l_{1i}l_{1j}\hat\tau(\vec l_1)\\
  &\times&
  \int\frac{\d^2l_2}{(2\pi)^2}l_{2i}\hat\psi(\vec l_2)
  \int\frac{\d^2l_3}{(2\pi)^2}l_{3j}\hat\psi(\vec l_3)
  \e^{\ii(\vec l_1+\vec l_2+\vec l_3)\cdot\vec\theta}\nonumber\;,
\label{eq:02-78}
\end{eqnarray}
whose Fourier transform is
\begin{equation}
  \frac{1}{2}\int\frac{\d^2l_1}{(2\pi)^2}\int\frac{\d^2l_2}{(2\pi)^2}
  (\vec l_1\cdot\vec l_2)[\vec l_1\cdot(\vec l-\vec l_1-\vec l_2)]
  \hat\tau(\vec l_1)\hat\psi(\vec l_2)\hat\psi(\vec l-\vec l_1-\vec l_2)\;.
\label{eq:02-79}
\end{equation}

Imagine now we Fourier transform Eq.~(\ref{eq:02-74}) using the expressions Eq.~(\ref{eq:02-77}) and Eq.~(\ref{eq:02-79}), square the result, keep terms up to second order in the lensing potential, and average them over the sky to obtain the power spectrum of the observed CMB temperature fluctuations. In increasing order of powers in the deflection angle, the first term squared yields the intrinsic CMB power spectrum. The second term squared, and the product of the first and the third terms, lead to a convolution of the power spectra for $\tau$ and $\psi$. The product of the first and the second terms vanishes because the mean deflection angle does, and other terms are of higher than second order in $\psi$. A somewhat lengthy, but straightforward calculation gives
\begin{equation}
  C_{l,\mathrm{obs}}^\tau=C_l^\tau\left(1-l^2R_\psi\right)+
  \int\frac{\d^2l_1}{(2\pi)^2}\left[\vec l_1\cdot(\vec l-\vec l_1)\right]^2
  C_{l_1}^\psi C_{|\vec l-\vec l_1|}^\tau\;,
\label{eq:02-80}
\end{equation}
where
\begin{equation}
  R_\psi=\frac{1}{3\pi}\int_0^\infty l^3\d l\,C_l^\psi\approx3\times10^{-7}
\label{eq:02-81}
\end{equation}
quantifies the total power in the deflection angle.

We thus see two effects of lensing on the CMB temperature power spectrum \cite{SE96.2}. First, its amplitude is lowered by a factor increasing quadratically with the multipole order. Second, it is convolved with the power spectrum of the lensing potential, which gives rise to a slight smoothing. For $l\gtrsim2000$, the factor $(1-l^2R_\psi)$ becomes negative, which signals the break-down of the Taylor approximation (\ref{eq:02-74}). There, however, the amplitude of $C_l^\tau$ is already so low that this is practically irrelevant. For large $l$, we can thus ignore the first term in (\ref{eq:02-80}) and approximate $(\vec l-\vec l_1)\approx\vec l$ in the second term. This gives
\begin{equation}
  C_{l,\mathrm{obs}}^\tau\approx l^2C_l^\psi R_\tau\quad\mbox{with}\quad
  R_\tau=\frac{1}{3\pi}\int_0^\infty l^3\d l\,C_l^\tau\approx10^{-9}\,\mu\mathrm{K}^2\;.
\label{eq:02-82}
\end{equation}
Lensing thus creates power in the Silk damping tail of the CMB \cite{ME97.1}.

\subsection{Gravitational lensing of the CMB polarisation}

Since Thomson scattering does not produce circular polarisation, the polarisation of the CMB can be described by a symmetric, trace-free, rank-2 tensor field $\mathcal{P}$ whose components in the coordinate basis are the usual Stokes parameters $Q$ and $U$ as given in Eq.~(\ref{eq:01-28}). As described before, and entirely analogous to the shear, we thus decompose the polarisation into $E$ and $B$ modes which, in the flat-sky approximation, are given by Eq.~(\ref{eq:01-51}), which can be combined into the form
\begin{equation}
  \hat\mathcal{E}\pm\ii\hat\mathcal{B}=\e^{\pm2\ii\varphi}\left(\hat Q\pm\ii\hat U\right)
\label{eq:02-83}
\end{equation}
To avoid confusion with the $E$ and $B$ modes of the gravitational shear, we use curly symbols for the $\mathcal{E}$ and $\mathcal{B}$ modes of the polarisation. Since no $\mathcal{B}$-mode polarisation is expected to have been produced when the CMB was released, we can write
\begin{equation}
  \hat Q\pm\ii\hat U=\e^{\mp2\ii\varphi}\hat\mathcal{E}\;.
\label{eq:02-84}
\end{equation}
The following steps are conceptually simple, but require lengthy calculations. First, we replace the temperature contrast $\tau$ by the combinations $Q\pm\ii U$ of Stokes parameters in Eq.~(\ref{eq:02-74}). According to Eq.~(\ref{eq:02-84}), this implies replacing $\hat\tau(\vec l_1)$ by $\e^{\mp2\ii\phi_1}\hat\mathcal{E}(\vec l_1)$ in the expressions Eq.~(\ref{eq:02-77}) and Eq.~(\ref{eq:02-79}). The results need to be multiplied by $\e^{\pm2\ii\phi}$ to find the observed, i.e.~lensed, combinations $(\hat\mathcal{E}\pm\ii\hat\mathcal{B})_\mathrm{obs}$. Having obtained those, we can form the power spectra
\begin{equation}
  \left\langle(\hat\mathcal{E}+\ii\hat\mathcal{B})(\hat\mathcal{E}+\ii\hat\mathcal{B})^*\right\rangle=
  C_l^\mathcal{E}+C_l^\mathcal{B}
\label{eq:02-120}
\end{equation}
and
\begin{equation}
  \left\langle(\hat\mathcal{E}+\ii\hat\mathcal{B})(\hat\mathcal{E}-\ii\hat\mathcal{B})^*\right\rangle=
  C_l^\mathcal{E}-C_l^\mathcal{B}\;.
\label{eq:02-121}
\end{equation}
Their sum and difference yield the separate power spectra $C_l^\mathcal{E}$ and $C_l^\mathcal{B}$,
\begin{eqnarray}
  C_{l,\mathrm{obs}}^\mathcal{E}&=&(1-l^2R_\psi)C_l^\mathcal{E}\nonumber\\&+&
  \int\frac{\d^2l_1}{(2\pi)^2}\left[\vec l_1(\vec l-\vec l_1)\right]^2
  \cos^22(\phi_1-\phi)C_{|\vec l-\vec l_1|}^\psi C_{l_1}^\mathcal{E}\;,\nonumber\\
  C_{l,\mathrm{obs}}^\mathcal{B}&=&
  \int\frac{\d^2l_1}{(2\pi)^2}\left[\vec l_1(\vec l-\vec l_1)\right]^2
  \sin^22(\phi_1-\phi)C_{|\vec l-\vec l_1|}^\psi C_{l_1}^\mathcal{E}\;.
\label{eq:02-122}
\end{eqnarray}
Thus, the observable $\mathcal{E}$-mode polarisation power spectrum differs from the temperature power spectrum Eq.~(\ref{eq:02-80}) merely by the phase factor $\cos^22(\phi_1-\phi)$ in the integral, where $\phi$ and $\phi_1$ are the angles enclosed by $\vec l$ and $\vec l_1$ with the $x_1$ axis. This phase factor differs from unity when the Fourier modes of the lensing potential and the intrinsic $\mathcal{E}$-mode polarisation are not aligned. It is very important that any such misalignment will create $\mathcal{B}$-mode from $\mathcal{E}$-mode polarisation, as Eq.~(\ref{eq:02-122}) demonstrates: when $\phi_1\ne\phi$, the phase factor $\sin^22(\phi_1-\phi)$ differs from zero, and the observed $B$-mode power spectrum is different from zero although any intrinsic $\mathcal{B}$-modes were explicitly ignored. Thus, we have arrived at a third effect of gravitational lensing on the CMB: lensing will create $\mathcal{B}$- from $\mathcal{E}$-mode polarisation \cite{ZA98.1}.

Structures in the CMB are typically much larger than structures in the gravitationally-lensing density field. Thus, we can take the limit $l\ll l_1$ and approximate
\begin{equation}
  C_{l,\mathrm{obs}}^\mathcal{B}\approx\int\frac{\d^2l_1}{(2\pi)^2}l_1^4
  \sin^22(\phi_1-\phi)C_{l_1}^\psi C_{l_1}^\mathcal{E}=
  \int \frac{\d l_1}{4\pi}l_1^5\,C_{l_1}^\psi C_{l_1}^\mathcal{E}\;,
\label{eq:02-123}
\end{equation}
which becomes independent of $l$.

\subsection{Recovery of the lensing potential}

We do not know the appearance of the unlensed CMB, but lensing produces characteristic patterns on the CMB which may reveal its presence. In particular, lensing mixes Fourier modes and thus correlates the formerly unrelated modes
\begin{equation}
  \hat\tau_\mathrm{obs}(\vec l)\quad\hbox{and}\quad
  \hat\tau_\mathrm{obs}(\vec l+\vec L)\;.
\label{eq:02-124}
\end{equation}
To lowest order in the lensing potential $\psi$,
\begin{equation}
  \hat\tau_\mathrm{obs}(\vec l)=\hat\tau(\vec l)-
  \int\frac{\d^2l_1}{(2\pi)^2}\vec l_1(\vec l-\vec l_1)\hat\psi(\vec l_1)\hat\tau(\vec l-\vec l_1)\;,
\label{eq:02-125}
\end{equation} 
so the expected correlation of different modes turns out to be
\begin{eqnarray}
  \left\langle
    \hat\tau_\mathrm{obs}(\vec l)\hat\tau^*_\mathrm{obs}(\vec l-\vec L)
  \right\rangle&=&(2\pi)^2C_l^\tau\delta(\vec L)\\&+&\left[
    \vec L(\vec L-\vec l)C_{|\vec l-\vec L|}^\tau+\vec L\cdot\vec lC_l^\tau
  \right]\hat\psi(\vec L)\;.\nonumber
\label{eq:02-126}
\end{eqnarray}
To lowest order in $\hat\psi$, lensing creates off-diagonal terms in the CMB temperature power spectrum, i.e.~it couples different fluctuation modes which would otherwise be independent.

This motivates the construction of a filter $w(\vec l,\vec L)$, of which we require that the convolution
\begin{equation}
  N(\vec L)\int\frac{\d^2l}{(2\pi)^2}\left[
    \hat\tau_\mathrm{obs}(\vec l)\hat\tau^*_\mathrm{obs}(\vec l-\vec L)
  \right]w(\vec l,\vec L)=\hat\psi_\mathrm{est}(\vec L)
\label{eq:02-127}
\end{equation} 
return an unbiased estimate $\hat\psi_\mathrm{est}(\vec L)$ of the lensing potential. The requirement that the filter be unbiased,
\begin{equation}
  \left\langle\hat\psi_\mathrm{est}\right\rangle=\hat\psi\;,
\label{eq:02-128}
\end{equation}
puts a normalisation constraint on the prefactor $N(\vec L)$,
\begin{equation}
  N^{-1}(\vec L)=\int\frac{\d^2l}{(2\pi)^2}\left[
    \vec L(\vec L-\vec l)C_{|\vec L-\vec l|}^\tau+\vec L\cdot\vec lC_l^\tau
  \right]w(\vec l,\vec L)\;.
\label{eq:02-129}
\end{equation}
The shape of the filter $w(\vec l,\vec L)$ follows from the requirement that the significance of its detections be optimised, i.e.~that the variance
\begin{equation}
  \left\langle\left(\hat\psi_\mathrm{est}-\hat\psi\right)^2\right\rangle=
  \left\langle|\hat\psi_\mathrm{est}|^2\right\rangle+
  \left\langle|\hat\psi|^2\right\rangle\approx
  \left\langle|\hat\psi_\mathrm{est}|^2\right\rangle
\label{eq:02-130}
\end{equation}
be minimised. Inserting the expression Eq.~(\ref{eq:02-127}) yields
\begin{equation}
  \left\langle\hat\psi_\mathrm{est}^*(\vec L)\hat\psi_\mathrm{est}(\vec L')\right\rangle=
  N^2(\vec L)\delta(\vec L-\vec L')\int\frac{\d^2l_1}{(2\pi)^2}
  C_{l_1}^\tau C_{|\vec l_1-\vec L|}^\tau w^2(\vec l_1,\vec L)\;.
\label{eq:02-131}
\end{equation}
Varying this with respect to the filter $w(\vec l,\vec L)$ under the bias constraint Eq.~(\ref{eq:02-128}), taking into account that the normalisation factor $N(\vec L)$ also depends on $w(\vec l,\vec L)$, and setting the variation to zero returns the filter
\begin{equation}
  w(\vec l,\vec L)=\frac
  {\vec L(\vec L-\vec l)C_{|\vec l_1-\vec L|}^\tau+\vec L\cdot\vec lC_l^\tau}
  {C_{|\vec l_1-\vec L|}^\tau C_l^\tau}\;,
\label{eq:02-132}
\end{equation}
whose application to the squared temperature field will return an unbiased, optimised estimate of the lensing potential \cite{HU01.2}. This demonstrates one example for the recovery of the lensing effects on the CMB (see also \cite{HI03.1, OK03.1}).

These remarks conclude the formal part of the review. We shall now proceed discussing applications of gravitational lensing to cosmic mass distributions of increasing scale.

\section{Structure and contents of galaxies}

\subsection{MACHOs and planets}

We know that the most of the matter in the universe is dark, but we do not know what this dark matter consists of. We know that it must not interact electromagnetically, because otherwise the cosmic microwave background would show temperature fluctuations on the level of $10^{-3}\,\mathrm{K}$ rather than $10^{-5}\,\mathrm{K}$. We also know that the dark matter must be cold in the sense that the velocity of its constituents must be small compared to the speed of light, because otherwise the large-scale distribution of the galaxies would be different. It is likely that this cold dark matter is composed of weakly-interacting elementary particles, but it could equally well consist of compact objects like, e.g.~low-mass black holes.

Gravitational lensing provides one way to test this possibility. Our galaxy, the Milky Way, is expected to be embedded into a halo which also predominantly consists of dark matter. If that dark matter was composed of compact objects rather than elementary particles, lines-of-sight through the Galaxy would occasionally pass close to one of those. They would act as point-mass lenses on sources in their background \cite{EI36.1}. Although their image splitting would be substantially below the detection threshold, they would cause a well-measurable magnification \cite{PA86.1}.

Quite independent of the mass spectrum of these hypothetical compact objects (called MACHOs\footnote{acronym for massive compact halo objects}), the probability of any one of them causing a microlensing event at any instant of time is of order $(v/c)^2$, where $v\simeq220\,\mathrm{km\,s^{-1}\,Mpc^{-1}}$ is a typical velocity for the stars in the Galaxy. Consequently, this microlensing optical depth is of order $10^{-6}$. Finding its magnification signature thus requires of order $10^6$ light curves to be monitored. Originally perceived more as science fiction, projects were carried out which observed sufficient numbers of stars in the Large and Small Magellanic Clouds (LMC and SMC, respectively) with sufficient accuracy and time sampling for detecting some microlensing events among the overwhelming signal from variable stars.

\begin{figure}[ht]
  \hfill\includegraphics[width=0.7\hsize]{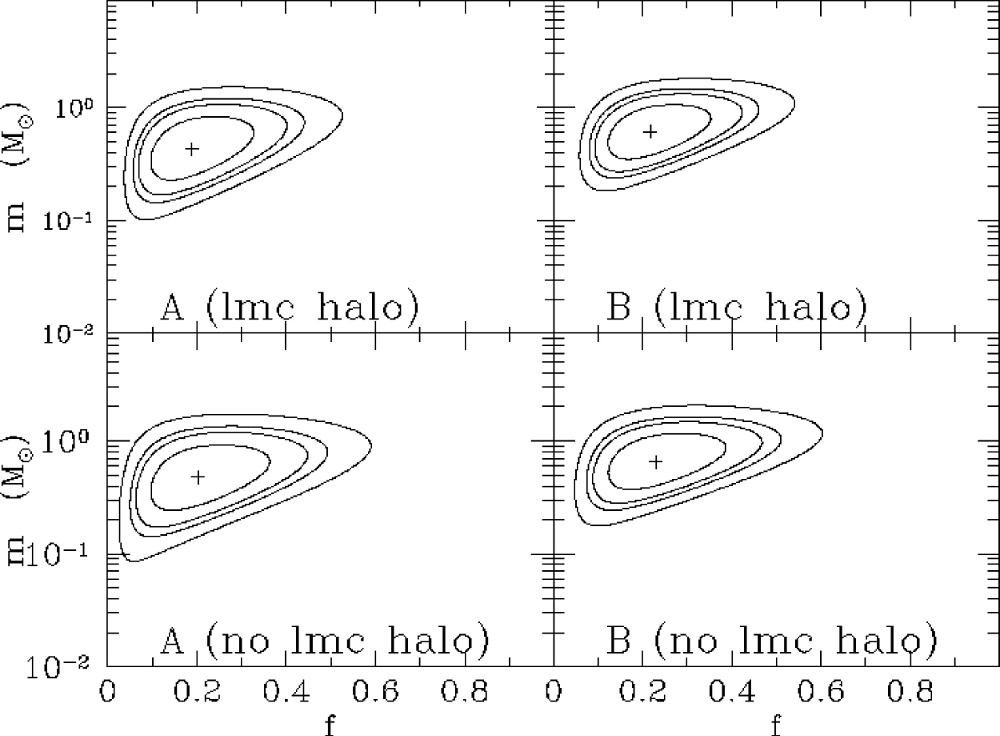}
\caption{Example of likelihood contours obtained from the MACHO experiment for one specific model for the Milky Way halo, with different cuts applied to the candidate microlensing events (left and right columns). The abscissa is the fraction of the halo mass contained in MACHOs, the ordinate is the MACHO mass. The contours show the 60\%, 90\%, 95\%, and 99\% confidence levels. (from \cite{AL00.1})}
\label{fig:06}
\end{figure}

Analysing data taken from $11.9$ million stars over $5.7$~years, the MACHO project \cite{AL00.1} found 13-17 events, while between 2 and 4 were expected from known stellar populations in the Milky Way and the LMC. The microlensing optical depth deduced from lensing events lasting between 2 and 400 days is $\tau=1.2^{+0.4}_{-0.3}\times10^{-7}$. This implies that between $8\%$ and $50\%$ of the Milky Way's halo can be composed of MACHOs (at 95\% confidence), whose most likely mass ranges between $0.15$ and $0.9\,M_\odot$ (cf.~Fig.~\ref{fig:06}). Consistently, the EROS project \cite{LA00.1} found, based on observations of LMC and SMC, that MACHOs cannot dominate the Galactic halo if their masses are $\lesssim1\,M_\odot$. They find that the halo mass fraction in MACHOs is $<20\%$ for MACHO masses between $10^{-7}\,M_\odot$ and $0.1\,M_\odot$ (at 95\% confidence).

Thus, although MACHOs have been detected between us and the Magellanic Clouds, they are insufficient for explaining all of the Milky Way's dark mass. These MACHOs can in principle be anywhere between the source stars and the observer, i.e.~in the dark halos of the Milky Way or of the Magellanic Clouds \cite{GO93.1, SA94.1, WU94.1, GO95.1, EV00.1}. It had been speculated that self-lensing of stars in the LMC by its own stars might suffice for explaining the observed optical depth \cite{AU99.1}, but later studies showed that certainly not all of the LMC lensing events can be explained as being due to stars in the LMC. Rather, the LMC needs to be embedded into an extended halo \cite{GY00.1, JE02.1, MA04.1}. It seemed thus safe to conclude that microlensing experiments have confirmed that the Galaxy and the Magellanic Clouds must have extended dark haloes, only a fraction of which can be composed of compact objects of stellar and sub-stellar mass.

With microlensing monitoring programs targeting the Andromeda galaxy M~31, the situation became less clear. Several early detections of microlensing candidates \cite{PA03.1, RI03.1, JO04.1, UG04.1} proved the feasibility of such surveys and were tentatively interpreted as demonstrating microlensing in the halo of M~31. Detailed analyses of meanwhile more than a dozen candidates arrive at different conclusions regarding the contribution by self-lensing. While \cite{IN06.1} argued that self-lensing should only occur near the centre of M~31, \cite{JO06.1} found that all candidates are consistent with self-lensing and rejected the hypothesis at the 95~\% level that $\gtrsim30\,\%$ of the halo mass could be contributed by MACHOs. Conversely, \cite{CA05.1} had used six short-duration events to conclude that the MACHO contribution should be $\gtrsim20\,\%$ if their mass was around $1\,M_\odot$. The main differences between these analyses are the models for the distributions of mass and stars in M~31 and the selection of microlensing events. Taking finite source sizes into account makes self-lensing less probable \cite{RI08.1}. Low variability in individual light curves of 29 multiply imaged quasars supports the impression that MACHOs contribute at most little to microlensing in galactic halos \cite{ME09.1}.

Even MACHOs in distant lensing galaxies have been constrained by means of microlensing. If a quasar is multiply imaged by a galaxy, stars and hypothetical MACHOs in that galaxy can act as microlenses on all images individually, giving rise to independent brightness changes in them. From the amplitude and the time scales of these brightness variations, assuming a plausible model for the size of the quasar source, upper limits can be placed on the fraction of the galaxy mass in compact objects within a given mass range \cite{WA00.2}. Turning this argument around, the size and the structure of the quasar engine can be constrained assuming microlensing by the galactic stars \cite{AN08.1, EI08.1}.

\begin{figure}[ht]
  \hfill\includegraphics[width=0.7\hsize]{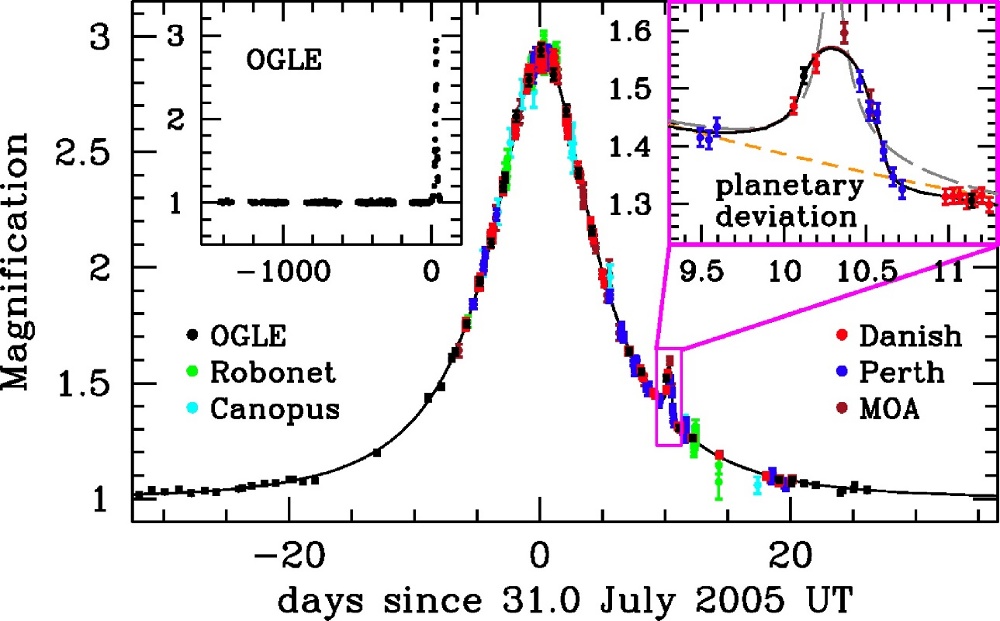}
\caption{Microlensing lightcurve showing the signature of a planet with 5.5 Earth masses. (from \cite{BE06.1})}
\label{fig:07}
\end{figure}

A fascinating application of microlensing was first discussed by \cite{MA91.1} who showed that binary lens systems may leave detectable signatures in microlensing lightcurves even if one of the companions is a planet. This is strongly favoured if the planet's distance from its host star approximately coincides with the star's Einstein radius \cite{GO92.1}. While the first planets detected in this way had Jupiter-like masses \cite{BO04.2, UD05.1}, even Earth-like planets have been found (\cite{BE06.1}, see Fig.~\ref{fig:07}). To date, ten planets in nine planetary systems have been detected by means of microlensing\footnote{see the data base at \href{http://exoplanet.eu/}{http://exoplanet.eu/}}.

\subsection{Galaxy density profiles}

Well over 100 cases of strong lensing by galaxies are now known. Most of them have two or four images, but a few have higher image numbers. Image splittings, typically of order an arc second, allow the projected lens mass to be constrained which is enclosed by the images. However, it turns out to be surprisingly difficult to constrain the mass profile from strong gravitational lensing alone. Essentially, multiple images constrain the average surface-mass density in an annulus bounded by the images.

Using many lens systems, and assuming their mass profiles to be self-similar, it becomes possible to trace the average surface mass density at different radii, and thus to map out the density profile. Analysing 22 galaxy lenses, and adapting a mass model composed of a concentrated component representing the light and a power-law component representing the dark matter, \cite{RU05.1} found that the slope of the density profile is very nearly isothermal, with a double-logarithmic slope of $n=2.06\pm0.17$ (isothermal has $n=2$). Models in which the mass traces the light and is therefore more centrally concentrated fail at the $99\%$ confidence level \cite{RU03.1}. If the dark matter follows the NFW density profile, $(22\pm10)\%$ of the matter inside two effective radii has to be dark. A weak trend is also seen in the mass-to-light ratio, $M/L\propto L^{0.14^{+0.16}_{-0.12}}$, consistent with the fundamental plane of elliptical galaxies. \cite{TR04.1} agree that the density profiles are nearly isothermal, but find a somewhat larger scatter. They confirm that lensing galaxies in which light traces mass are ruled out, and find a dark-matter fraction of between $15\%$ and $65\%$ within the effective radius. Combining stellar kinematics with lensing in 15 lensing galaxies of the Sloan Lens ACS Survey (SLACS, \cite{BO06.1}), \cite{TR06.1} find isothermal mass distributions within the Einstein radii. The density profiles found do not depend on environment \cite{TR09.1}.

The picture becomes more complete when constraints from strong and weak lensing are combined. While multiple images produced by strongly lensing galaxies support isothermal mass distributions when interpreted with power-law density profiles, they probe the density profiles only in the narrow range of radii enclosing the multiple images. Weak lensing can be observed to much larger distances from galaxy centres. The combined analysis of 22 elliptical lensing galaxies by \cite{GA07.1} is one example further discussed below. Weak lensing was used to improve the mass models for the double quasar 0957+561 \cite{FI97.1} and to refine estimates of the Hubble constant from its time delay \cite{NA09.1, FA10.1}. \cite{FA04.1} studied the environment of eight lensed quasars by means of weak lensing and found indications for nearby galaxy groups in five of them.

We have seen in the introduction that gravitational lenses are expected to produce an odd number of images. In contrast, all but very few observed galaxy-lens systems have an even image number, most of them either two or four. The missing images are expected to be faint if the central density profile of the lensing galaxies is steep enough, thus their absence can be used for constraining the central concentration of the lensing mass distributions. Based on this argument, \cite{RU01.1} find that inner mass distributions of lensing galaxies cannot be much shallower than isothermal. Conversely, \cite{WI04.1} use a lens system in which a faint, central image has been found to constrain the mass of the central black hole in the lensing galaxy to be $<2\times10^8\,h^{-1}\,M_\odot$. From the general absence of faint, central images, \cite{KE01.1} concludes that the central mass profiles of lensing galaxies must be more concentrated than CDM alone predicts. Central black holes may reconcile CDM density profiles with even image numbers only if they are about an order of magnitude more massive than expected from the relation between black-hole and bulge masses.

Time-delay measurements in multiple-image systems promise constraints on the Hubble constant, provided a sufficiently accurate mass model for the lens is known. Conversely, considering the Hubble constant as known, time-delay measurements can be used as further constraints on the lensing density profile. Values for the Hubble constant derived this way tended to be lower than those, e.g.~obtained from the HST Key Project (e.g.~\cite{FA02.1}), but lens models constructed upon a large number of constraints yield values which are very well in agreement with other determinations. For instance, \cite{KO03.2} find $H_0=75^{+7}_{-6}\,\mathrm{km\,s^{-1}\,Mpc^{-1}}$ from time delays measured in the quadruply lensed quasar B~1608$+$656. From detailed modelling of the same lens system, \cite{SU10.1} derive $H_0=70.6\pm3.1\,\mathrm{km\,s^{-1}\,Mpc^{-1}}$ for a standard cosmological model, and $H_0=69.0^{+4.9}_{-5.0}\,\mathrm{km\,s^{-1}\,Mpc^{-1}}$ when combined with the WMAP constraints. In the latter case, the equation-of-state parameter is constrained to be $w=-0.94^{+0.17}_{-0.19}$. Non-parametric mass models for lensing galaxies provide an interesting approach alternative to the parameterised density profiles. With this technique, \cite{SA06.1} construct non-parametric mass distributions for 10 galaxy-lens systems and derive $H_0=72.8^{+7.0}_{-13.5}\,\mathrm{km\,s^{-1}\,Mpc^{-1}}$, while \cite{PA10.1} find $H_0=66^{+6}_{-4}\,\mathrm{km\,s^{-1}\,Mpc^{-1}}$ from 18 lens systems.

\begin{figure}[ht]
  \hfill\includegraphics[width=0.7\hsize]{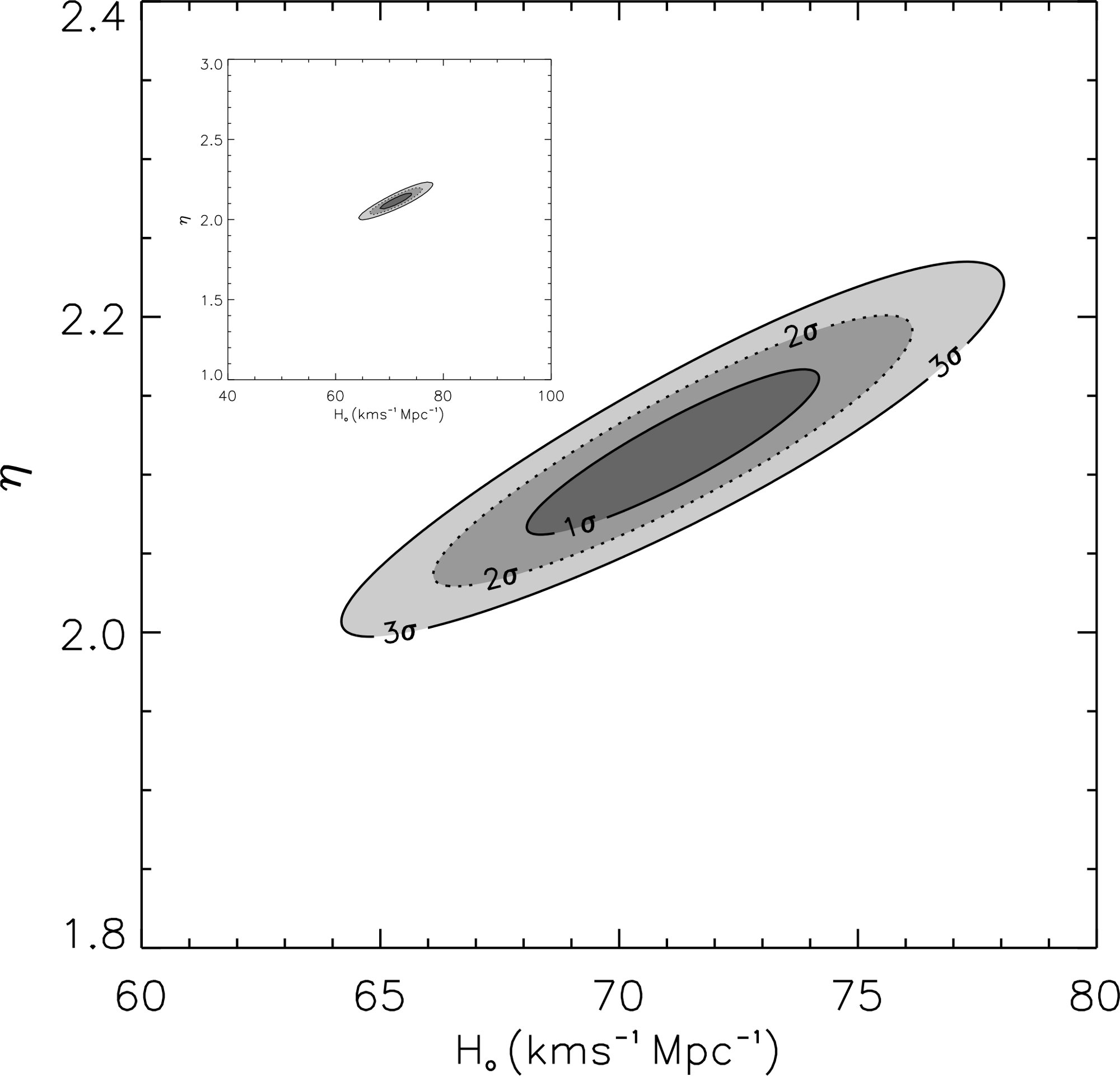}
\caption{Likelihood contours in the plane spanned by the Hubble constant and the double-logarithmic slope of the density profile assumed for the lens models, obtained from fitting 14 time-delay measurements. (from \cite{DO06.1})}
\label{fig:08}
\end{figure}

An interesting problem with interpreting time delays in four galaxy-lens systems was pointed out by \cite{KO03.1}. It turns out that the time delays between multiple images is essentially determined by the mean surface-mass density in an annulus around the lens centre bounded by the images \cite{KO02.1}. An estimate for that mass density can also be obtained converting the visible light to mass, assuming typical values for the fraction $f_\mathrm{b}$ of matter that condenses into stars. Adopting $f_\mathrm{b}\simeq0.02$ in accordance with local observations works well with near-isothermal mass models, but yields substantially too low values for the Hubble constant, $H_0=(48\pm5)\,\mathrm{km\,s^{-1}\,Mpc^{-1}}$. Conversely, values for the Hubble constant agreeing with the HST Key Project result, $H_0=(72\pm8)\,\mathrm{km\,s^{-1}\,Mpc^{-1}}$, are compatible with the measured time delays only if lens models with constant mass-to-light ratios are adopted, which are otherwise ruled out \cite{KO03.1}. Using 14 time delays and a prior for the Hubble constant derived from other cosmological observations, \cite{DO06.1} find that the density profiles of the lensing galaxies can only barely be compatible with being isothermal (see Fig.~\ref{fig:08}). In a detailed analysis including various deviations from idealised lens models, \cite{OG07.1} finds $H_0=(68\pm10)\,\mathrm{km\,s^{-1}\,Mpc^{-1}}$ including systematic errors.

There is thus the apparent problem that multiple-image configurations in galaxy lenses favour isothermal density profiles, while values for the Hubble constant derived from time delays are only in agreement with independent measurements if the density profiles are steeper than isothermal. A possible solution was pointed out by \cite{RE07.1} who showed that measurements of time delays and the Hubble constant can be reconciled if perturbations of the density profiles around isothermality are allowed.

\begin{figure}[ht]
  \hfill\includegraphics[width=0.7\hsize]{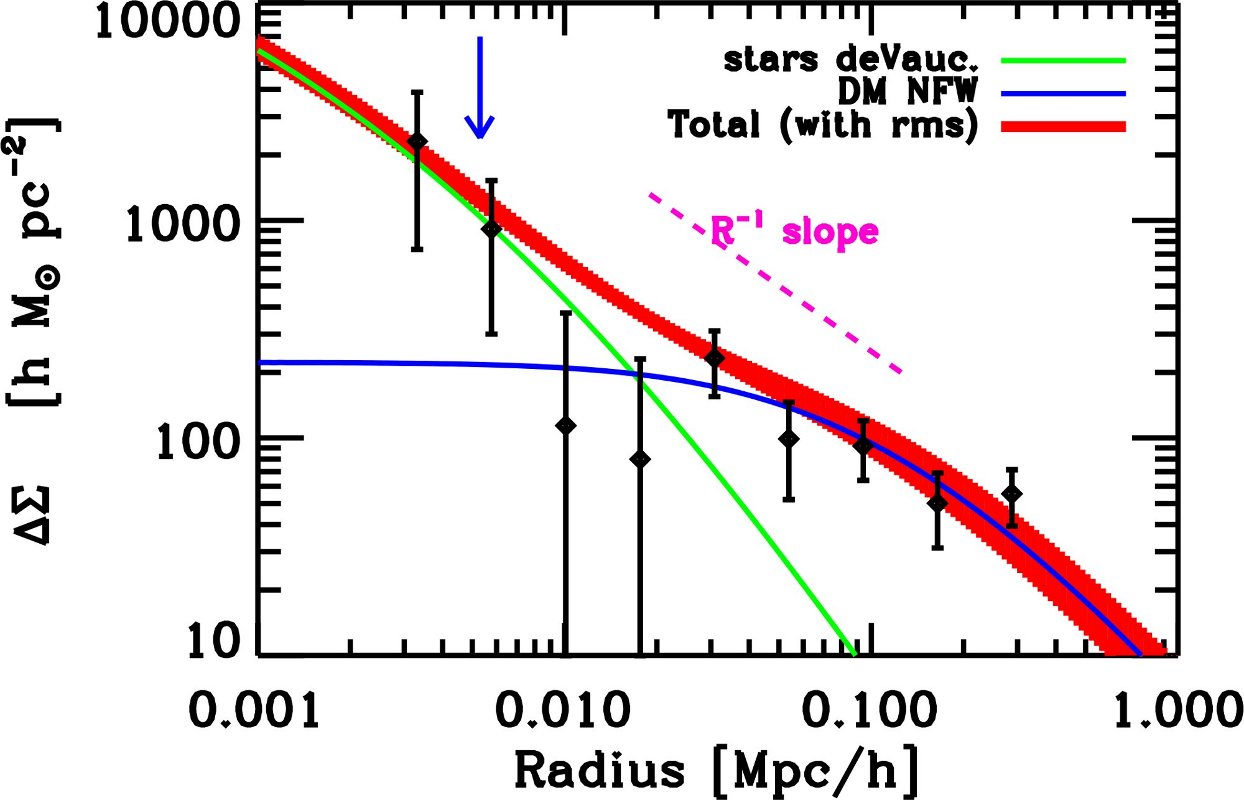}
\caption{Surface-density profiles obtained from 22 strongly-lensing elliptical galaxies, compared to expected stellar and dark-matter components. (from \cite{GA07.1})}
\label{fig:09}
\end{figure}

Altogether, the picture emerges that galaxy mass distributions approach isothermal or slightly steeper density profiles where baryonic physics dominates, and turn into the NFW profiles expected from numerical simulations for stable dark-matter dominated structures \cite{MA06.1, GA07.1, MA08.1} (see Fig.~\ref{fig:09}). Selecting galaxies by strong lensing may lead to substantially biased samples, though \cite{MA09.1}, while the spectroscopic selection defining SLACS \cite{BO06.1} identifies an unbiased galaxy sample \cite{AU08.1, GR09.1}.

\subsection{Substructure in lensing galaxies}

Interestingly, the observed lenses with four images (quadruples) are about as abundant as such with two images (doubles), while they should only contribute 25\% to 30\% of the galaxy lenses. The fraction of quadruples can be enhanced by satellite galaxies orbiting the main lens galaxies \cite{CO04.1} or by matter in their larger-scale environments \cite{KE04.1}, although the latter explanation is potentially problematic because it also tends to lower inferred values of the Hubble constant.

Axially-symmetric lens models are insufficient for modelling observed multiple-image systems. At least elliptical lens models are necessary. Embedding the lenses into additional external shear fields helps fitting observed image configurations, but typically more shear is required (10\%-15\%) than the average large-scale structure can provide (1\%-3\%, \cite{KE97.1}). This hints at the presence of asymmetries and substructures in the lensing galaxies.

\begin{figure}[ht]
  \hfill\includegraphics[width=0.7\hsize]{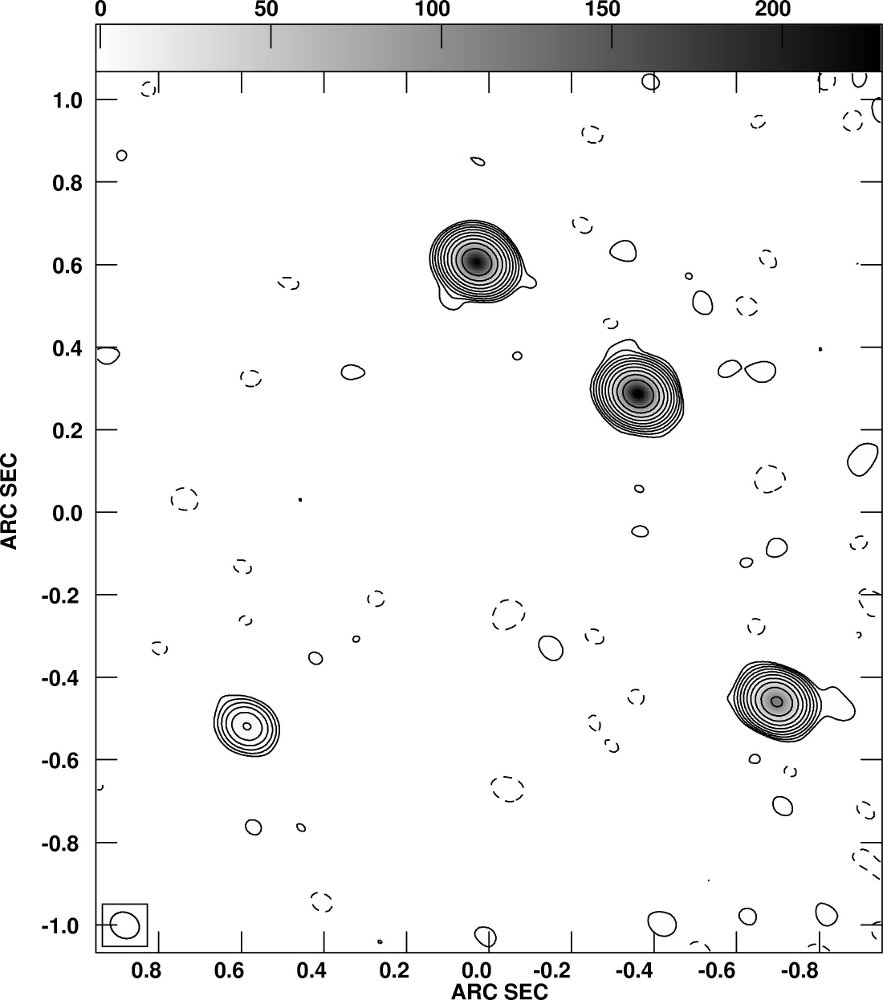}
\caption{MERLIN radio image of the multiply-imaged quasar B~1422$+$231 (courtesy of the JVAS/CLASS team, 
\href{http://www.jb.man.ac.uk/research/gravlens/lensarch/}
{http://www.jb.man.ac.uk/research/gravlens/lensarch/}).
If image by a single, unperturbed lens, the two outer of the three bright images should together be as bright as the one between them.}
\label{fig:10}
\end{figure}

It is an interesting problem which caused much recent discussion that lens models are typically very successful in reproducing image positions, but fail in a large fraction of lens systems to explain the flux ratios between different images. A particularly obstinate and well-known case is B~1422$+$231 \cite{KO94.1, HO94.1}, see Fig.~\ref{fig:10}. This is most striking in situations where the source falls just inside a cusp of the caustic curve, in which case the sum of the signed magnifications of the three related images should vanish exactly. This expectation is frequently violated in real lens systems.

\cite{MA98.1} first suggested that substructure in the lensing galaxy could account for these anomalous flux ratios. While microlensing by the stars in the lens would less affect radio than optical fluxes because of their larger size, lensing by larger-scale substructures would equally change radio and optical flux ratios. Alternatively, CDM galaxy halos should contain sub-halos which may also account for anomalous flux ratios. \cite{BR02.1} found similar modelling problems for images produced by a simulated lensing galaxy as for B~1422$+$231. \cite{CH02.1} discussed that the sub-halo population of CDM halos produces perturbations of the magnitude required for explaining anomalous flux ratios. \cite{ME02.1} estimated that $\gtrsim5\%$ of the lensing halo mass must be contained in substructures and argued that elliptically deformed power-law models embedded into external shear are insufficient for most lenses. Similarly, \cite{DA02.1} concluded that substructure comprising $0.6\%$ to $7\%$ of the lens mass, with a median at $2\%$, was necessary for reproducing the observed anomalous flux ratios, in excellent agreement with CDM halo simulations. They also estimated that the sub-halos should have masses in the range $(10^6\ldots10^9)\,h^{-1}\,M_\odot$. \cite{ME02.2} found that halo substructures with masses within $(10^5\ldots10^7)\,h^{-1}\,M_\odot$ may explain the curved radio jet in B~1152$+$199.

In contrast to these arguments, \cite{EV03.1} explicitly constructed smoothly deformed lens models which could well reproduce image configurations and flux ratios for most lens systems and argued that substructure in the lensing galaxies and smoothly deformed lenses are both viable explanations for the anomalous flux ratios. Along the same line, \cite{MO03.1} and \cite{QU03.1} showed that disks in lens galaxies can alter image magnification ratios considerably, while \cite{CH03.1} pointed out that halos projected onto the main lens galaxy may also cause the observed magnification perturbations.

From a somewhat different perspective, \cite{SC02.1} discussed that decomposing lensing galaxies into microlenses has the most prominent effect when part of the lensing mass remains smooth. Specifically, they showed that so-called saddle-point images can be substantially demagnified in presence of microlenses. \cite{SC04.1} added that the magnification distribution of the macro-images depends on the mass spectrum of the microlenses, in contrast to earlier expectations.

\cite{KO04.1} investigated various alternative explanations for the anomalous flux ratios, such as absorption, scattering, scintillation, uncertainties in the macro-model, and stellar microlensing, and arrived again at the conclusion that halo substructures remain as the most likely reason. The parities of the three images formed near a cusp point in the caustic is an important argument to exclude other alternatives to lensing. \cite{BR04.1} verified that numerically simulated galaxies can produce anomalous flux ratios as observed and emphasised the importance of the demagnification of saddle-point images. Although it appears doubtless that CDM halos contain sufficient substructure for sufficiently perturbing image flux ratios, such sub-halos must also appear projected onto at least one of the images. The probability for that is low. \cite{MA04.2} find in numerical simulations that the probability of finding suitably massive sub-halos in front of macro-images is only $\lesssim0.5\%$. Newer simulations all agree that the expected level of substructures in lensing-galaxy haloes is insufficient for explaining the anomalous flux ratios, even if high mass resolution and baryonic physics are taken into account \cite{MA06.2, AM06.1, MA06.3, XU09.1}. Halos unrelated to the lensing galaxies, but projected onto them, can have a substantial effect on flux-ratio anomalies \cite{ME05.2, MI07.2}. There is agreement that perturbing masses must be compact \cite{CH05.1, SH08.1}, even though smooth perturbations appear in sufficiently flexible lens models \cite{SA07.1}. In some cases, detailed modelling based on all lens components visible in detailed observations reproduces the measured flux ratios well without any further substructure \cite{MK07.1, ML09.1}.

The situation thus remains interestingly confused. It seems likely that the combination of dark-matter substructures, the tidal field of the environment, mergers and random projections provides a way out. \cite{KE09.1} suggest that time-delay perturbations may add further constraints on halo substructures.

\subsection{Lens statistics}

The abundance of galaxy lenses has often been used for constraining the cosmological constant $\Lambda$. While early studies typically found \emph{upper} limits of $\Lambda\lesssim0.7$ (e.g.~\cite{KO96.1, FA98.1}), more recent investigations find values which are better compatible with other determinations, (e.g.~\cite{CH99.1, CH02.2}), finding spatially-flat model universes with low matter density ($\Omega_0\simeq0.3$) preferred. The reason for this change is that gradually more realistic galaxy luminosity functions were used for estimating the expected number of lenses, rather than error-prone extrapolations of local galaxy number densities towards high redshift \cite{KE02.1}.

The Sloan Digital Sky Survey has allowed the definition of a homogeneously selected quasar sample \cite{IN08.1} from which cosmological parameters were derived \cite{OG08.1}. Assuming a spatially flat universe, a value of $\Omega_{\Lambda0}=0.74^{+0.17}_{-0.16}$ was derived for the cosmological constant, where statistical and systematic errors were combined in quadrature. Allowing a free equation-of-state parameter gave $w=-1.1^{+0.67}_{-0.78}$ and a matter-density parameter of $\Omega_\mathrm{m0}=0.26\pm0.08$ when combined with independent cosmological constraints.

Halos are expected to have a continuous mass spectrum in universes dominated by cold dark matter, which is described by mass functions such as those derived by \cite{PR74.1}, \cite{SH02.1} and \cite{JE01.1}. Thus, one would expect a continuous distribution of splitting angles between fractions of an arc second to several ten arc seconds. \cite{NA88.1} investigated whether the observed image-splitting distribution was consistent with expectations from CDM. They found observation and theory agreed if selection effects were taken into account. \cite{KO95.1} found that the splitting-angle distribution in CDM is grossly incompatible with microwave-background constraints in a model universe with high matter density and vanishing cosmological constant, but that both could be comfortably reconciled in a spatially-flat, low-density CDM model.

Occasionally, therefore, lens systems should be detected with splitting angles of ten or more arc seconds. \cite{PH01.1} interpreted the absence of wide-separation lenses in the CLASS survey as a being due to low central mass concentrations in group- and cluster-sized halos. It was perceived as a further confirmation of the CDM paradigm when a quadruply imaged quasar was detected in the Sloan Digital Sky Survey with a splitting angle of $14.62$ arc seconds \cite{IN03.1}, for which \cite{WI04.2} derived a lens mass of $(5\pm1)\times10^{13}\,h^{-1}\,M_\odot$ within a radius of $100\,h^{-1}\,\mathrm{kpc}$ based on a non-parametric lens model. \cite{OG04.1} noted that the triaxiality of CDM halos must be taken into account in probability and mass estimates for the formation of wide-separation lens systems, a theme which is familiar from studies of strong lensing in galaxy clusters.

Clearly, cosmological parameters from the statistics of strong gravitational lensing by galaxies are generally no longer competitive compared to those based on observations of the cosmic microwave background because the uncertainties in lens models and sample selection are considerable. It should be kept in mind, however, that the cosmic microwave background does not independently measure the Hubble constant, but the expansion rate during the time of recombination. Hence, independent measurements in particular of the Hubble constant are and remain most important.

\subsection{Weak lensing of galaxies by galaxies}

Less distant galaxies can act as weak gravitational lenses on more distant galaxies. Their shear imprints a feeble tangential distortion pattern on the images of background galaxies which appear projected close to them. This weak signal is superposed on the intrinsic ellipticities and irregularities of the background-galaxy images and thus requires statistical techniques for its extraction. \cite{BR96.1} first discussed the principal features of this effect and searched for it in a sample of galaxies, in which they separated background from foreground galaxies according to their apparent brightness. They could already infer that the shear profile of brighter galaxies was compatible with an isothermal mass profile with a circular velocity of $v_\mathrm{c}=(220\pm80)\,\mathrm{km\,s^{-1}}$. They also placed a lower limit $r_*\gtrsim100\,h^{-1}\,\mathrm{kpc}$ on the halo size of the lensing galaxies. \cite{DE96.1} searched for galaxy-galaxy lensing in the Hubble Deep Field (North) and found a mean velocity dispersion for the lensing halos of $\sigma_\mathrm{v}=185^{+30}_{-35}\,\mathrm{km\,s^{-1}}$, and a weak lower limit on the halo radius.

\cite{SC97.1} devised a maximum-likelihood technique for efficient analysis of galaxy-galaxy lensing data which specifically took the redshift distributions of foreground and background galaxies into account. They applied this technique to numerically simulated data and calibrated its performance. \cite{NA97.2} and \cite{GE98.1} developed methods for detecting the weak-lensing signal of galaxies embedded in galaxy clusters. Applying their technique to the cluster Cl~0939$+$4713, \cite{GE99.1} detected the shear signal of individual massive cluster galaxies. More recently, \cite{NA02.1} compared the weak-lensing signal of early-type, $L_*$ galaxies in clusters and in the field and found evidence for the cluster galaxies to be truncated, with a truncation radius shrinking with the density of the environment. \cite{HO04.1} combined weak-lensing data on galaxy halos to show that they are flattened.

Recent wide-field surveys also triggered an exciting development of galaxy-galaxy lensing. \cite{FI00.1} used the Commissioning Data of the SDSS to infer that the tangential shear profile is compatible with a power law with exponent between $0.7$ and $1.1$, i.e.~close to isothermal. They found a best-fitting circular velocity of $v_\mathrm{c}=(150\ldots190)\,\mathrm{km\,s^{-1}}$ and a lower limit to the physical halo radius of $260\,h^{-1}\,\mathrm{kpc}$. From the Las Campanas Redshift Survey, \cite{SM01.1} deduced an isothermal tangential shear profile within $200\,h^{-1}\,\mathrm{kpc}$ and a circular velocity of $v_\mathrm{c}=(164\pm20)\,\mathrm{km\,s^{-1}}$ for $L_*$ field galaxies. They found a virial mass for the dark halo of a typical $L_*$ galaxy of $(2.7\pm0.6)\times10^{11}\,h^{-1}\,M_\odot$. \cite{WI01.1} used data taken with the UH8K camera at the Canada-France-Hawaii telescope to measure galaxy-galaxy lensing. They also found tangential shear profiles compatible with an isothermal slope and a rotation velocity of $v_\mathrm{c}=238^{+27}_{-30}\,\mathrm{km\,s^{-1}}$ for $L_*$ galaxies. They concluded that the mass-to-light ratio of $L_*$ galaxies in the $B$ band is $M/L\simeq(121\pm28)\,h\,M_\odot/L_\odot$. According to the analysis of 22~square degrees of CFHT Legacy Survey data by \cite{PA07.1}, $L_*$ galaxies at redshift $0.3$ reside in haloes of $(1.1\pm0.2)\times10^{12}\,h^{-1}\,M_\odot$ and have an $R$-band mass-to-light ratio of $173\pm34\,h\,M_\odot/L_\odot$.

Combining the galaxy-galaxy weak-lensing signal obtained from SDSS data with the Tully-Fisher and fundamental-plane relations for late- and early-type galaxies, respectively, \cite{SE02.1} found that the galaxy velocity profile must drop substantially towards the virial radius, which indicates a steep dark-matter profile. \cite{GU02.1} compared theoretically motivated CDM halo models with SDSS data and constrained the halo properties of galaxies with luminosities $\gtrsim L_*$. They constrained the virial mass of an early-type $L_*$ halo to $M_{200}=(5\ldots10)\times10^{11}\,h^{-1}\,M_\odot$, and somewhat less for late-type galaxies, depending on the colour. They found a gentle increase of the mass-to-light ratio with luminosity, with $M/L\simeq17\,h\,M_\odot/L_\odot$ for late-type and $M/L\simeq45\,h\,M_\odot/L_\odot$ for early-type $L_*$ galaxies.

\cite{SH04.1} studied the cross-correlation between galaxies and mass from the galaxy-galaxy lensing signal detected in SDSS data. The wide area covered by the survey allowed constraining the correlation function out $10\,h^{-1}\,\mathrm{Mpc}$. They find a power law with a correlation length of $r_0\simeq(5.4\pm0.7)\,h^{-1}\,\mathrm{Mpc}$ and an exponent of $1.79\pm0.05$. The bias parameter turns out to be approximately scale-independent (see also \cite{HO01.1}), while \cite{HO02.1} find the bias parameters to be gently increasing from Mpc to larger scales. Comparisons with theoretical expectations for the galaxy distribution relative to the dark matter find overall good agreement \cite{WE04.1}, except that the simulated mass-to-light ratio is somewhat too high \cite{YA03.1}. Satellite galaxies orbiting the lensing galaxies could be physically aligned with their hosts and thus mimic a weak galaxy-galaxy lensing signal. \cite{HI04.1} estimated this possible contamination and constrained it to less than $15\%$ at the relevant scales.

The availability of huge surveys with sufficient depth and image quality for weak-lensing studies has opened new applications also for galaxy-galaxy lensing. The statistical results obtained (e.g.~by \cite{MA06.1, GA07.1, MA08.1}) on the density profiles of galactic haloes were already described above. Exciting examples of more detailed studies enabled this way are the measurement of mean masses of halos hosting active galactic nuclei, which found that radio-loud AGN reside in host haloes which are typically $\approx20$ times more massive than for radio-quiet AGN \cite{MA09.2}, and the combination of galaxy-galaxy lensing with galaxy correlations to specify the mean mass-to-light ratio of the galaxies \cite{CA09.1}.

\section{Galaxy clusters}

\subsection{Strong lensing}

Strong lensing in galaxy clusters was first detected by \cite{SO87.1} and \cite{LY89.1}. They found extended, arc-like images in the galaxy clusters A~370 and Cl~2244. Several explanations were proposed for these objects, among them gravitational lensing of background galaxies \cite{PA87.1}, which was confirmed when the redshift of the arc in A~370 was measured and found to be substantially higher than the cluster's \cite{SO88.1}.

\begin{figure}[ht]
  \hfill\includegraphics[width=0.7\hsize]{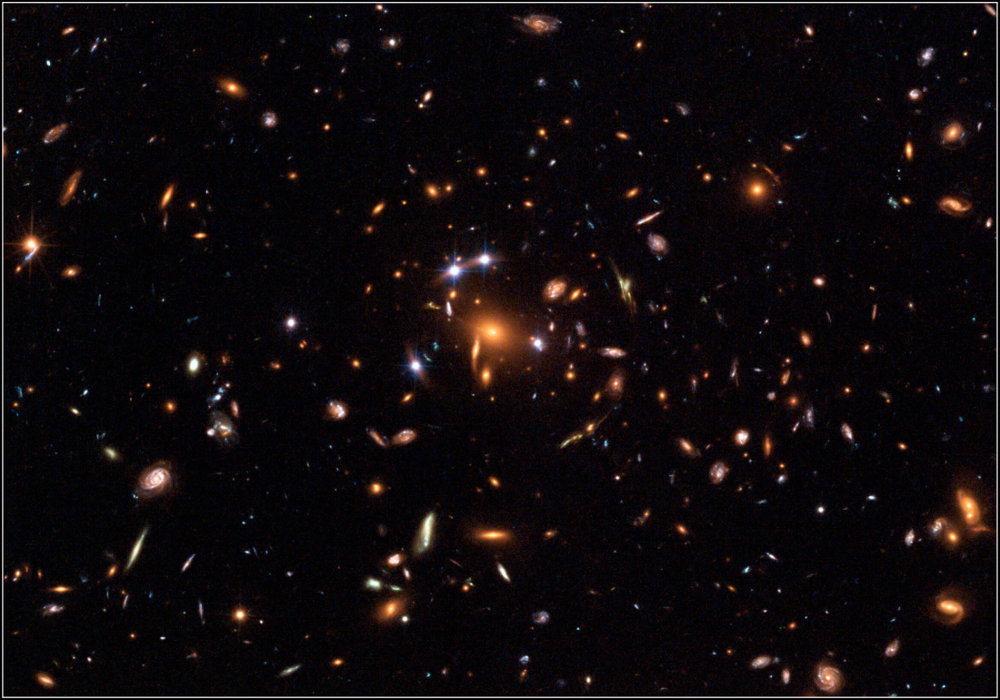}
\caption{This HST image of the cluster SDSS J~1004+4112 \cite{SH05.1} shows arcs, multiply imaged galaxies and a quadruply lensed quasar with a maximum image separation of $14.62$ arc seconds \cite{IN03.1}.}
\label{fig:11}
\end{figure}

It was quickly recognised that gravitational arcs provided important information on the structure of galaxy clusters. It was unclear at the time how the dark matter was distributed and whether the X-ray surface-brightness profiles, which typically show a flat core of $\simeq200\,h^{-1}\,\mathrm{kpc}$ radius, were representative for the dark-matter profiles. Arcs were soon found to reveal the following about clusters: (1) Cluster mass distributions cannot typically be axially symmetric, because large counter-arcs would otherwise be expected \cite{GR88.1, KO89.1}. (2) The substantial amounts of dark matter in galaxy clusters cannot be attached to the galaxies because arcs would then have much smaller curvature radii \cite{HA89.2, BE90.1}. Particularly striking were the detections of ``straight arcs'' in two clusters \cite{PE91.1, MA92.1, PI96.1} because they visually demonstrated the need for substantial concentrations of dark matter with very high mass-to-light ratio \cite{KA92.1}. (3) Clusters need to have steep density profiles, because arcs would be substantially thicker otherwise \cite{HA89.1}. For clusters to be strong lenses, their central convergence $\kappa$ has to be close to unity, but for arcs to be thin, the convergence at their locations has to be around $0.5$. From cluster centres to the arc radii of typically $10''\ldots30''$, the $\kappa$ profile must thus fall by approximately a factor of two. Cluster core radii, if they exist, must thus be substantially smaller than the X-ray core radii, which was also confirmed by the detection of ``radial arcs'' \cite{FO92.1, MI93.1, ME93.1}.

Arcs allow cluster masses to be easily estimated. It was soon discovered that the masses obtained this way are very close to mass estimates derived from the X-ray temperature and surface-brightness profile. This is not obvious because gravitational lensing is sensitive to the mass irrespective of its physical state, while the interpretation of X-ray data requires assumptions on symmetry and hydrostatic equilibrium of the gas with the gravitational potential well, if not on isothermality of the intracluster gas. This overall agreement being reassuring, a systematic discrepancy was soon revealed in the sense that masses derived from strong lensing were typically higher by factors of $\simeq2\ldots3$ than X-ray masses \cite{WU94.2, MI95.2, WU96.1}. There are many more recent examples. Many find substantially discrepant mass estimates based on X-ray and strong-lensing observations \cite{PR05.1, GI07.1, MI08.1, HA08.1, EB09.1}, while good agreement is found in other clusters \cite{RZ07.1, BR08.1}.

\cite{BA96.2} used numerical simulations to show that X-ray mass estimates can be systematically lower in merging clusters because their X-ray gas is still cooler than expected from their total mass, which is already seen by the lensing effect. This seems to explain the mass discrepancy at least in some clusters (e.g.~\cite{SM95.1, OT98.1}). Asymmetries and cluster substructures also play an important role. Due to their relatively larger shear, asymmetric and substructured clusters are more efficient lenses at a given mass. Mass estimates based on axially symmetric models are thus systematically too high \cite{BA95.1, HA98.1}.

\cite{AL98.1} distinguished clusters with and without cooling flows and found an appreciable mass discrepancy in clusters without, but good agreement of X-ray and lensing mass estimates in clusters with cooling-flow. This supports the concept that well-relaxed clusters which had sufficient unperturbed time to develop a cooling flow are well-described by simple, axially-symmetric models for lensing and the X-ray emission, while dynamically more active clusters tend to give discrepant mass estimates; this was confirmed by \cite{WU00.1}. \cite{MA99.1} noted that the mass discrepancy is reduced if cluster density profiles are steeper than inferred from the X-ray emission. It thus appears that mass discrepancies can commonly be traced back to dynamical activity in unrelaxed clusters (see also \cite{SM05.1}), but at least part of the disagreement also occurs because of model restrictions which, if removed, generally lead to better agreement \cite{GA05.1, DO09.1, MO10.1}.

\subsection{Cluster mass profiles}

Assuming mass profiles with cores, tangential arcs require small core radii as described above, but radial arcs require the cores to be finite \cite{LE94.1, LU99.1}. Numerical simulations of CDM halos, however, show that density profiles flatten towards the core, but do not develop flat cores \cite{NA96.1, NA97.1}. \cite{BA96.1} showed that radial arcs can also be formed by halos with such ``cuspy'' density profiles, provided the central cusp is not too steep.

In principle, the relative abundances and positions of radial compared to tangential arcs in clusters provide important constraints on the central density profile in clusters \cite{MI95.1, MO01.1, OG01.1}. Radial arcs are still too rare for successfully exploiting this method. Being much closer to the cluster cores than tangential arcs, they are also more likely to be confused with or hidden by the light of the cluster galaxies. Following \cite{MI95.1}, \cite{SA04.1} compiled a sample of clusters containing radial and tangential arcs and added constraints on the central mass profile from velocity-dispersion measurements in the central cluster galaxies. They demonstrated that, assuming axially-symmetric mass distributions, central density profiles have to be substantially flatter than those found in CDM simulations. However, even small deviations from axial symmetry can invalidate this conclusion and establish agreement between these observations and CDM density profiles \cite{BA04.1, ME07.1}.

Attempts at modelling arcs with isothermal mass distributions are typically remarkably successful (see \cite{KN96.1} for an impressive early example). This is all the more surprising as numerical simulations consistently find density profiles which are flatter than isothermal within the scale radius and steeper outside. In a very detailed analysis, \cite{GA03.1} find that an isothermal core profile for the cluster MS~2137 is preferred compared to the flatter NFW profile. \cite{SM01.1} constrain the core density profile in A~383 using X-ray, weak-, and strong-lensing data and find it more peaked than the NFW profile, but argue that this may be due to the density profile of the cD galaxy. Similarly, \cite{KN03.1} find in a combined weak- and strong-lensing analysis of Cl~0024$+$1654 that an isothermal mass profile can be rejected, while the NFW profile fits the data well. \cite{SH08.2} show that the strong-lensing effects in two clusters A~370 and MS~2137 can be explained similarly well by isothermal and NFW density profiles, leading to substantial uncertainties in derived cluster properties and magnifications. Strong lensing alone constrains cluster density profiles only close to cluster centres, leaving considerable freedom in the mass models. Certainly, the innermost cluster density profiles can be significantly influenced and steepened by baryonic physics.

It is a puzzling and potentially highly important problem that strong-lensing analyses of galaxy clusters in many cases find that NFW density profiles well reproduce the observed image configurations, but with concentration parameters that are substantially larger than expected from numerical simulations \cite{BR05.1, CO07.1, HE07.1, UM08.1, BR08.3}. This so-called overconcentration problem is currently much debated, in particular because other studies find concentration parameters in the expected range, sometimes in the same clusters
\cite{HA06.1, LI08.1}. Due to selection biases and projection effects, strongly lensing clusters should be among the most concentrated clusters, but some clusters such as A~1689 seem to be extraordinarily concentrated. Whether this is a significant contradiction to the $\Lambda$CDM model remains to be clarified.

\subsection{Arc abundance and statistics}

The mean density profile of galaxy clusters can also be constrained statistically because the probability for a cluster to become a strong lens depends sensitively on the mass concentration in its core \cite{WU93.1}. \cite{MI93.1} suggested that the core densities of strongly lensing clusters could be enhanced by projection of elongated clusters along the line-of-sight. \cite{BA94.1} used a numerically simulated galaxy cluster to show that asymmetric, substructured cluster models are significantly more efficient strong lenses than axially-symmetric mass distributions because of their enhanced tidal field. Averaging over a sample of simulated clusters, \cite{BA95.2} quantified that the cross sections for arc formation could be up to two orders of magnitude larger for asymmetric than for axially symmetric cluster models of the same mass.

\cite{HA97.1} confirmed that structured lenses help understanding the observational results of \cite{LE94.1}, who detected six arcs in a sample of 16 clusters selected for their high X-ray luminosity as measured by the EMSS satellite, but argued that even more concentrated mass profiles than those used by \cite{BA95.2} are necessary for explaining them quantitatively. \cite{BA98.1} used samples of numerically simulated clusters to estimate the total arc-formation probability in different cosmological models. Comparing their results with the data from \cite{LE94.1}, they concluded that only their cluster sample taken from a simulation with low matter density ($\Omega_0=0.3$) and no cosmological constant could well reproduce the measured high arc abundance, but the other three models failed badly. In particular, a flat cosmological model with $\Omega_0=0.3$ and $\Omega_\Lambda=0.7$ produced an order of magnitude less arcs than observed.

This so-called arc-statistics problem was disputed based on calculations using analytic models for cluster lenses \cite{CO99.1, KA00.1}, which failed to reproduce the strong dependence on the cosmological constant claimed by \cite{BA98.1}. The possible influence of cluster galaxies on the arc-formation efficiency of cluster lenses was investigated by \cite{FL00.1} and \cite{ME00.1}, but found to be negligible. \cite{MO99.1} confirmed that axially-symmetric mass models adapted to the X-ray emission do not produce a sufficient number of arcs. They found that using NFW profiles for the dark-matter profile helped, but the profiles required too high masses, and proposed that substructured mass distributions could be the solution. \cite{ME03.2} adapted elliptically distorted lenses with NFW mass profile (see also \cite{GO02.1}) to numerically simulated clusters and found the analytic models inadequate for quantitative arc statistics despite the asymmetry, demonstrating the importance of substructures.

\cite{OG03.1} studied the strong-lensing properties of triaxial (rather than ellipsoidal) halos and found that they may well explain the high arc abundance \cite{LE94.1, LU99.1}, provided their central density slopes are steep enough, with a double-logarithmic slope near $-1.5$. \cite{WA04.1} simulated the magnification probability for light rays propagating through a section of the Universe and found that the abundance of high-magnification events depends strongly on the source redshift. They attributed this to the exponential mass function of massive halos, which leads to a steep increase with source redshift in the number of halos suitable for strong lensing. Identifying the probability for highly magnified light bundles on random patches of the sky with the probability for finding arcs in massive galaxy clusters, they suggested this result as the resolution for the arc statistics problem. \cite{DA04.1} used numerical cluster simulations to estimate arc cross sections and found reasonable agreement with the earlier results of \cite{BA98.1}, but arrived at a higher expected arc abundance because they inserted a higher normalisation for the number density of both X-ray clusters and background sources.

\cite{WI99.1} noted that the arc radii in clusters depend only weakly on clusters mass and suggested that massive cD galaxies may be the reason. However, \cite{ME03.1} studied the effect of cD galaxies on the overall arc abundance and found it insufficient to remove the arc statistics problem. If the cosmological constant is replaced by some form of dynamical dark energy, structures tend to form earlier during cosmic history. Since cluster core densities reflect the mean cosmic density at their formation time, clusters thus tend to be more concentrated in dark-energy compared to cosmological-constant models. \cite{BA03.1} estimated the effect of higher cluster concentrations on arc statistics by analytic means. They found that dark energy may in fact increase arc abundances noticeably, but again not sufficiently for solving the arc statistics problem.

Galaxy clusters at high redshifts are found to be remarkably efficient lenses \cite{GL03.1, ZA03.1} even though they should be by far not massive enough for producing large arcs. A particularly impressive example is the cluster RX~J105343$+$5735 at $z=1.263$ which contains a large arc from a source at $z=2.577$ \cite{TH01.1}. In this respect, it is interesting that the strong-lensing efficiency of clusters can be increased substantially and on a short timescale during a major merger \cite{TO04.1}. As a subcluster approaches a cluster, the tidal field is increased, leading to a first maximum of the cross section approximately when the two virial regions touch. The cross section then slightly decreases and approaches a second maximum when the separation of cluster and subcluster is minimal. A third peak corresponding to the first is formed when the subcluster leaves the virial region again after the merger. During that process, the arc cross section can change by an order of magnitude or more on a time scale of $\simeq0.1\,\mathrm{Gyr}$. It thus appears that strong lensing can be a transient phenomenon at least in some clusters which would otherwise be not massive or concentrated enough. The dependence of the main merger epoch on cosmic history would then establish an interesting link between high-redshift, strong cluster lenses and the cosmological framework model.

So far, all effects studied, including baryon cooling in cluster cores \cite{PU05.1, WA08.1} and line-of-sight projection effects \cite{PU09.1}, returned moderate enhancements of the expected arc abundance. The main problem is now that recent measurements converge on a low normalisation parameter $\sigma_8\approx0.8$ for the dark-matter power spectrum, which drastically lowers the expected arc abundance \cite{LI06.1, FE08.1}. Thus, while the strong-lensing properties of individual clusters seem to be sufficiently understood \cite{HO05.1}, the arc-statistics problem persists. It is certainly related to the overconcentration problem and the problem of large Einstein radii \cite{BR08.2} and indicates that there is something fundamental we do not understand in the population of galaxy clusters.

Besides improved methods for predicting arc abundances \cite{FE06.1}, selection biases in existing samples of strongly lensing clusters must be understood \cite{FE07.1} and the sizes of observed arc samples need to be enlarged \cite{HE08.1} before further progress will be made. Automatic search algorithms for arcs \cite{LE04.2, HO05.1, SE07.1, CA07.1} calibrated on simulated images \cite{ME08.1} and applied to upcoming wide-field surveys will improve the situation in the near future.

\subsection{Other applications of strong cluster lensing}

If a cluster produces arcs from multiple sources at different redshifts, the lensing mass distribution remains the same, but the geometrical lensing efficiency is different for the arcs. Since this depends on cosmological parameters, these can thus be purely geometrically constrained from multiple-arc systems \cite{LI98.1, GA00.1}. \cite{SO04.1} applied this technique to multiple arcs in the cluster A~2218 and found that a universe with critical matter density and no cosmological constant is excluded at $>4\,\sigma$ confidence from this single cluster. In principle, the abundance of arcs in clusters and the geometry of multiple arc systems in individual clusters is sensitive to the possible time evolution of the dark energy, but in view of the uncertainties in cluster mass models, substantive conclusions will be hard to arrive at \cite{ME05.4, ME05.3, MA05.4}.

For alleviating potential problems e.g.~with the abundance of satellite galaxies, it was proposed that the dark-matter particles might interact with each other in another way than through gravity. Such a self-interaction would act as a source of isotropic pressure and thus symmetrise and smooth mass distributions \cite{MI02.1}. Strong gravitational lensing, being very sensitive to cluster asymmetries, places a tight limit on the interaction cross section. Using numerical simulations, \cite{ME01.1} showed that the strong-lensing efficiency of galaxy clusters would abruptly disappear if the specific self-interaction cross section was $\gtrsim0.1\,\mathrm{cm^2\,g^{-1}}$.

Finally, lensing clusters are frequently being used as cosmic telescopes, magnifying distant sources above the limits for photometry or spectroscopy. To give a few examples, \cite{CA96.1} used the magnification by A~370 to detect CO lines in arc sources. \cite{FR98.1} and \cite{PE99.1} identified sources at $z=4.04$ lensed by A~2390, \cite{KN04.1} found an object with $z\simeq7$ lensed by A~2218, and \cite{PE04.1} claim to have detected an object with $z=10.0$ magnified by A~1835. The magnification by galaxy clusters as cosmic telescopes has recently been used in systematic searches for galaxies at very high redshift \cite{ST07.1, SM07.1, RI08.2, BR09.1} and for increasing the resolution in detailed studies of distant galaxies \cite{SW06.1, SW07.1}.

\subsection{Weak cluster lensing}

Apart from the occasional spectacular strong-lensing effects, clusters imprint a coherent weak distortion pattern onto the many faint and distant galaxies in their background. Those galaxies reach number densities of $\simeq40$ per square arc minute in typical images taken with large ground-based telescopes. The virial region of a typical galaxy cluster thus covers of order $10^3$ galaxies. Due to their intrinsically irregular shapes, lensing-induced distortions cannot be inferred from individual galaxies. Averaging over a few galaxies, however, the intrinsic ellipticities should average to zero, leaving the ellipticity caused by the gravitational shear as the average signal.

As shown in the introduction, shear and convergence are both related through the scalar lensing potential. Knowing the shear thus allows the scaled surface-mass density to be reconstructed. \cite{KA93.1} were the first to show that cluster convergence maps could be obtained by convolving the measured shear signal with a simple kernel, opening the way to systematic, parameter-free, two-dimensional cluster studies. Their technique was immediately applied to the cluster MS~1224, for which \cite{FA94.1} found a surprisingly high mass-to-light ratio of $\simeq800\,h\,M_\odot/L_\odot$ in solar units, about a factor of four times the typical cluster value.

Weaknesses in the convolution algorithm by \cite{KA93.1}, such as its limitation to formally infinite data fields and weak shear, were discussed and removed by, e.g.~\cite{SE95.1, SE96.1}. Another technique for recovering cluster mass maps based on a maximum-likelihood approach was proposed by \cite{BA96.3}, later augmented with maximum-entropy regularisation \cite{SE98.1} and further developed by \cite{MA02.1}. Based on this approach, algorithms have been developed for combining constraints from weak and strong gravitational lensing into unique, non-parametric reconstructions of cluster mass distributions \cite{BR05.2, BR06.1, CA06.1, ME09.2}.

An algorithm for measuring the weak shear signal from data fields was first described and implemented by \cite{KA95.1}. The Shear Testing Programme (STEP, \cite{HE06.1, MA07.1}) was launched to test and improve the accuracy of shear measurements from weakly distorted images of distant galaxies in the presence of several perturbing effects.

\begin{figure}[ht]
  \hfill\includegraphics[width=0.7\hsize]{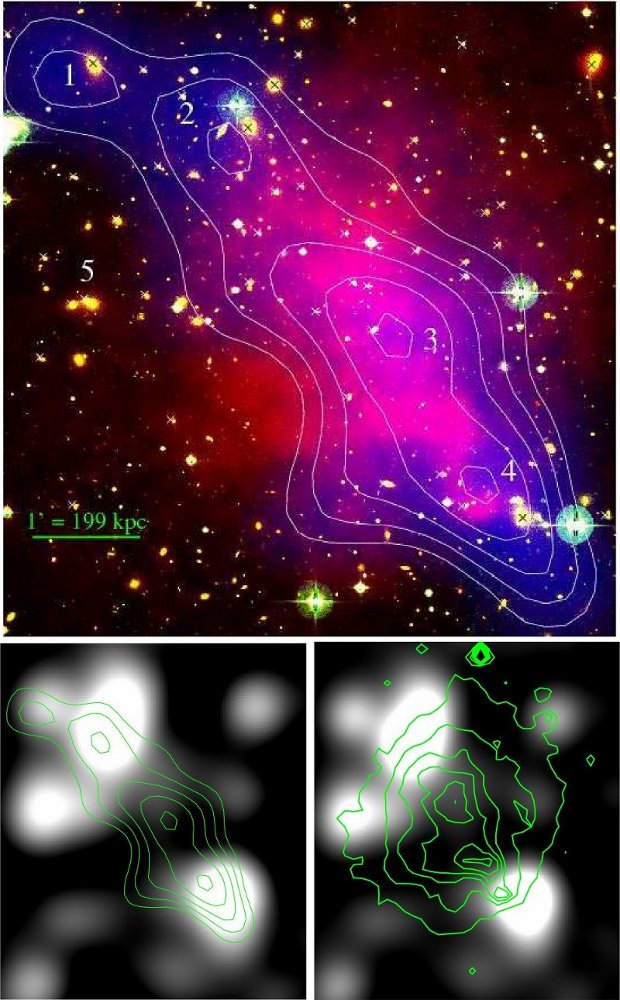}
\caption{The galaxy cluster A~520 (the cosmic train wreck), in which the galaxies are seen displaced from the dark matter (lower left panel) and the X-ray gas (lower right panel). In the top panel, the lensing signal (blue) and its contour lines are superposed on the X-ray image (red) and the cluster galaxies (orange). (from \cite{MA07.2})}
\label{fig:12}
\end{figure}

These inversion techniques for cluster lenses have by now been applied to numerous clusters, far too many to mention them all here. For most of them, the mass-to-light ratios turned out to be quite normal, i.e.~$M/L\simeq250\ldots300$ in blue and $M/L\simeq150\ldots200$ in red colour bands, respectively. Some examples are \cite{CL98.1, HO02.2} and \cite{GA04.1}. Statistical analyses of large samples of galaxy groups and clusters have allowed extending the mass range in which mass-to-light ratios can be probed. Values of $M/L=185\pm28$ \cite{PA05.1} in blue and $M/L\simeq250$ \cite{LI09.1} in red spectral ranges have been obtained for galaxy groups with masses around $(10^{13}\ldots10^{14})\,M_\odot$. Cross-correlating the weak-lensing signal around a sample of $\simeq130,000$ brightest cluster galaxies from the Sloan digital Sky Survey with red light, \cite{SH09.1, SH09.2} found an average mass-to-light ratio of $M/L=(362\pm54)\,b^2$, where the bias factor $b$ should be of order unity. Mass and light generally appear well correlated in weak-lensing clusters. The finding that mass followed light well in two of the three clusters in the A~901/902 super-cluster field while the third cluster showed a significant offset between the mass and the light \cite{GR02.1} could not be confirmed by a follow-up study with HST data \cite{HE08.2}. A very peculiar case is the cluster A~520, where the galaxies are seen displaced from the coincident dark-matter distribution and X-ray emission \cite{MA07.2}, see Fig.~\ref{fig:12}.

\begin{figure}[ht]
  \hfill\includegraphics[width=\hsize]{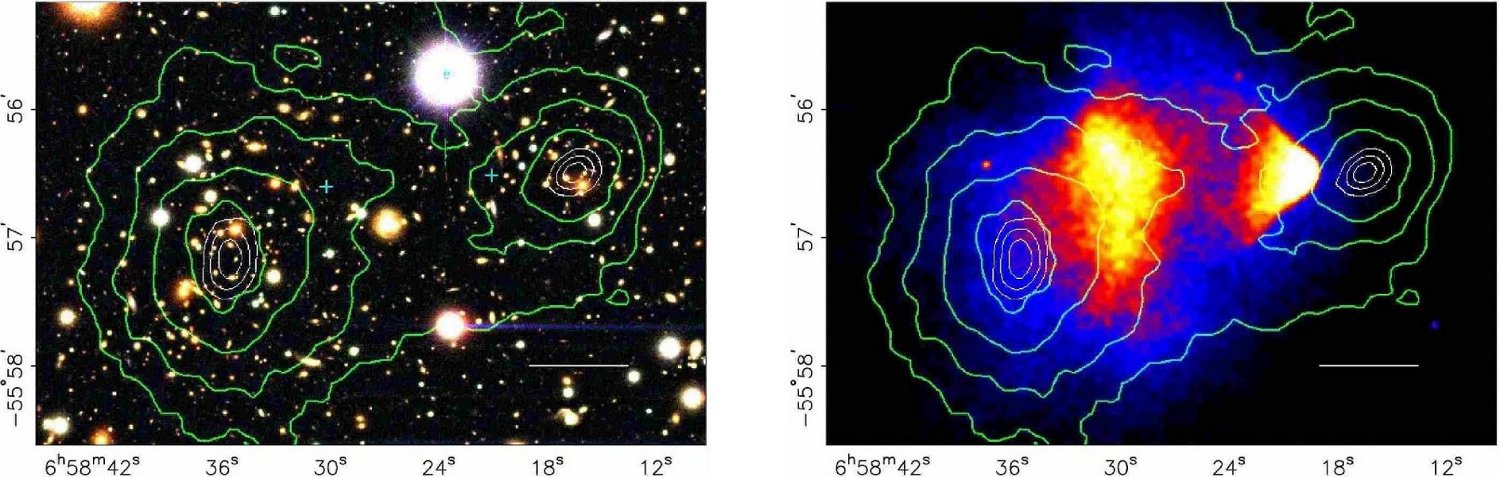}
\caption{The galaxy cluster 1E~0657$-$558 (the bullet cluster) whose galaxies and dark matter (left panel) are displaced from the X-ray gas (right panel). (from \cite{CL06.1})}
\label{fig:13}
\end{figure}

Interesting phenomena appear in comparisons between the X-ray surface-brightness and the weak-lensing mass contours. While the X-ray emission follows the mass in many clusters (see \cite{GI99.1, CL00.1, HO00.1, CL02.1} for examples), instructive deviations have been discovered. \cite{MA02.2} find good agreement between surface-density and X-ray contours in the outer parts of A~2218, but deviations near the cluster centre, which they interpret as a sign of dynamical activity in the cluster. Several recent studies find the X-ray gas lagging behind the dark matter in merging clusters \cite{CL04.1, MA04.3, JE05.1}, as expected for hot gas embedded into collision-less dark-matter halos. A particularly interesting case is the cluster 1E~0657$-$558, called the bullet cluster, whose X-ray emission appears in between two galaxy concentrations and dark-matter distributions recovered from weak lensing \cite{CL04.2, CL06.1}, see Fig.~\ref{fig:13}. Its appearance suggests the interpretation that two clusters have lost their gas by friction while passing each other in the course of a merger. Other such clusters have since been found \cite{BR08.4}.

If the dark-matter particles interacted with each other, such a separation between gas and dark matter would be suppressed. Thus, from gas lagging behind the dark matter in merging clusters, and from small dark-matter core radii, limits could be obtained for the self-interaction cross section of the dark-matter particles, typically finding values $\lesssim(0.1\ldots1)\,\mathrm{cm^2\,g^{-1}}$ \cite{AR02.1, NA02.2, MA04.3}, comparable to what \cite{ME01.1} concluded from strong cluster lensing. Clusters like the bullet cluster have also been used to constrain the specific cross section for the self-interaction of dark-matter particles to $\sigma/m\lesssim1\,\mathrm{cm^2\,g^{-1}}$ \cite{RA08.1}.

Although different mass estimates agree well in some clusters (e.g.~\cite{CL02.1, IR02.1, JI04.1, JE05.1, MA05.1, HO07.1, ZH08.1}), significant discrepancies between cluster masses derived from weak lensing and X-ray observations are often found \cite{LE99.1, AT02.1, HO02.3, PR05.1, GI07.1} and interpreted as signalling dynamical processes in unrelaxed cluster cores. Of the 38 clusters in the X-ray selected sample studied by \cite{DA02.2}, $\simeq30\%$ show signs of dynamical activity, and more than $50\%$ are strong lenses. Based on a sample of 24 clusters between redshifts $0.05$ and $0.31$, \cite{CY04.1} claim that clusters with temperatures $\lesssim8\,\mathrm{keV}$ show good agreement between different mass estimates, while hotter clusters do not. Spectacular examples of massive, distant clusters are presented in \cite{LO05.1, JE09.1}. Despite their youth, the agreement between X-ray and weak-lensing mass estimates hints at established clusters.

Projected cluster mass profiles obtained from weak lensing are often well fit by the isothermal profile \cite{SH01.1} or by both the isothermal and the NFW mass profiles \cite{CL00.1, CL01.1, AT02.1}. As data have improved and samples were enlarged, the NFW profile was most often found to describe the lensing data very well \cite{CL01.1, MA06.1}. When based on weak lensing alone, NFW concentration parameters tend to be somewhat lower than theoretically expected \cite{CL02.1, HO02.2, JE05.1, MA08.1}, which may be due to intrinsically triaxial cluster halos \cite{CL02.1}. However, there is an increasing number of clusters for which NFW profiles with reasonably high concentration parameters are deduced (e.g.~\cite{CL01.1, AR02.1}). \cite{CL00.1} find the more massive of six high-redshift clusters less concentrated than the less massive ones, which is also expected from theory. \cite{DA03.1} fit the generalised NFW profile to six massive clusters at $z\simeq0.3$, finding a central double-logarithmic slope $\alpha=-0.9\ldots-1.6$ at 68\% confidence. Assuming $\alpha=-1$, the concentration parameters are well in the expected range, i.e.~$5\ldots10$ depending on cluster mass. The overconcentration problem mentioned earlier persists in many cases when constraints from strong and weak gravitational lensing are combined.

Large-scale structure in front of and behind galaxy clusters is projected onto them and can affect weak-lensing mass determinations. Using large-scale structure simulations, \cite{ME99.1} estimate that weak-lensing mass estimates exceed real cluster masses by several tens of per-cents due to the added large-scale structure. \cite{HO03.1} estimated that projected large-scale structure approximately doubles the error budget for weak-lensing cluster mass estimates. However, cluster mass profiles are affected by cluster substructures and asymmetries only at the per-cent level \cite{KI01.1, CL04.1}.

We have seen in the discussion of strong cluster lensing that clusters at moderate and high redshifts, $z\gtrsim0.8$, are already remarkably efficient strong lenses. The first weak-lensing mass map of a cluster at such high redshift (MS~1054$-$03 at $z=0.83$) was produced by \cite{LU97.1}. The weak-lensing signal of many similarly distant clusters was measured since, typically confirming the presence of well-developed, massive and compact clusters at that epoch \cite{CL98.1, GI99.1, LO05.1, MA05.1, JE09.1}, but also frequently indicating violent dynamical activity in cluster cores \cite{HO00.1, HO02.3, JE05.1}.

Occasionally, detections of clusters with very high mass-to-light ratios (e.g.~\cite{FA94.1, FI99.1}) are claimed and raise the question whether cluster-sized dark-matter halos may exist which are so inefficient in producing stellar or X-ray emission that they are invisible for anything but gravitational lensing. \cite{ER00.1} detected a peak in the weak-lensing signal 7 arc minutes south of the cluster A~1942 where no optical or infrared emission could be found \cite{GR01.1}. A more recent analysis \cite{LI06.2} revealed a discrepancy between ground- and space-based data which remains unresolved. A similarly dark weak-lensing signal peak was discovered next to the high-redshift cluster Cl~1604$+$4304 by \cite{UM00.1}. However, another tangential shear alignment potentially revealing a dark halo \cite{MI02.2} was meanwhile found to be spurious \cite{ER03.1}, and a new analysis of the A~901/902 supercluster by \cite{HE08.2} did not confirm the anomalous mass-to-light ratio in one of the three matter concentrations found by \cite{GR02.1}.

\cite{SC96.1} introduced the aperture-mass statistic specifically for detecting dark-matter halos through their weak-lensing signal. The aperture mass is a weighted integral within a circular aperture over the shear component tangentially oriented with respect to the aperture centre. When applied to numerical simulations, the aperture-mass statistic turned out to be highly efficient in finding group- and cluster-sized halos, although the completeness of the resulting halo catalogues has to be balanced against the frequency of spurious detections by carefully choosing the signal-to-noise threshold \cite{RE99.1, WH02.1}. Spurious detections caused by projections of large-scale structures are frequent, but can be substantially suppressed by optimal filtering \cite{MA05.2, PA07.2}. \cite{WI01.2} report the first detection of a galaxy cluster through weak lensing, which was confirmed later through its optical signal. \cite{SC04.2} use the aperture-mass technique for confirming the weak-lensing signal of clusters found optically in the ESO Imaging Survey. Recent applications of cluster-detection techniques based on weak lensing \cite{HE05.1, WI06.1, DI07.1, GA07.2, MI07.1, SC07.1} have returned catalogues with varying degrees of contamination by spurious detections. Their statistical analysis is likely to provide important information on the evolution of non-linear cosmic structures in the near future \cite{MA09.3, DI10.1}.

\cite{BA01.1} showed that the detection efficiency of the aperture-mass technique varies strongly with the density profile of the dark-matter halos, allowing a statistical discrimination between isothermal and NFW profiles. \cite{MI02.3} found $4.9\pm2.3$ dark-matter halos in a field of 2.1 square degrees taken with the Subaru telescope, which is consistent with expectations based on CDM models and NFW density profiles \cite{KR99.1}.

As mentioned before, cosmological models with dynamical dark energy cause dark-matter halos to be more concentrated compared to models with cosmological constant. While this should in principle lead to a higher number of weak-lensing halo detections in dark-energy cosmologies and thereby provide a way for discriminating cosmological-constant from dark-energy models, the expected sensitivity is very weak due to competing effects \cite{BA02.1, WE03.1}. \cite{WE02.1} argue that clusters in formation, which are not virialised yet and thus under-luminous, may be detected through weak lensing. They suggest this as an explanation for the potential dark clusters found by \cite{ER00.1} and \cite{UM00.1} and argue that cosmological constraints could be placed by comparing the numbers of visible and dark clusters.

\section{Large-scale structures}

Weak gravitational lensing by large-scale structures is covered by several dedicated reviews, highlighting different aspects of this rich and quickly developing subject \cite{ME99.2, BA01.2, HO02.5, ME02.4, RE03.1, HO08.1, MU08.1}. We can only summarise the most important aspects here and refer the interested reader to those reviews for further detail.

\subsection{Expectations and measurements}

Being inhomogeneously distributed in the Universe, matter on scales even larger than galaxy clusters must also gravitationally lens background sources. Early studies \cite{BL91.1, MI91.1, KA92.2} calculated the ellipticities and ellipticity correlations expected to be imprinted on the images of background galaxies, and found them to be of order a few per cent on arc-minute angular scales. In a first attempt at measuring this weak cosmological lensing signal, \cite{MO94.1} could place an upper limit in agreement with theoretical expectations.

Since weak cosmological lensing is highly sensitive to the non-linear evolution of the large-scale structures \cite{JA97.1}, numerical simulations had to be carried out for precisely estimating the amplitude of the signal and the shape of the ellipticity correlation function (e.g.~\cite{BA92.1, JA00.1, HA01.1, VA03.1, HI09.1}). The cosmological potential of large weak-lensing surveys was quickly pointed out \cite{BE97.1, KA98.1, WA99.1}, emphasising the possibility of measuring in particular the matter density parameter $\Omega_0$ and the amplitude $\sigma_8$ of the dark-matter power spectrum.

\cite{SC98.1} announced the detection of a coherent shear signal in the field of the radio galaxy PKS~1508$-$05 which they interpreted as being caused by large-scale structure lensing. The breakthrough came soon thereafter, when several different groups almost simultaneously reported the measurement of the cosmic-shear correlation function \cite{BA00.1, WA00.1, WI00.1, MA01.1}. Given the difficulty of the measurement and the different telescopes, cameras, and analysis techniques used, the agreement between these results and their compatibility with theoretical expectations was exciting and encouraging.

Cosmological parameters were soon derived from these first cosmic-shear measurements \cite{WA01.2}, finding $\sigma_8\gtrsim0.7$ and $\Omega_0\lesssim0.4$ for spatially-flat cosmological models. Two-point statistics of the cosmic shear are approximately proportional to the product $\sigma_8\Omega_0^2$, i.e.~they are degenerate in these two parameters. This degeneracy can be lifted using third-order statistics such as the skewness \cite{WA01.1, KI05.1}, which arises because the non-linear evolution of cosmic structures leads to non-Gaussianity in the weak-lensing signal. Non-Gaussianities were first detected by \cite{BE02.1} in the Virmos-Descart weak-lensing survey.

Much effort was devoted to calibrating weak-lensing measurements, to designing optimal cosmic-shear estimators and studying their noise properties. \cite{ER01.1} used numerical simulations to show that relative accuracies of $10\%\ldots15\%$ can be reached by cosmic-shear measurements. A method for estimating the weak-lensing power spectrum inspired by the CMB data analysis was proposed by \cite{HU01.1}. \cite{CO01.1, SE07.2, HA09.1} investigated how non-Gaussianity can affect cosmological parameter estimates from the cosmic-shear power spectrum. Different estimators for the two-point statistics of cosmic shear and their correlation matrices were dicussed by \cite{SC02.3, JO08.1}.

Numerous weak-lensing surveys have meanwhile been conducted. A non-exhaustive selection of the results obtained on $\sigma_8$ for fixed $\Omega_0=0.3$ is given in Tab.~\ref{tab:02}. Although most values of $\sigma_8$ agree within the error bars, the scatter is still substantial. This is at least partially due to remaining systematics in the data analysis, as will now be discussed.

\begin{table}[ht]
\caption{Non-exhaustive selection of results for $\sigma_8$ extracted from weak-lensing surveys. $\Omega_0=0.3$ is adopted throughout. $\sigma_8$ scales with $\Omega_0$ approximately like $\sigma_8\propto\Omega_0^{-0.5}$.}
\label{tab:02}
\begin{center}
\begin{tabular}{|rl|l||rl|l|}
\hline
\multicolumn{2}{|l|}{$\sigma_8$} & reference &
\multicolumn{2}{|l|}{$\sigma_8$} & reference \\
\hline
$1.04$ & $\pm0.05$ & \cite{MA01.1} & $0.86$ & $\pm0.05$ & \cite{SE06.1} \\
$0.81$ & $^{+0.14}_{-0.19}$ & \cite{HO02.4} & $0.85$ & $\pm0.06$ & \cite{HO06.1} \\
$0.97$ & $\pm0.13$ & \cite{BA03.2} & $1.06$ & $^{+0.17}_{-0.16}$ & \cite{KI07.1} \\
$0.72$ & $\pm0.09$ & \cite{BR03.1} & $0.52$ & $^{+0.13}_{-0.17}$ & \cite{SC07.2} \\
$0.97$ & $\pm0.35$ & \cite{HA03.1} & $0.80$ & $\pm0.10$ & \cite{HE07.2} \\
$0.71$ & $^{+0.12}_{-0.16}$ & \cite{JA03.1} & $0.87$ & $^{+0.09}_{-0.07}$ & \cite{MA07.4} \\
$1.02$ & $\pm0.16$ & \cite{RH04.1} & $0.74$ & $\pm0.04$ & \cite{BE07.1} \\
$0.83$ & $\pm0.07$ & \cite{WA05.1} & $0.70$ & $\pm0.04$ & \cite{FU08.1} \\
$1.02$ & $\pm0.15$ & \cite{MA05.3} & $0.65$ & $^{+0.09}_{-0.14}$ & \cite{HA09.1} \\
$0.68$ & $\pm0.13$ & \cite{HE05.2} & & & \\
\hline
\end{tabular}
\end{center}
\end{table}

\begin{figure}[ht]
  \centerline
   {\includegraphics[width=0.49\hsize]{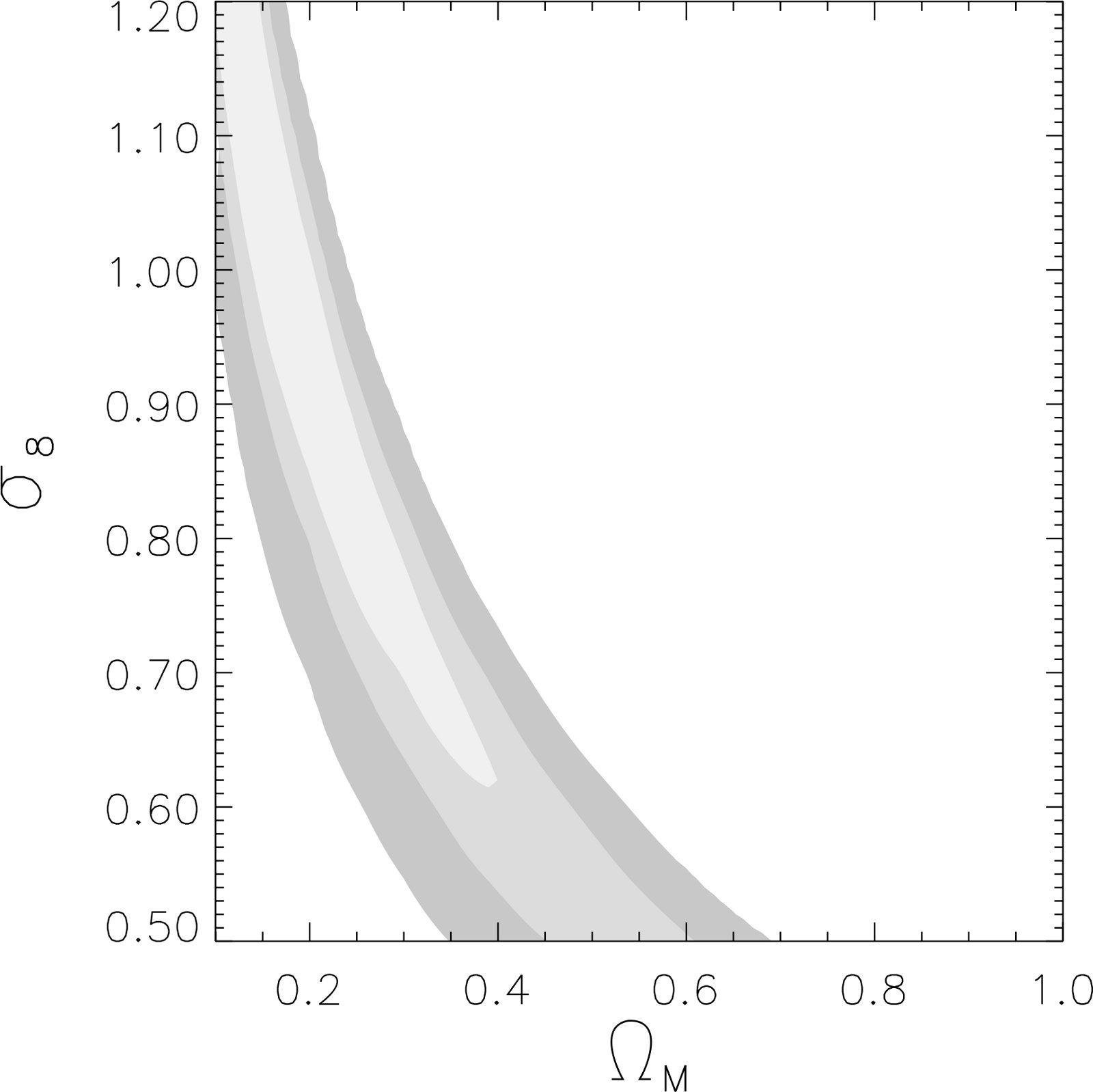}
    \includegraphics[width=0.49\hsize]{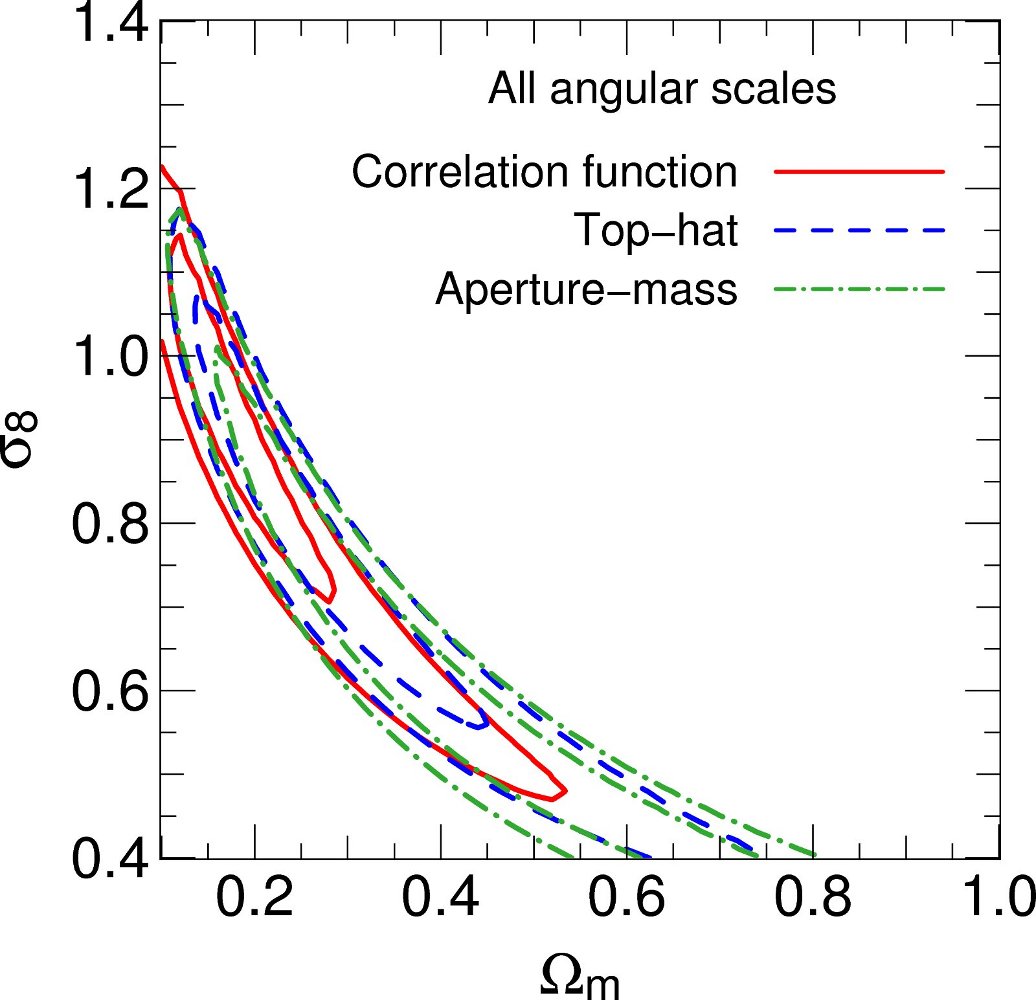}}
\caption{Constraints on two cosmological parameters, the normalisation parameter $\sigma_8$ and the matter-density parameter $\Omega_\mathrm{m0}$, obtained from cosmic-shear correlation functions in two different surveys of gravitational lensing (left panel: from \cite{BE07.1}, right panel: from \cite{FU08.1})}
\label{eq:14}
\end{figure}

\subsection{Systematics}

Numerical simulations show that the Born approximation is valid to very good approximation \cite{DO05.1, SH06.1, HI09.1}. While this is so, the effects of weak lensing can be summarised by a scalar potential. Then, only such distortion patterns can be caused by weak lensing which can be described by derivatives of a scalar potential. This leads to the $E$ and $B$-mode decomposition discussed in the introduction. Significant $B$ modes in the data are interpreted as remainders of undetected or incompletely removed systematics.

More or less significant $B$ modes have been found in almost all weak-lensing surveys. Until recently, their origin was unclear. \cite{SC02.2} showed that source clustering could cause a $B$-mode contribution to the distortion, but not of sufficient strength on arc-minute scales to fully account for the observations. \cite{HO04.2} showed that incomplete correction for the anisotropies in the point-spread function of the imaging system could give rise to a substantial $B$ mode. Application of an improved model for the point-spread function to the Virmos-Descart weak-lensing survey caused the $B$ mode to disappear \cite{WA05.1}. It thus seems that much of the $B$-mode problem, which was discussed at length in the literature, was due to insufficient correction for the distortions imprinted by the imaging system. However, incomplete knowledge of cosmic-shear correlation functions, which are necessarily always limited to finite fields, can also lead to production of $B$ modes \cite{KI06.1}. Methods for removing this $B$-mode contamination were developed by \cite{SC07.3, FU10.1}.

There are (at least) five important sources of systematic error in weak-lensing measurements: distortions by the telescope, miscalibrated distortion measurements, insufficient knowledge of the non-linear power spectrum, insufficient information on the redshift distribution of the background galaxies, and intrinsic galaxy alignments. Extensive literature exists on all of these effects. Distortions by the telescope, summarised as anisotropies of the point-spread function, are typically corrected by measuring the shapes of stars in a data field and fitting the measured distortions by low-order polynomials. More recently, \cite{JA06.1} showed how the correction for anisotropic point-spread functions can be improved. The calibration of shear measurements has been addressed by several large-scale efforts, i.e.~the STEP programme \cite{HE06.1, MA07.1} and the GREAT08 challenge \cite{BR09.2}. Spatially varying calibration errors were addressed by \cite{GU05.1}, and \cite{WA06.1} studied that sampling errors in the redshift distribution have a similar effect as shear calibration errors. Precise knowledge of the non-linear power spectrum is important on angular scales smaller than $\sim10'$ \cite{HI09.1}, where baryons can also change the dark-matter power spectrum noticeably \cite{JI06.1}.

Analyses of cosmic-shear measurements assume that intrinsic galaxy ellipticities are uncorrelated such that they average to zero when several images are combined. However, galaxies being physically close to each other are also expected to have their shapes aligned, e.g.~by the tidal field of the large-scale matter distribution into which they are embedded. The potential effect of intrinsic rather than lensing-induced galaxy alignments depends obviously on the depth of the survey. Deep surveys project galaxy images along light paths which are substantially longer than any large-scale structure correlation scale and thus suppress any spurious signal due to intrinsinc alignments of physically neighbouring galaxies. In shallow surveys, however, intrinsic source alignments may substantially contaminate any weak-shear signal \cite{CR00.1, HE00.1, CA01.1, CR01.1}. Measurements of intrinsic alignments were found to agree well with these predictions \cite{LE02.1, BR02.2}. Recent numerical \cite{JI02.1} and analytic studies \cite{LE04.1} claim strong intrinsic alignments of galaxy halos. Comparing simulations and data, \cite{FA09.1} specified that theoretical predictions based on the orientations of dark-matter haloes overpredict the alignment by a factor of $\sim2$, but agree well with measurements if the central orientations of galaxies placed in these haloes are being used instead.

Possibilities for removing the signal contamination due to intrinsic alignments were discussed extensively. They advocate using photometric redshifts to remove physically close pairs of source galaxies from the analysis \cite{KI02.2, HE03.1}. An application of this technique to the multi-colour Combo-17 survey \cite{HE04.1} showed intrinsic alignments near the lower end of the theoretical predictions and slightly lowered the value of $\sigma_8$ previously derived from these data \cite{BR03.1} to $0.67\pm0.1$ for $\Omega_0=0.3$. \cite{HI04.2} point out that foreground galaxies are aligned with the large-scale structures lensing background galaxies, thus giving rise to a higher-order alignment between galaxies at different redshifts. They quantified this systematic effect \cite{MA06.5, HI07.1} and showed that it can substantially reduce the measured shear signal, leading to a likely underestimate of the $\sigma_8$ parameter by several per cent.

\subsection{Perspectives}

For several years now, cosmological parameters are obtained with high accuracy mainly from combined analyses of CMB data, large-scale galaxy surveys and the dimming of type-Ia supernovae with redshift. What is the role of weak gravitational lensing in this context? It should be emphasised that parameter constraints from the CMB alone suffer from degeneracies which can only be broken using additional information. Measuring directly the dark-matter density and the normalisation of its fluctuations, gravitational lensing adds constraints which substantially narrow the parameter ranges allowed by the CMB, as \cite{HU99.1} pointed out. \cite{CO03.1} combined weak-lensing data from the Red-sequence Cluster Survey with CMB data from the WMAP satellite to find $\sigma_8=0.89\pm0.05$, $\Omega_0=0.30\pm0.03$ and a Hubble constant of $H_0=(70\pm3)\,\mathrm{km\,s^{-1}\,Mpc^{-1}}$. \cite{TE05.1} predicted the accuracy of joint cosmological parameter estimates using CMB data in combination with the weak-lensing constraints from the CFHT Legacy Survey.

The exploitation of higher than second-order statistics will become increasingly important for breaking degeneracies in the weak-lensing parameter estimates and the investigation of non-Gaussianity developing due to non-linear structure growth \cite{ZH03.1, TA03.1, DO04.1}. Much valuable information must also be contained in the morphology of the two-dimensional weak-lensing pattern.

Among the most exciting perspectives of weak lensing is its potential to study the three-dimensional distribution of dark structures. Although lensing observables measure the two-dimensional, projected density distribution, selecting sources at different distances allows structures along the line-of-sight to be resolved. Sufficiently accurate information on the distance to the source galaxies is provided by photometric redshifts. \cite{SI04.1} show that the accuracy of cosmic-shear parameter estimates can be improved by a factor of $5\ldots10$ by splitting the source galaxies into only four redshift bins. \cite{HE03.2} developed an elegant formalism for extracting three-dimensional information from weak-lensing data (see also \cite{PE04.2}), and \cite{TA04.1} recovered the three-dimensional matter distribution in the field of the Combo-17 survey, using photometric redshifts as distance indicators for the source galaxies.

The possibility to extract three-dimensional information from weak-lensing data opens a way for studying the growth of cosmic structures along the line-of-sight from the distant and past universe. This, in turn, is sensitive to the change of the dark-energy density with time. Thus, sufficiently accurate and wide-field weak-lensing surveys will allow the dark energy to be constrained \cite{HU02.1}. The sensitivity of such experiments is very promising \cite{MU03.1, BE04.1}, in particular if higher-order statistics are considered. This is one of the strongest motivations for a weak-lensing survey from space \cite{MA04.4, RE04.1}. The impact of several sources of systematic error on the determination of dark-energy properties is of great interest in this context  \cite{KI05.2, HE06.2, BR07.1, KI08.1}. Enormous progress is expected regarding the control of systematics and the suppression of noise from planned space-based surveys of large fractions of the sky, such as the survey to be undertaken by the proposed DUNE-Euclid satellite mission \cite{RE09.1}.

Planned wide-area surveys of the sky in radio wavebands are expected to yield information on both cosmology and the distribution of neutral hydrogen during reionisation \cite{PE04.3}. Weak lensing in the data of the FIRST radio survey undertaken at 20~cm wavelength was recently detected by \cite{CH04.1}.

\subsection{Cosmic magnification}

Besides distorting the images of distant galaxies, large-scale structures also magnify background sources. To linear order, the power spectrum of cosmic magnification is simply four times the power spectrum of the cosmic shear, i.e.~both contain the same amount of cosmological information. Cosmic shear, however, is much more easily measurable than cosmic magnification because it can reasonably be assumed that the ellipticities of distant galaxies average to zero. The intrinsic flux of any given source being unknown, cosmic magnification is much harder to identify.

Currently the most promising (and perhaps the only) method for detecting cosmic magnification is the magnification bias. If a population of distant sources, e.g.~quasars, is observed within a solid angle $\delta\Omega$ on the sky where the magnification is $\mu$, fainter sources become visible there, but their number density is reduced because the solid angle is stretched by the magnification. The net effect depends on how many more sources the magnification lifts above the flux threshold of the observation. If the number-count function of the sources is sufficiently steep, the dilution is outweighed and the magnification causes more sources to be visible, while sources with flat number counts are diluted and appear less numerous than without the magnification.

The spatial galaxy distribution follows the dark-matter structures which act as lenses on background sources. Bright quasars, for example, have a steep number-count function and thus appear more numerous behind magnifying dark-matter overdensities. The correlation of foreground galaxies with the lensing matter then leads to a positive cross-correlation on angular scales of arc minutes and larger between distant QSOs and foreground galaxies which are physically separated by cosmological distances.

The cross-correlation function between background sources and foreground galaxies was derived in linear approximation by \cite{BA95.3} and \cite{DO97.1}; see also \cite{GU01.1}. However, magnification is non-linear in shear and convergence. Computing the theoretical expectation of the cosmic magnification accurately is thus considerably more complicated than for the cosmic shear. \cite{ME03.3} derived second-order corrections to the linear cosmic-magnification statistics and found that the linear approximation was systematically low by $20\%\ldots30\%$, which was confirmed by \cite{TA03.2} in a fully non-linear treatment based on the halo model of the dark-matter distribution. The accuracy of analytic calculations can be calibrated using numerical simulations of light propagation through large-scale structures (e.g.~\cite{BA03.3, ME03.3}). Sufficiently accurate analytic and numerical calculations of cosmic magnification yield typical magnifications of $\lesssim10\%$ for point sources at redhifts $z_\mathrm{s}\simeq1$. While most emphasis was put on cross-correlations between QSOs and galaxies, cosmic magnification may also induce detectable cross-correlations between background and foreground galaxies \cite{MO98.1, MO98.2}. \cite{JA02.1} discussed survey strategies for detecting cosmic magnification.

\cite{BA93.2} confirmed that the QSO-galaxy correlations detected by \cite{SE79.1} and \cite{FU90.1} can indeed be explained in terms of gravitational lensing by large-scale structures. Subsequent more detailed correlation studies confirmed that the correlations showed the significance variations expected from the lensing hypothesis \cite{BA93.1}, and revealed cross-correlations of distant QSOs with infrared \cite{BA94.3, BA97.1} and X-ray emission \cite{BA94.2}. While the \emph{existence} of QSO-galaxy cross-correlations was thus firmly established, their \emph{amplitude} and \emph{angular scale} was typically found to be substantially too high \cite{WI98.1, NO00.1, BE01.1, GA03.2}, although some analyses obtained the theoretically expected results \cite{RO94.1, NO01.1}. The observational situation was thus utterly confused. \cite{FO96.1} found evidence for cosmic shear in the vicinity of five distant QSOs. Dust absorption in foreground galaxies may be responsible for similar anti-correlations as expected from weak lensing of faint background QSOs (e.g. \cite{CR01.2}), and selection effects may cause correlations as well as anti-correlations (e.g. \cite{FE97.1}). The detailed relation between dark-matter halos and their galaxy content adds further uncertainty \cite{JA03.3}.

Due to the weakness of the signal and the possibility of confusing it with other effects, large multi-colour surveys such as the SDSS were expected to be necessary for an unambiguous detection of cosmic magnification and its exploitation in terms of cosmological parameters \cite{BE99.1, ME02.3}. The degeneracy between the density and bias parameters $\Omega_0$ and $b$ can be broken determining the three-point cross-correlation between QSOs and galaxy pairs \cite{ME03.5}. Clear and unambiguous evidence for cosmic magnification was finally detected by \cite{SC05.1} who measured the cross-correlation between a colour-selected sample of $\simeq200,000$ distant QSOs and foreground galaxies in $\simeq3,800$ square degrees of the SDSS data and found a signal significant at the 8-$\sigma$ level and in complete agreement with theoretical expectations.

A very promising manifestation of cosmic magnification was recently discussed by \cite{ME03.4}. Distant QSOs magnified by intervening matter are more likely to show absorption features in their spectra due to the gas associated with the lensing structures. Using the QSO sample from the 2dF survey, \cite{ME03.4} demonstrated evidence for lensing by MgII and FeII absorbers along the lines-of-sight to the QSOs. This was further studied by \cite{ME05.1} and discovered in the SDSS data by \cite{ME08.2}.

\cite{WA10.1} pointed out that cosmic shear and magnification provide complementary cosmological information. Using deep data of the CFHLS, \cite{HI09.2} detected the effect of cosmic magnification by cross-correlating distant, so-called Lyman-break galaxies with foreground galaxies.

\subsection{Gravitational lensing of the CMB}

This subject has been comprehensively reviewed by \cite{LE06.1}. Pioneering studies \cite{CA93.1, CA93.2, SE96.2, SE96.3, CA97.1} pointed out that the CMB is expected to be weakly gravitationally lensed at a measureable level and developed methods for computing the effect of lensing on the CMB power spectra. Lensing of the CMB can be seen as a diffusion process whereby structures are blurred, but also created on small angular scales in the Silk-damping tail \cite{ME97.1}. The effect of lensing on the CMB cannot be identified in the CMB power spectrum, but in higher-order statistical measures \cite{BE97.2, BE98.1}. Distortions of the probability distribution function of CMB temperature fluctuations, higher-order correlations, quadratic filtering and maximum-likelihood approaches were developed to identify lensing in the CMB at the statistical level and in two-dimensional reconstructions \cite{GU00.1, HU01.2, KE02.2, HI03.1, HI03.2}. It was pointed out that lensing of the CMB must be treated in spherical geometry on the full sky \cite{HU00.1, CH02.3, OK03.1, CH05.1} if percent-level accuracy is to be achieved. Since lensing modifies the CMB power spectrum from which cosmological parameters are derived at high precision, undiscovered or uncorrected lensing leads to biases in cosmological parameters \cite{LE05.1}. On the other hand, lensing can break cosmological parameter degeneracies in the CMB data \cite{ST99.1}. Gravitational lensing of the CMB by fully non-linear cosmological structures was simulated on the full sky by \cite{CA08.1, CA09.2}, who showed how even large-scale $B$-mode CMB polarisation can be created by small-scale non-linear structures. Simulations were also used to study how well CMB lensing can be recovered in presence of systematic effects \cite{PE09.1}. So far, there are no published direct detections of CMB lensing, but there is indirect evidence with significance around 3-$\sigma$ obtained by cross-correlating the CMB with different samples of distant foreground sources used for tracing the potentially lensing large-scale structure \cite{SM07.2, HI08.1}. Direct evidence for CMB lensing should soon be discovered in the data of the Planck CMB mission.

\section{Summary\label{sec:6}}

Many are the applications of gravitational lensing to cosmology, and the results are numerous, as the preceding discussion has shown. A review like this must be based on a subjective selection which is necessarily biased to some degree. Within these limitations, I summarise the results as follows:

\begin{itemize}

\item Microlensing experiments in the Galaxy have shown that, although massive compact objects exist in its halo, they are insufficient to make up all the dark matter in the Galactic halo. These studies have extended towards the Andromeda galaxy M~31. It is not clear yet what fraction of the observed microlensing must be attributed to self-lensing by the visible stars. Low-mass planets have been detected by means of microlensing.

\item Central density profiles of lensing galaxies are well described as isothermal within the radial range where they produce multiple images. Their cores are thus more concentrated than CDM predicts. This indicates that galaxy density profiles have been steepened by baryonic physics. At larger radii, weak galaxy-galaxy lensing shows that the isothermal density profiles steepen and approach the NFW density profile shape.

\item Galaxy-galaxy lensing finds large halo sizes with radii of $\gtrsim200\,h^{-1}\,\mathrm{kpc}$. Halos of cluster galaxies seem to be smaller, as expected. The biasing of galaxies relative to the dark-matter distribution is found by galaxy-galaxy lensing to be almost scale-independent, or gently increasing with scale.

\item Galaxies have to be structured in order to explain multiple-image geometries and the high fraction of quadruple compared to double images. Anomalous flux ratios of quadruple images seem to be best explained by lensing, but simulations show that the expected level of substructure is insufficient to explain the observed anomalies.

\item Measured time delays between multiple images lead to an interesting conflict between the lensing mass distribution and the Hubble constant: Isothermal profiles yield Hubble constants which are substantially too low, and lens models giving compatible Hubble constants have too steep mass profiles. It seems that this conflict can be resolved allowing perturbations of the density profiles.

\item The statistics of distant sources multiply imaged by galaxies is sensitive to the cosmological parameters. Recent applications of this method showed agreement with a low-density universe with cosmological constant.

\item Galaxy clusters have to be asymmetric, and they must be dominated by dark matter which is more broadly distributed than the cluster light. Cores in the dark-matter distribution must be small or absent. Frequent and substantial discrepancies between lensing and X-ray mass determinations are most likely signalling violent dynamical activity in clusters.

\item It seems that galaxy clusters in the ``concordance'', low-density spatially-flat cosmological models cannot explain the observed abundance of gravitational arcs. Clusters need to be highly substructured and asymmetric, and their dynamics temporarily boosts their strong-lensing efficiency. Yet, theoretical expectations fall substantially below extrapolations of the observed number of arcs. Surprisingly massive and compact clusters which are significant weak and powerful strong lenses exist at redshifts $z\simeq0.8$ and above.

\item Although cluster density profiles inferred from strong and weak lensing do typically not contradict expectations from CDM, isothermal density profiles are not ruled out by strong gravitational lensing. Claims of flat central profiles are not supported by reasonably asymmetric models.

\item Typical mass-to-light ratios derived from weak cluster lensing range around $\simeq200$ in solar units, but very high values have occasionally been found. While this may indicate a separation of gas from dark matter in cluster mergers, the possible existence of dark clusters is intriguing.

\item Cosmic shear, i.e.~the distortion of background-galaxy images due to weak gravitational lensing by large-scale structures, has been detected and found to be in remarkable agreement with theoretical expectations. It has enabled constraints on the matter-density parameter and the normalisation parameter $\sigma_8$ of the dark-matter fluctuations.

\item Systematic effects such as image distortions in the telescope, calibration errors on shape measurements, insufficient knowledge of the non-linear matter power spectrum and the redshift distribution of background galaxies and possible intrinsic alignments of source galaxies are important and substantial and need to be carefully corrected.

\item Joint analyses of weak lensing and CMB data allow parameter degeneracies in both types of experiment to be lifted. When combined with photometric redshifts of source galaxies, three-dimensional reconstructions of the large-scale matter distribution become possible. This will also allow constraints on the dark energy.

\item Cosmic magnification, which is more complicated to measure than cosmic shear, can be quantified by the magnification bias. It has been detected, and most recent measurements are also in excellent agreement with theoretical expectations.

\item Gravitational lensing of the Cosmic Microwave Background is inevitable and affects cosmological parameter estimates obtained from the CMB at the per cent level, if uncorrected. It broadens the peaks in the CMB power spectra, creates small-scale temperature fluctuations in the Silk damping tail and converts part of the $E$-mode polarisation of the CMB into $B$ modes. CMB lensing has been marginally detected at the expected level by cross-correlating the CMB with distant foreground sources.

\end{itemize}

\section*{Acknowledgements}

I am most grateful to discussions with numerous colleagues who helped improving this review substantially. In particular Andrea Macciò, Robert Schmidt, Peter Schneider and Dominique Sluse gave detailed and most appreciated comments.

\appendix

\section{Light deflection in Newtonian approximation}

Usually, the equations of gravitational lensing are derived from the deflection angle of a point mass in Newtonian approximation. To establish the link between this derivation and that based on the equation of geodesic deviation given above, we show here two alternative derivations for the Newtonian deflection angle. One is based on the geodesic equation, the other on Fermat's principle.

\subsection{Geodesic equation}

The geodesic equation asserts that the null wave vector $k$ is parallel transported along the light ray,
\begin{equation}
  \nabla_kk=0\quad\Rightarrow\quad k(k^\alpha)+k^\beta\omega^\alpha_\beta(k)=0\;.
\label{eq:a1-1}
\end{equation}
For evaluating it in the Newtonian metric (\ref{eq:01-28}), we conveniently return from the dual basis (\ref{eq:01-29}) to the coordinate basis. Then, $k=\omega(1, \vec e)$, and we find
\begin{eqnarray}
  \left(\partial_\eta+e^i\partial_i\right)\omega+\omega e^j\phi_j&=&0\;,\nonumber\\
  \left(\partial_\eta+e^i\partial_i\right)e^j&=&-2\left[\phi^j-e^j(e^i\phi_i)\right]
\label{eq:a1-2}
\end{eqnarray}
for the time and space components of the geodesic equation, respectively. The second Eq.~(\ref{eq:a1-2}) can be re-written in the conventional way
\begin{equation}
  \d_\eta e^i=-2\partial^i\phi\;.
\label{eq:a1-3}
\end{equation}
Identifying the tangent vector $e^i$ with
\begin{equation}
  e^i=\frac{\d x^i}{\d\eta}
\label{eq:a1-4}
\end{equation} 
Eq.~(\ref{eq:a1-3}) already reproduces Eq.~(\ref{eq:01-40}).

\subsection{Fermat's principle}

A static metric can be written in the form
\begin{equation}
  \d s^2=g_{00}c^2\d t^2+g_{ij}\d x^i\d x^j\;,
\label{eq:a1-5}
\end{equation}
where the $g_{00}$ and $g_{ij}$ are independent of time. Let $x^\mu(\lambda)$ be a light path parameterised by the affine parameter $\lambda$ and $\dot x^\mu=\d x^\mu/\d\lambda$, then
\begin{equation}
  \left\langle\dot x, \dot x\right\rangle=0
\label{eq:a1-6}
\end{equation}
and the light paths are given by the variational principle
\begin{equation}
  \delta\int_{\lambda_1}^{\lambda_2}\left\langle\dot x, \dot x\right\rangle\d\lambda=0\;.
\label{eq:a1-7}
\end{equation}
In this stationary space-time, we can choose the affine parameter such that
\begin{equation}
  g_{00}\frac{\d(ct)}{\d\lambda}=-1\;.
\label{eq:a1-8}
\end{equation} 
If we carry out the variational principle giving up the condition that $\delta t=0$ at the end points of the paths,
\begin{equation}
  0=\delta\int_{\lambda_1}^{\lambda_2}\left\langle\dot x, \dot x\right\rangle\d\lambda=
  -2\delta t\,\Big|_{\lambda_1}^{\lambda_2}=-2\delta\int_{\lambda_1}^{\lambda_2}\d t\;.
\label{eq:a1-9}
\end{equation}
Since
\begin{equation}
  -g_{00}c^2\d t^2=g_{ij}\d x^i\d x^j=\d\sigma^2
\label{eq:a1-10}
\end{equation}
for light rays, Eq.~(\ref{eq:a1-5}) implies
\begin{equation}
  \delta\int_{\lambda_1}^{\lambda_2}\frac{\d\sigma}{\sqrt{-g_{00}}}=0\;.
\label{eq:a1-11}
\end{equation}
This gives Fermat's principle for a static space-time in general relativity. For the Newtonian metric Eq.~(\ref{eq:01-28}),
\begin{equation}
  -g_{00}=1+2\phi\;,\quad\d\sigma=\left(1-2\phi\right)^{1/2}\left|\dot{\vec w}\,\right|\,\d\lambda\;,
\label{eq:a1-12}
\end{equation}
such that
\begin{equation}
  \delta\int_{\lambda_1}^{\lambda_2}\sqrt{\frac{1-2\phi}{1+2\phi}}\left|\dot{\vec w}\,\right|\d\lambda=
  \delta\int_{\lambda_1}^{\lambda_2}n(\vec w)\left|\dot{\vec w}\,\right|\d \lambda=0\;,
\label{eq:a1-13}
\end{equation}
where
\begin{equation}
  n(\vec w)=\sqrt{\frac{1-2\phi}{1+2\phi}}\approx1-2\phi
\label{eq:a1-14}
\end{equation}
is the effective index of refraction. The Euler-Lagrange equation of the variational principle (\ref{eq:a1-13}) is
\begin{equation}
  n\frac{\d\vec e}{\d\lambda}=\vec\nabla n-\left(\vec e\cdot\vec\nabla n\right)\vec e=\vec\nabla_\perp n
\label{eq:a1-15}
\end{equation}
if we scale the affine parameter $\lambda$ such that $\dot{\vec w}=\vec e$ has unit length, $\vec e^2=1$. Writing
\begin{equation}
  \frac{1}{n}\vec\nabla_\perp n=\vec\nabla_\perp\ln n=\vec\nabla_\perp\ln(1-2\phi)\approx-2\vec\nabla_\perp\phi\;,
\label{eq:a1-16}
\end{equation} 
we reproduce the equation of motion
\begin{equation}
  \d_\lambda\vec e=-2\vec\nabla_\perp\phi
\label{eq:a1-17}
\end{equation}
derived above directly from the geodesic equation. Note that, to first order in $\phi$, we can identify $\d\vec e/\d\lambda$ with $\d\vec e/\d\eta$; see Eq.~(\ref{eq:a1-3}).

\subsection{Deflection angle of a point mass}

For a point mass $M$ at the coordinate origin,
\begin{equation}
  \phi=\frac{GM}{c^2r}=\frac{R_\mathrm{s}}{2r}\;,
\label{eq:a1-18}
\end{equation} 
with the Schwarzschild radius $R_\mathrm{s}$. Anticipating a small deflection, we can use Born's approximation, integrate $\vec\nabla_\perp\phi$ along the unperturbed, straight light path and thus write Eqs.~(\ref{eq:a1-3}) or (\ref{eq:a1-17}) in the form
\begin{equation}
  \frac{\d\vec e}{\d z}=-\frac{R_\mathrm{s}}{r^3}\cvector{x\\y}=
  \frac{R_\mathrm{s}}{\left(x^2+y^2+z^2\right)^{3/2}}\cvector{x\\y}\;,
\label{eq:a1-19}
\end{equation}
if we introduce the spatial coordinate system such that the light ray propagates into the $\vec e_z$ direction. Integration over $z$ gives the deflection angle
\begin{equation}
  \vec\alpha=-\frac{2R_\mathrm{s}}{b^2}\vec b\;,\quad\vec b=\cvector{x\\y}\;.
\label{eq:a1-20}
\end{equation}
Since the field equation is linear in the Newtonian limit, the deflection angle of more complicated mass distributions is a linear superposition of point-mass deflection angles.

\end{document}